\documentclass[12pt]{article}
\usepackage{titling}
\usepackage{setspace} 
\usepackage[flushmargin]{footmisc} 

\renewcommand{\footnotelayout}{\doublespacing}

\usepackage{titlesec}
\titlespacing*{\section}
{0pt}{2.5ex}{0.3ex}
\titlespacing*{\subsection}
{0pt}{1.5ex}{0.3ex}
\titlespacing*{\subsubsection}
{0pt}{1.5ex}{0.3ex}

\usepackage{outlines}
\usepackage{url}
\usepackage{graphicx}
\usepackage{placeins}
\usepackage{float}
\usepackage{extpfeil}
\usepackage{mathtools}
\usepackage{soul}
\usepackage{eurosym}
\usepackage{natbib}
\usepackage{rotating}
\usepackage{ltablex}
\usepackage[para]{threeparttable}
\usepackage{multirow}
\setcitestyle{authoryear,open={(},close={)}}
\usepackage{geometry}

\usepackage{comment}
\usepackage[colorlinks,citecolor=blue,filecolor=black,linkcolor=blue,urlcolor=blue]{hyperref} 
\usepackage{pdflscape}
\makeatletter
\usepackage[toc,page]{appendix}
\usepackage{booktabs}

\makeatother
\usepackage{array}
\newcolumntype{C}[1]{>{\centering\arraybackslash\hspace{0pt}}p{#1}}
\usepackage{booktabs}
\usepackage{fancyhdr}
\usepackage{caption}
\usepackage{subcaption}

\usepackage{color}
\usepackage{xcolor}
\definecolor{green}{HTML}{008069}
\definecolor{greenpastel}{HTML}{B8D8BE}
\hypersetup{colorlinks,linkcolor={green},citecolor={green},urlcolor={green}}

\usepackage{threeparttable}
\usepackage{epigraph}
\usepackage{changepage}
\usepackage{fancyhdr}
\usepackage{centernot}
\pretitle{\begin{flushleft}\large}
\posttitle{\end{flushleft}}
\preauthor{\begin{flushleft}}
\postauthor{\end{flushleft}}
\predate{\begin{flushleft}\setlength{\parskip}{0ex}}
\postdate{\end{flushleft}}
\usepackage{color}
\usepackage{xcolor}
  {\end{tabular}\par\medskip}
\usepackage{booktabs}
\usepackage{array} 
\newcolumntype{M}[1]{>{\centering\arraybackslash}m{#1}}

\usepackage{mathptmx}
\usepackage{newtxtext,newtxmath}

\usepackage{cleveref}

\newcommand{\tabfnhighschooltypeall}{\setstretch{1} \textbf{Note:} Standard deviation in brackets. This table shows the proportion and characteristics of students enrolling into the three types of high school. The measure of Socio-Economic Status is a continuous variable with higher values indicating a higher socio-economic background. See Section \ref{subsec:overview_schoolsystem} for details. Note that the total number of observations for the 2018 and 2019 cohorts is 997,555 but data about high-school track is missing for 7,368 students.}

\newcommand{\tabfnhighschooltyperestricted}{\setstretch{1} \textbf{Note:} Standard deviation in brackets. This table shows the proportion and characteristics of students enrolling into the three types of high school in the restricted sample. The measure of Socio-Economic Status is a continuous variable with higher values indicating a higher socio-economic background. See Section \ref{subsec:overview_schoolsystem} for details.}

\newcommand{\regfootnotesrepmwdmw}{\setstretch{1} \textbf{Note:} Robust standard errors in parentheses. *** $p<0.01$, ** $p<0.05$, * $p<0.1$. Clustering at the middle school (grade 8) or high school (grade 10) level for Columns 1 to 4 and at the primary school (grade 5) class level for Columns 5 to 7. This Table reports estimates from Equation \eqref{eq:mw} (Column 1) and \eqref{eq:dmw} (Columns 2 to 5). Female (Column 5) and Immigrant (Column 6) are dummy variables, SES (Column 7) ranges from 1 to 100. Data are stacked by subject (Italian or math), so there are two observations per student.}

\newcommand{\regfootnotesrepmwdmwplacebo}{\setstretch{1} \textbf{Note:} Robust standard errors in parentheses. *** $p<0.01$, ** $p<0.05$, * $p<0.1$. Clustering at the primary school (grade 2) level for Columns 1 to 2 and at the primary school (grade 5) class level for Columns 3 to 5. This Table reports estimates from Equation \eqref{eq:mw} (Columns 1 and 2) and \eqref{eq:dmw} (Columns 3 to 7).  Female (Column 5) and Immigrant (Column 6) are dummy variables, SES (Column 7) ranges from 1 to 100. Data are stacked by subject (Italian or math) and are from the companion dataset covering grades 2, 5 and 8.}

\newcommand{\regfootnotesmainspec}{\setstretch{1} \textbf{Note:} Robust standard errors in parentheses. *** $p<0.01$, ** $p<0.05$, * $p<0.1$. Clustering at the middle school (grade 8) or high school (grade 10) level. Panel A reports estimates $\hat{\beta}$ and $\hat{\delta}$ from Equation \eqref{eq:both_ranks_uncond}. Panel B reports estimates $\hat{\beta}$ and $\hat{\delta}$ from Equation \eqref{eq:main_specification}. Data are stacked by subject (Italian or math).}

\newcommand{\regfootnotesrobcheckabilitycorr}{\setstretch{1} \textbf{Note:} Robust standard errors in parentheses. *** $p<0.01$, ** $p<0.05$, * $p<0.1$. Clustering at the primary school (grade 2) or middle school (grade 8) level. Panel A reports estimates $\hat{\beta}$ and $\hat{\delta}$ from Equation \eqref{eq:both_ranks_uncond}. Panel B reports estimates $\hat{\beta}$ and $\hat{\delta}$ from Equation \eqref{eq:simple_specification}. Panel C reports estimates $\hat{\beta}$ and $\hat{\delta}$ from Equation \eqref{eq:main_specification}. Data are stacked by subject (Italian or math) and are from the companion dataset covering grades 2, 5 and 8.}

\newcommand{\regfootnotesrankvsclasseffects}{\setstretch{1} \textbf{Note:} Standard errors in parentheses. *** $p<0.01$, ** $p<0.05$, * $p<0.1$. This Table reports correlations between observables at the primary school class level and class effects on performance in middle school (Column 1) and high school (Column 2). All variables are standardized. See Section \ref{sec:rankvsclass} for details.}

\newcommand{\regfootnotescorrperfscore}{\setstretch{1} \textbf{Note:} Robust standard errors in parentheses. *** $p<0.01$, ** $p<0.05$, * $p<0.1$. Clustering at the middle school (grade 8) or high school (grade 10) level. This Table reports correlations between primary school class mean SES and primary school class mean student achievement and student's standardized test scores in middle school (Column 1) or high school (Column 2). Class mean SES and class mean test scores are standardized. Controls for SES, gender, immigration status, grade 5 invisible rank, and grade 5 individual test score are included. In Columns 2 and 3, control for individual test score is replaced by the visible rank. Data about SES is missing for 11,052 students in our main sample as indicated in Table \ref{tab:cohort_students_chara}. See Section \ref{sec:rankvsclass} for details.}

\newcommand{\regfootnotesmechaschoolsorting}{\setstretch{1} \textbf{Note:} Robust standard errors in parentheses. *** $p<0.01$, ** $p<0.05$, * $p<0.1$. Clustering at the middle school (grade 8) or high school (grade 10) level. This Table reports estimates $\hat{\beta}$ from Equation \eqref{eq:main_specification} with a measure of school quality used as the dependent variable. Data are stacked by subject (Italian or math). Note that 40,902 students of our main sample are missing (see footnote \ref{footnote_schoolsorting}). See \ref{subsec:mecha_school_sorting} for details.}

\newcommand{\regfootnotesmechasubjectinterest}{\setstretch{1} \textbf{Note:} Robust standard errors in parentheses. *** $p<0.01$, ** $p<0.05$, * $p<0.1$. Clustering at the high school (grade 10) level. This Table reports estimates $\hat{\beta}$ from Equation \eqref{eq:motivation_spec} (Column 1) and Equation \eqref{eq:main_specification} (Column 2) using a measure of subject interest as dependent variable.  Data are stacked by subject (Italian or math). See Section \ref{subsec:mecha_motivation} for details.}

\newcommand{\regfootnotesmechaunstacked}{\setstretch{1} \textbf{Note:} Robust standard errors in parentheses. *** $p<0.01$, ** $p<0.05$, * $p<0.1$. Clustering at the high school (grade 10) level. This Table reports estimates $\hat{\beta}$ from Equation \eqref{eq:motivation_spec}. The dependent variable is a measure of parental support (Columns 1 and 2), of self-esteem (Columns 3 and 4), of peer recognition (Columns 5 and 6) and of career expectations (Columns 7 and 8). The variable of interest is the primary school class rank in math (odd Columns) or in Italian (even Columns). See Section \ref{subsec:mecha_motivation} for details. }

\newcommand{\regfootnotesmechaunstackedschoolperception}{\textbf{Note:} Robust standard errors in parentheses. *** $p<0.01$, ** $p<0.05$, * $p<0.1$. Clustering at the high school (grade 10) level. This Table reports estimates $\hat{\beta}$ and $\hat{\delta}$ from Equation \eqref{eq:motivation_spec} using a measure of school perception as a dependent variable. The variable of interest is the Primary school class rank in math (Columns 1 and 2) or in Italian (Column 3 and 4). See Section \ref{subsec:mecha_motivation} for details.}

\newcommand{\regfootnotesrobcheckcoveragefull}{\textbf{Note:} Robust standard errors in parentheses. *** $p<0.01$, ** $p<0.05$, * $p<0.1$. Estimates from Equation \eqref{eq:main_specification} on the subset of classes in which all students are observed. Clustering at the current grade (8 or 10) level. Data are stacked by subject (Italian or math).}

\newcommand{\regfootnotesrobcheckcoveragedrop}{\textbf{Note:} Robust standard errors in parentheses. *** $p<0.01$, ** $p<0.05$, * $p<0.1$. Estimates from Equation \eqref{eq:main_specification} on the subset of classes in which all students are observed but after dropping the the 10\% weakest students in terms of class grades. Clustering at the middle school (grade 8) or high school (grade 10) level. Data are stacked by subject (Italian or math).}

\newgeometry{top=1in, bottom=1in, left=1in, right=1in}

\title{\centering \Large 
\textsc{What is essential is visible to the eye} \\ 
\vspace{-.65cm}
\textsc{\normalsize Saliency in primary school ranking and its effect on academic achievements}\thanks{ Ladant: Northwestern University, Harvard University (email: fladant@g.harvard.edu); Hédou: Stanford University; Sestito: Bank of Italy; Bargagli-Stoffi: Harvard University. We greatly benefited from comments and suggestions from Isaiah Andrews, Giulia Bovini, Michela Carlana, Raj Chetty, David Deming, Marta De Philippis, Carola Frydman, Dyani Gaudillière, Oliver Hart, Caroline Hoxby, Clémence Idoux, Larry Katz, Richard Murphy, Ashesh Rambachan, Stefanie Stantcheva, Davide Viviano and participants of the Public Finance and Labor Economics Workshop at Harvard and at the Economics of Education workshop at KU Leuven. We thank Patrizia Falzetti and Michele Cardone from Invalsi for providing data access. The data underlying this article were provided by Invalsi by permission. Data will be shared on request to the corresponding author with the permission of Invalsi. A previous version of this paper was circulated under the title ``\textit{Rather first in a village or second in Rome? The effect of students’ class rank in primary school on subsequent academic achievements}''.}}
\author{\centering François-Xavier Ladant \hspace{.3cm} Julien Hédou \hspace{.3cm}  Paolo Sestito \hspace{.3cm}  Falco J. Bargagli-Stoffi \\
\vspace{.5cm}
First Version: November 2022 \hspace{.6cm} This Version: January 2024} 

\date{}

\begin{document}

\defcitealias{mw20}{MW}
\defcitealias{dmw21}{DMW}

\newgeometry{top=.2in, bottom=1in, left=1in, right=1in} 
\doublespacing
\maketitle

\thispagestyle{empty}
\vspace{-1.2cm}
We propose a new strategy to identify the impact of class rank, exploiting a ``visible'' primary school rank from teachers' exam grades, and an ``invisible'' rank from unreported standardized test scores. Leveraging a unique panel dataset on Italian students, we show that the visible rank has a substantial impact on students' perceptions, which affects subsequent academic performance. However, the effect of being surrounded by higher-SES or higher-achieving peers remains positive even accounting for the decrease in rank. Higher-ranked students self-select into high schools with higher average student achievements. Finally, exploiting an extensive survey, we identify psychological mechanisms channeling the rank effect. \\



\newpage

\setcounter{page}{1}
\onehalfspacing
\renewcommand{\footnotelayout}{\doublespacing}

\newgeometry{top=1in, bottom=1in, left=1in, right=1in} 
\section*{Introduction}
Is it better to be first in a village than second in Rome, as Caesar famously quipped?\footnote{\textit{Parallel Lives}, Plutarch.} In other words, how important is one’s relative position within a group? Recent research across various fields in economics has highlighted the substantial role of peer comparison,\footnote{For instance, it has been shown to affect labor market decisions \citep{deutscher2020firm,dube2019fairness,card_et_al_2012}, productivity \citep{mas2009peers}, consumption patterns \citep{muggleton2022workplace}, academic achievement \citep{zarate2023uncovering,mw20,dobbie2014impact}, fairness consideration and motives for redistribution \citep{hvidberg_etal_2022,easterlin2022easterlin} or behavior \citep{pop2013going}.} but to accurately identify its impact is challenging.\footnote{Most theories on inequalities and intergenerational mobility focus on levels of human capital accumulation and parental wealth or income, rather than relative positions. See \cite{mogstad2023family} for a review.} Indeed, what constitutes an appropriate reference group and an adequate comparison metric is not obvious in many settings. It also requires a seemingly contradictory exercise --- i.e., to change the perception of ranks within a given group while holding absolute positions constant.

In this paper, we make two main contributions. First, we study this question in the context of the Italian public school system by leveraging extensive and detailed data, including a survey covering all Italian students in grade 10 in 2018 that is analyzed here for the first time. The school setting is particularly pertinent to investigate this question as the reference group --- the class --- is well-defined and class ranks provide students with a salient metric to assess their relative position within their group. Second, we propose and implement a new strategy to identify the effect of rank. 

Fundamentally, we seek to estimate how a student’s rank affects her future performance through a change in her perception and behavior. The crux of the problem is that the rank reflects underlying ability, which naturally also impacts performance. Identification thus requires variations in rank conditional on individual ability. 

So far, the literature has employed standardized test scores both as a proxy for ability and as the basis for computing students' ranks. To make rank vary conditional on ability within this framework, it is necessary to exploit variations in class peers’ abilities. Since its proposal by \cite{mw20}, virtually all subsequent papers in the literature have utilized this variation for identification. Nonetheless, this strategy fails to account for higher-order ability peer effects,\footnote{For a detailed discussion about ability peer effects in school, see \cite{sacerdote2011peer}.} as individual ability still affects the rank through its interaction with the moments of the class ability distribution, thereby leading to omitted variable bias.\footnote{See, e.g., \cite{delaney_devereux_2022}. To illustrate how this impacts ranking, consider two students with similar scores relative to their class averages: the student in the class with a smaller variance in scores will rank higher if above average, but lower if below average, compared to the student in the class with a larger variance.} 

To circumvent this issue, we propose an alternative strategy that involves using two sets of scores: teachers' grades, reported to students, and standardized test scores, unreported to them. This brings two key benefits. On the one hand, noise in teachers' grading generates variations in rank that are orthogonal to ability and that can be theoretically harnessed for identification. Relying on such variations is expedient because it avoids intrinsic omitted variable bias --- as ranks would still vary conditional on both individual and class peers' abilities. On the other hand, if the two scores are similarly correlated with students' abilities, we can exploit the difference in the association between the two resulting ranks and later outcomes. Intuitively, the rank based on class grades, \textit{visible} to students, may affect their performance by changing their own perceptions and behaviors, while the rank based on test scores, \textit{invisible} to students, can serve as a placebo test to ensure that other controls adequately capture any remaining relationship between students' ranks and later outcomes.

In this regard, Italy provides an advantageous quasi-laboratory. First, students are regularly evaluated through national standardized tests. The National Institute for the Evaluation of the Italian Education System (Invalsi) administers yearly standardized tests at multiple points in a student’s education: twice in primary school (grades 2 and 5), once at the end of middle school (grade 8), and twice in high school (grades 10 and 13). The subjects tested are Italian and math. These tests are proctored and anonymously graded by a different teacher than the one instructing students in the specific subject. Grading follows a precise rubric, consistent across Italian school districts. The resulting standardized test scores are comparable across the country and provide us with a proxy for ability in primary school and a measure of academic performance in middle and high school. Importantly, students are not informed about their performance on the test. Second, teachers have ample discretion in choosing how to grade their students in their courses, and class grades are extremely salient to students: not only do they appear on semester reports that are sent to them and their parents, but admission to the next grade also depends upon these grades exclusively. Lastly, grades are often publicly disclosed by teachers to the class, and students talk to each other about their performance, making it likely that the ensuing rank is known. 

Our strategy assures identification under weaker assumptions than in existing literature since it only requires (i) that one of the scores be indeed unreported so that the invisible rank can serve as a placebo and (ii) that the two scores be similarly correlated to ability. In particular, intricate ability peer effects and students' selective sorting into classes are no longer a threat to identification, as they would symmetrically impact both ranks.

Yet, two main concerns about (ii) arise in our context. First, teachers' grades might reflect students' abilities better than scores from a one-time standardized test. There is thus a risk that, by construction, the visible rank could be inherently more predictive of ability. Second, the discrepancy between the two ranks might result from a difference in non-cognitive skills, such as participation or grit, if they were captured by class grades but not test scores. We allay both concerns with a suite of placebo checks. 

Our main panel dataset comprises several hundred thousand Italian public school students from two successive cohorts. We observe these students during their last year of primary school, last year of middle school, and second year of high school. Our data is at the class level and includes each student's scores in Italian and math on national standardized tests and first-semester grade averages as reported to students and their parents. To our knowledge, we are the first to harness panel data covering entire student cohorts at the class level. This allows us to construct ranks within the relevant peer group, as well as to put estimated effects in perspective by adequately measuring class value added and discussing its potential drivers. We also match our main panel with an extensive survey including all students in their second year of high school in 2018\footnote{This detailed questionnaire (\textit{Questionario di Contesto}) was administered as part of the 2018 standardized test. Its questions were deemed too intrusive, causing a scandal in the Italian media and resulting in its subsequent discontinuation.} asking them about their motivations and aspirations, from which we explore potential psychological channels for the rank effect. Finally, we avail ourselves of two additional panels --- one covering students twice in primary school and once in middle school (grades 2, 5, and 8), and the other once in middle school and twice in high school (grades 8, 10, and 13) --- that we use in our placebo checks and additional analyses. 

We begin by implementing the standard strategies of the literature to estimate the effect of grade 5 rank: \cite{dmw21} (hereafter \citetalias{dmw21}) and \cite{mw20} (hereafter \citetalias{mw20}). They rely on the lack of correlation between the rank and students' observable predetermined characteristics (e.g., gender, race) to tackle concerns about omitted variable bias. We show that, even in that case, the rank remains significantly correlated with a previous, unreported and unbiased measure of ability (grade 2 standardized test scores). This demonstrates that usual balance checks do not suffice to reject bias, thereby warranting the search for a new identification strategy.

Our main specification includes both the visible and the invisible ranks in grade 5. We first ensure that both ranks are similarly correlated to underlying ability: we show that, unconditionally, the two ranks have quasi-symmetric impacts on future test scores and we then add controls so that both ranks are uncorrelated with past test scores. Once these controls are included, the invisible rank's impact on future performance is consistently small and imprecisely estimated, providing additional evidence that the visible rank is operating through a perception channel. However, this comes at a price, as it necessitates the inclusion of student-level fixed effects. The rank effect is therefore estimated off of within-student variations, which precludes us from exploring whether it might differ by subject.   

Our findings reveal that the rank effect is sizable: our main specification estimates that ranking at the top of class compared to the bottom in primary school leads to a gain of 8.1 percentiles in the national standardized grade distribution in middle school and 7.5 in high school, which is equivalent to moving up by about 36,000 ranks nationally. By contrast, using standard identification strategies leads to a significant underestimation of the rank effect: \citetalias{mw20} and \citetalias{dmw21} specifications yield estimates that are respectively about 30\% and 70\%  significantly lower than ours.

We then explore non-linearities and heterogeneities in the rank effect. We find that it has a non-negative impact on performance in middle school: while the impact increases linearly with the rank above the median, students ranking below do not perform significantly worse. In contrast, we find a linear impact of the rank on performances in high school, albeit less precisely estimated around the median, which suggests that the rank might be less salient to students ranking in the middle. We further investigate how gender, immigration status, and socio-economic status influence the rank effect, but find no significant difference across groups. We also show that class size, and peers' or individual abilities (as measured by standardized test scores) do not cause the rank effect to change.

Additionally, we compare this relative position effect with the effect of peer group quality. We define ``class quality'' as class value added. In the vein of \cite{chettyetal_2011}, we use class fixed effects from our main specification to estimate class quality, which is thus a measure of all class inputs (e.g., teacher and peer effects, size, and amenities) affecting subsequent academic performances. We find that the impact on future academic performances of increasing class quality by one standard deviation is six-fold larger than the effect of increasing the rank by one standard deviation. Besides, the effect of increasing the average in peers' standardized test scores by one standard deviation remains largely positive, even accounting for the decrease in rank. We thus conclude that Ceasar was misguided, and that it is better to trade a lower rank for a more competitive environment. 

To understand what may explain such a lasting effect of rank on academic outcomes, we explore the mechanisms that may be at play. First, we show that higher-ranked students self-select into high schools with significantly higher average student achievements. No such self-selection is observed in middle school, where enrollment is mostly based on residency criteria and, as such, much more constrained. Second, we exploit the 2018 questionnaire in which students were asked about a wide range of different topics related to school and beyond, such as their self-esteem, parental support, career expectations, and relationships with peers. Students had to state how much they agreed with a series of statements related to each topic, using a six-point Likert scale. 
We find that a higher visible rank enhances students' interests in the subjects studied. As the remaining variables do not vary by subject, we have to drop student fixed effects from our specification. We offer evidence that this does not significantly alter the results and, if anything, would lead us to estimate a lower bound of the effect. We show that a higher visible rank improves students' perception of school and positively affects their self-esteem and perceived recognition by their peers. Evidence suggests that the rank in Italian has a larger impact than the rank in math.

\textbf{\textit{Related Literature.}} Our paper first speaks to the inequality literature. Most theories focus either on absolute levels of resources, such as human capital accumulation, parental wealth, or income \citep{lee_seshadri_19, becker_kominers_2018,becker_tomes_79}. Evidence on the various impacts of the quality of the reference group abounds: neighborhood quality impacts future earnings and college attendance rates \citep{chetty_hendren_2018,chetty_hendren_2018ii}, kindergarten class quality influences future earnings \citep{chettyetal_2011}, and inter-generational mobility in the US is strongly dependent on geographical location \citep{chetty_landopportunity}. The crucial role of parental income for children's educational attainment and future income has also been documented \citep{carneiro_et_al_2021,caucutt2017correlation}. In this regard, our paper suggests a way to measure the impact of relative position in groups that are clearly defined, with transparent comparison metrics. We are also able to compare this impact to that of the quality of the reference group while exploring plausible pathways through which the rank effect may have an impact.

Our work also contributes to the literature examining the effects of relative position, which have been analyzed across fields in economics. For instance, \cite{dube2019fairness} show that separations are mostly explained by wage comparison with peers rather than with the market. \cite{muggleton2022workplace} demonstrate that wage rank within the workplace affects consumption patterns, conditional on the wage level. \cite{hvidberg_etal_2022} argue that one's relative income within certain reference groups influences one's views about the fairness of inequalities in society. \cite{easterlin2022easterlin,easterlin2010happiness} suggest that social comparison is key to happiness and is mainly driven by one's relative income rather than by income itself. In the school context, the \textit{Big-Fish-Little-Pond} phenomenon, whereby rank improves non-cognitive skills and lowers the effort costs in studying, has long been documented \citep{marsh2008big,marsh1987big,marsh1984determinants}. More recently, \cite{pop2013going} show that students perceiving themselves as academically weaker tend to be marginalized and have more frequent negative interactions with their peers. \cite{zarate2023uncovering} find that peer effects vary by gender and skills and that higher-achieving peers in primary school decrease girls' test scores. 

Our study adds to the literature exploring the rank effect in school.  The first strand of rank papers uses randomized experiments to underpin the impact of rank disclosure \citep[e.g.][]{goulas_2015,azmat_2010}. These studies find that revealing rank to students has a positive impact on the performance, achievements, and earnings of those at the top of the ranking distribution. A second line of research exploits across-cohort variations to identify a rank effect \citep{elsner2018rank,elsner2017big} based on a strategy propounded by \cite{hoxby_2000}. Rank is found to positively impact students’ performance and negatively affect anti-social behaviors.

Our work most closely relates to the recent literature spawned by the seminal work of \citetalias{mw20}. Their strategy has become the gold standard in the literature and virtually all subsequent papers have utilized it \citep[see, e.g.,][]{rury_2022,goulas2022comparative,carneiro2022,dmw21,elsner2021achievement,pagani_et_al2021,delaney2021high}. It can be summarized as follows: as classes vary in size and students' abilities, two similarly able students in two different classes will end up with a different rank.\footnote{In practice, however, most of these papers observe data at the school level, which they assimilate to a class.} These papers find that rank in the early stages of education affects subsequent performance and achievements and influences personality traits and behavior. In their review of the literature, \cite{delaney_devereux_2022} however point out a major limitation inherent to this strategy: namely, the difficulty to sort out the rank effect from other ability peer effects even ``\textit{randomizing identical students into many classes that are identical in every way except that they differ in terms of the distribution of human capital of their students}'', which is, in their view, ``\textit{the best} [one] \textit{could do.}'' Our paper is an attempt at overcoming this limitation by proposing an alternative strategy that exploits the difference in saliency between two scores and the resulting ranks. The importance of students' awareness of their relative ranks has already been underlined in the rank literature \citep{megalokonomou2022good,yu_2020} but has never been leveraged for identification.\footnote{This might also be due to data limitation: Both papers observe a ``\textit{perceived rank}'' on a 5-point scale, which is self-reported by students in survey samples in China respectively comprising 3,596 and 2,472 individuals. Their identification strategy is the same as in \citetalias{mw20}.}

Finally, beyond our new identification strategy, this paper also stands out in two important respects. First, we offer evidence that we can more credibly control for individual ability and other ability peer effects, thereby increasing confidence in interpreting the rank effect. Furthermore, our unique administrative data at the class level makes it possible to account for potential confounding variables, such as teacher quality, or class size, as well as estimate the quality of the group of reference. This is especially true in the Italian context where the classes are substantially fixed within the same school level. We can then more thoroughly compare the rank effect to the group quality effect, allowing us to put the former's magnitude in perspective. Second, we explore the mechanisms channeling the rank effect by exploiting a rich questionnaire administered to all Italian students in a given year, thus shoring up the external validity of our conclusions.

\textbf{\textit{Paper organization.}} The remainder of this paper is structured as follows. Section \ref{sec:institutional_context} describes the Italian school system and Invalsi tests; Section \ref{sec:data_descstat} presents our data, our study sample characteristics as well as details on the computation of key variables; Section \ref{sec:empirical_strategy} lays out our empirical strategy; Section \ref{sec:main_results} discusses our main results; Section \ref{sec:rankvsclass} compares the rank effect with the effect of class value added; Section \ref{sec:mechanisms} explores the mechanisms through which rank may impact student performance; and Section \ref{sec:conclusion} concludes.

\section{Institutional Context}\label{sec:institutional_context}

\subsection{Overview of the Italian School System}\label{subsec:overview_schoolsystem}

The Italian school system consists of three different sections. Primary school (grades 1 to 5) lasts five years and begins at age six. Middle school (grades 6 to 8) lasts three years and begins at age 12. High school (grades 9 to 13) lasts five years and begins at age 14. While every primary and middle school student shares the same core curriculum across the country, there exist three main different types of high schools into which students can self-select: academic (``Liceo''), technical (``Istituto Tecnico''), and vocational (``Istituto Professionale''). The academic track prepares students to go to a university and is typically chosen by students coming from higher socio-economic backgrounds \citep{brunello2007does}. The technical and vocational tracks include students from lower socio-economic backgrounds as well as a higher share of children of immigrants and fewer women (Table \ref{tab:vocational_vs_academic_all}). 

Some important features of the Italian educational system are worth highlighting. First, in primary and middle school, class composition is determined by school principals and according to guidelines set by law. Specifically, within a school, classes should be balanced in terms of gender and socio-economic status (SES).\footnote{The SES is computed by Invalsi and is standardized so that, at the national level, the mean is zero and the standard deviation is one \citep{masci_etal_2018}.}\label{footnote_ses_def} Principals usually seek to ensure comparability across classes and heterogeneity within classes \citep{carlana_2019}. 

Second, class composition is stable within sections: from grades 1 to 5, grades 6 to 8, and grades 9 to 13 students remain with the same peers in most cases.\footnote{Exceptions to class composition stability arise when a student repeats a year, moves in or out of the area, or is transferred to another school by their parents \citep{carlana_2019}, which are marginal events until high school \citep{salza_2022}.} The reshuffle occurs when going from primary to middle school and from middle school to high school and is substantial.\footnote{A student has an average of 64\% (IQR = 57\%-78\%) new peers in middle school and 82\% (IQR = 78\%-90\%) in high school.} 

Third, principals assign a math and an Italian teacher to each class. The teacher-class match is also stable within sections \citep{barbieri2011determinants}. Each individual teacher typically has a large amount of discretion in how to organize educational activities and grade their pupils. The grades we observe in grade 5 are the average of the grades obtained on a series of written and oral teacher-specific exams over the course of the first semester and range from 1 to 10. Importantly, attendance, participation, and a student's general behavior are evaluated by a distinct grade and should, in principle, not be factored in the Italian or math grade.

\paragraph{Saliency of class grades.} Class grades are extremely salient to each student, as they not only are directly disclosed to them and their parents on the semester report but also determine admission to the next grade. The class grades we observe are the ones determining how students are ranked within classes at the end of the first semester. In that sense, the rank computed from these scores is therefore the ``true'' one. The extent to which students perceive it exactly is unclear but, given our context, this proxy is likely very close.

Indeed, students are likely to know of their classmates' grades. The grades we observe are an average of oral and written exams. Oral exams are conducted during daily lectures and grades are announced publicly. Although written exams are not necessarily public in the same way, it is common for teachers to read the list of all results to the entire class and students talk to each other about their performance. For that matter, there is evidence that, in a context where class grades are disclosed to students and their parents, students have a good sense of their relative ranking \citep{megalokonomou2022good,yu_2020}.  

\subsection{Invalsi Test and Student Questionnaire}\label{sec:invalsi_test_desc}

In the spring of each year, the National Institute for the Evaluation of the Italian Education System (Invalsi) administers standardized tests in Italian and math in grades 2, 5, 8, 10, and 13. The tests are presented to students as ability tests and are not compulsory. Hence, we we only observe students who took the test.  

Except in grade 8,\footnote{In the period we cover, grade 8 Invalsi tests are higher stakes as they count towards one-sixth of students' final grade at the end of middle school and results are thus known to students.} students are not supposed to be informed about their performance on the test and scores do not play a role in middle school or high school enrollment determination \citep{carlana_2019}. Further, in grade 5, the timing of the test makes it materially impossible for students to know about their performances. Indeed, results are sent back to schools at the beginning of the subsequent school year, i.e., once students have left primary school. As students are identified by an anonymous number, schools could not even reach out to students to disclose individual results.

The Invalsi test content is similar to that of the OECD-Pisa test and consists of multiple-choice as well as open-ended questions \citep{angristetal_2017}. Tests are proctored and graded by a different
teacher than their usual instructor in the specific subject \citep{lucifora_2020}. Grading is anonymous and follows a precise rubric that is consistent across the country (\citealp{carlana_2019}). 

Since its implementation in 2009/2010, the Invalsi test has been widely covered in the media and is considered by parents when selecting schools, as results have been made public since 2012. Primary and middle school choice is principally based on residency criteria, while high schools are free to compete to attract students \citep{lucifora_2020}.

In 2018, Invalsi administered a questionnaire in addition to the test. It asked students about many topics related to school: e.g., parental support received in relation to their studies, confidence in Italian or math, relationships with peers, career expectations, and self-esteem. The answers are numerical on a six-point Likert scale, with 1 indicating students strongly disagree with the statement and 6 that they strongly agree. The questions and answers are described in Online Appendix \ref{sec:app_questionnaire}.

\section{Data and Descriptive Statistics}\label{sec:data_descstat}

Our main dataset covers two successive cohorts: one was in primary school (grade 5) in 2013, in middle school (grade 8) in 2016, and in high school (grade 10) in 2018. The other was in grade 5 in 2014, grade 8 in 2017, and grade 10 in 2019.

\subsection{Overview of the Dataset}\label{subsec:overview_dataset}
The dataset includes all students satisfying the following two conditions: (i) being in high school (grade 10) in 2018 or 2019 and (ii) taking the Invalsi test that same year. As explained in Section \ref{sec:invalsi_test_desc}, this test is not mandatory: 93.8\% of Italian students in grade 10 took it in 2018 and 93.7\% in 2019 (Table \ref{tab:invalsi_attendance}). The dataset is then constructed in reverse: data about their performances in middle school (grade 8) and primary school (grade 5) are retrieved so that a student is always observed in high school but may or may not be observed in earlier grades (see Online Appendix \ref{subsec:app_retention_rate}).

A variable recording the proportion of students who took the test by class allows us to recover the actual number of students that should have been observed in grade 5 and grade 8. Table \ref{tab:invalsi_attendance} shows that students in our dataset accounted for 89.8\% of grade 5 students in 2013 and 87.8\% in 2014.

\paragraph{Class and School Size.} In Italy, class size is set by law. The cohorts we observe are subject to the regulation established by the \textit{Decreto Ministeriale 331/98}.\footnote{The \textit{Decreto del Presidente della Republica 81/2009} set new bounds on class size from the 2009-2010 school year but was rolled out one grade per year starting in first grade \citep{angristetal_2017}.} Primary school classes should have between nine and 28 students, but exceptions are possible (see Online Appendix \ref{app:subsec_class_size_distribution}).  
 
\paragraph{Sample Selection.}\label{subsec:sample_selection} To ensure the validity of our results, we make some adjustments. We first restrict our sample to classes for which we observe more than 90\% of the students in grade 5. The resulting sample has a mean coverage of 95.8\%, which means that we miss fewer than one student on average (Table \ref{tab:coverage_class}). Second, we discard classes for which suspicions of cheating were high (see Online Appendix \ref{subsec:app_cheating}). Third, to use as accurate a rank as possible, we remove classes in which we lack either Italian or math class grades in grade 5 for any observed student. Fourth, we keep students for whom we observe standardized test scores in both grade 8 and 10. We ultimately select 205,123 students in the 2018 Cohort and 182,491 students in the 2019 Cohort (Table \ref{tab:selected_sample}).

The remaining descriptive statistics and the main results are computed after sample selection. As shown in Table \ref{tab:cohort_students_chara}, students in the restricted sample do not differ substantially from other students observed in grade 5, alleviating concerns that our restriction could have led to the selection of a biased sub-sample of students.


\subsection{Scores}\label{subsec:scores}
In this paper, we use two sets of scores: class grades, assigned by teachers and reported to students, and standardized test scores, unreported to them.

\textit{\textbf{Class grades.}} Class grades range from 1 to 10 (see Table \ref{tab:desc_stat_cohort_class_score}). We premise that missing students are at the bottom of the class grade distribution, a reasonable assumption as they either repeated at least a year over the course of their studies or decided not to take standardized tests. To shore up the validity of this assumption, we present a robustness check in Online Appendix \ref{sec:app_other_subset_classes}.

\textit{\textbf{Standardized test scores.}} We convert standardized test scores to percentiles, following the literature. To keep an accurate measure of where students fall in the national distribution, we compute the percentile standardized grades on the unrestricted sample. By construction, prior to sample selection, the distribution of percentile standardized grades exhibits a uniform distribution in grades 5, 8, and 10. Table \ref{tab:cohort_ipscore_chara} shows that, after sample selection, standardized test scores are still uniformly distributed in grade 5. This indicates that we do not leave out a specific portion of the ability distribution at the national level. For subsequent grades, our sample restriction leads to a slightly skewed distribution, which is coherent: to be included in the restricted sample, students must be observed in grade 5. This excludes any student who missed at least one later standardized test or repeated at least a year, i.e., those likely to be academically weaker.

\subsection{Rank Computation}\label{subsec:rank_computation}
In accordance with the literature, we construct $R^V_{ics}$, the visible rank based on class grades in grade 5, and $R^I_{ics}$, the invisible rank based on standardized test scores taken in grade 5, as follows:
\begin{align*}\label{eq:perc_rank}
    R^{t}_{ics} = \frac{N_c - n^t_{ics}}{N_c-1}, \quad t \in \left\{V,I\right\}
\end{align*}
where $n^t_{ics}$ is the ordinal rank based on reported or unreported scores of student $i$ in class $c$ and subject $s$ and $N_c$ is class $c$ size. The highest (lowest) performing student in class $c$ has thus $n = 1$ ($n = N_c$) and thus $R = 1$ ($R = 0$). In case of a tie, we assign the mean rank. This is the most neutral way of proceeding, as it warrants that the average rank of a class is the same as that of a class of similar size without any ties, while not arbitrarily breaking ties.\footnote{Online Appendix \ref{sec:app_rankandties} provides evidence that using a Mean tie-breaking rule is vindicated and does not drive our results.} Figures \ref{fig:qqplot_prank} and \ref{fig:qqplot_iprank} in Online Appendix \ref{subsec:app_qqplot} plot the distributions of the two rank measures $R^V_{ics}$ and $R_{ics}^I$, which are both very close to a uniform distribution as expected.

\section{Empirical Strategy}\label{sec:empirical_strategy}

\subsection{Problem Statement}
Our goal is to estimate the effect of a student's class rank in grade 5 on her subsequent academic achievements in middle school and high school. Since \citetalias{mw20}, the standard strategy has been to use standardized test scores both to measure ability and to construct the rank $R^I_{ics}$. This rank, invisible to students, is taken as a proxy for the visible rank $R^V_{ics}$ computed from class grades assigned by teachers \citepalias[see, e.g.,][]{dmw21}.

Within this framework, identification requires variations in classes' ability distributions. In class $c$, student $i$'s rank $R^I_{ic}$ is a function of her ability $a_i$ and the class ability distribution $F_c(.)$, i.e., $R^I_{ic} = g(a_i,F_c)$.
Identification relies on the fact that two students with the same $a_i$ can end up with different ranks depending on $F_c$. If $F_c$ is as good as random, $R^I_{ic}$ can vary even while holding $a_i$ constant.

Replicating the specification proposed by \citetalias{mw20} leads us to estimate:
\begin{equation}\label{eq:mw}
    T^g_{ics} = \alpha + \beta R^I_{ics}+ \sum_{t=1}^{100} 1[T^5_{ics} = t]  + \theta_{cs} + x_i'\gamma + \varepsilon_{ics} \tag{MW}
\end{equation}
where for student $i$ in class $c$ studying subject $s$, $T^g_{ics}$ is the percentile standardized test score in grade $g \in \left\{8,10 \right\}$, $\theta_{cs}$ a class-by-subject fixed effect, and $x_i$ are student-level observables (e.g., gender or SES). Ability is proxied by fixed effects in percentile standardized test grades in primary school $T^5_{ics}$.

This specification controls for the class average ability through the class-by-subject fixed effect but not higher-order ability peer effects. These effects impact students' ranks differently depending on individual ability, which leads to omitted variable bias as shown in Online Appendix \ref{sec:app_toyexample} and discussed in \cite{delaney_devereux_2022}.

In light of these concerns, \citetalias{dmw21} suggest constructing a specification to ensure that the rank is uncorrelated with students' predetermined characteristics (``balance checks''), thereby making it more likely to be as good as random.  To that end, they allow ventiles of standardized test scores to vary by quartiles of school mean and variance in standardized test scores.\footnote{Notice that their unit of observation is the school while ours is the class.} This exact specification does not pass the balance checks in our context but the following does:
\begin{equation}\label{eq:dmw}
    T^g_{ics} = \alpha + \beta R^I_{ics}+ \sum_{D = 1}^{216} \sum_{t=1}^{100} 1[T^5_{ics} = t] 1[d_c=D] + \theta_{cs} + x_i'\gamma + \varepsilon_{ics} \tag{DMW}
\end{equation}
where $T^5_{ics}$ denotes the percentile  standardized test scores in grade 5 for student $i$ in class $c$, and subject $s$. Classes are characterized by their sixtile in the distributions of mean, variance, and kurtosis in test scores, yielding 216 class distribution groups $D$. 

For the balance checks, we look at the same observables as in \citetalias{dmw21} to the extent possible given the few demographic variables we observe: dummies for gender and immigrant status and percentalized SES (i.e., ranging from 1 to 100). 

Columns 1 to 4 of Table \ref{replication_mw_dmw_balance} show the estimates of Equations \eqref{eq:mw} and \eqref{eq:dmw} using our data. Columns 5 to 7 show that the rank is uncorrelated to observables. 

\subsection{Limitations of Existing Approaches}

Our data allows us to run an additional balance check to more directly investigate whether the rank is as good as random. Indeed, if it were the case, the rank should be uncorrelated with a previous and similar measure of ability: standardized test scores in grade 2, which are unbiased and unreported to students.

To that end, we use a companion dataset, in which we observe students in grade 2 in 2013, grade 5 in 2016, and grade 8 in 2019, which allows us to retain grade 5 rank as a pivot while being able to check whether it is correlated with measures of ability in grade 2. We apply the same criteria of sample selection as for the main sample. 

We begin by verifying that, unconditionally, grade 2 test scores do reflect ability. Indeed, they exhibit a strong linear and positive relationship with the invisible rank in grade 5, as shown in Figure \ref{subfig:abilityrank_undisclosed_ts}.

We estimate Equations \eqref{eq:mw} and \eqref{eq:dmw} but our main outcome is now standardized test scores in grade 2. If the rank in grade 5 is indeed uncorrelated to ability, its coefficient should be small and imprecisely estimated. Table \ref{replication_mw_dmw_ability_placebo} shows that both specifications fail this placebo check (Columns 1 and 2). Strikingly, the fact that the invisible rank is not correlated to observables (Columns 3 to 5) does not preclude it from being strongly correlated to a previous measure of ability. We have to look for an alternative approach. 

\subsection{Motivation for our Approach}\label{subsec:strategy_motivation}

Our strategy stems from the intuition that, if a rank effect is at play, it will be more directly identified if we construct it from a score that is actually visible to students. In our context, we observe two scores for each student: one that is reported (class grade), yielding a \textit{visible} rank, and one that remains unreported (standardized test score), yielding an \textit{invisible} rank. Both scores are naturally correlated with students' underlying abilities and preparation, and hence may be directly related to students' later outcomes. 

Our strategy leverages the symmetry of the two scores in their correlation with underlying ability and their asymmetry in saliency. This allows us to estimate how the rank affects students' perceptions. To that end, we rely on variations in the visible rank arising from noise in teachers' grading: in theory, it is now possible to perfectly control for individual and peers' abilities and still observe variations in rank. 

This strategy permits identification under relaxed assumption as it only requires (i) that one score be indeed unreported so that the invisible rank can be used as a placebo, and (ii) that the two scores be similarly correlated to students' ability. Specifically, as long as these two conditions are met, students' selective sorting into classes or intricate ability peer effects are no longer a threat to identification since they would similarly impact both ranks. 

We first need to check that the visible and the invisible ranks are imperfectly correlated. We find a correlation of 0.6 and Figure \ref{fig:hist_rank_res} shows that, holding the invisible rank constant, there is a variation of at least one (two) visible rank for 80\% (50\%) of our sample. This indicates a fair amount of noise in how teachers grade, consistent with what has been observed in similar contexts \citep[see, e.g.,][]{dee_et_all_2019}.

We also have to verify that (ii) holds. This might not be the case for two main reasons. First, teachers may well have a more informed judgement about students' abilities. Class grades, and hence the visible rank, might thus reflect them better than a one-time standardized test. Second, class grades might also capture the impact of non-cognitive skills as observed and valued by teachers. This would be a problem if non-cognitive skills mattered more for the performance on standardized test --- which is, however, not obvious \citep[see, e.g., ][and references therein]{jackson2018test}.

To alleviate these concerns, we start by verifying that, absent any control for ability, the visible rank $R^V_{ics}$ and the invisible rank $R^I_{ics}$ comparably affect later outcomes. To that end, we estimate the following equation:
\begin{equation}\label{eq:both_ranks_uncond}
 T^g_{ics} = \alpha + \beta R^V_{ics}+ \delta R^I_{ics} + \theta_{cs} + \varepsilon_{ics}
\end{equation}
with the usual notations. $\hat{\beta}$ and $\hat{\delta}$ should be of the same magnitude --- we especially need to check that $\hat{\beta}$ does not substantially exceed $\hat{\delta}$.

Our specification will then have to pass the ability placebo check with standardized test scores in grade 2 --- a previous, unbiased and unreported measure of ability. This placebo check allows us to further address both concerns. First, if both ranks are uncorrelated with past test scores, we will have shown that neither rank captures ability better than the other at baseline. Second, if students displaying higher non-cognitive skills performed better on standardized tests, this should be detected on past as well as on future test scores. By contrast, the perception effect of the rank cannot impact past performance. If there is no correlation between the visible rank and past test scores, a difference in effect on future test scores between the two ranks will be evidence that this is due to a change in perception.\footnote{In fact, there would be a problem if the following were simultaneously true: (i) the effect of non-cognitive skills on test scores strengthens over time and (ii) conditional on ability, the remaining differences in rank are solely attributable to differences in a defined set of non-cognitive skills. We believe it is not a threat to identification for several reasons: first, as explained in Section \ref{sec:institutional_context}, behavior, participation and attendance are evaluated through a distinct grade and are not to be included in Italian or math grades. Therefore, differences in ranks conditional on ability are much more likely to stem from noise in teachers' grading. Second, non-cognitive skills are multi-dimensional and each teacher values some positively and others negatively in an idiosyncratic fashion. Thus, differences in ranks conditional on ability cannot reflect a clear hierarchy in terms of the same non-cognitive skills, making a bias unlikely.}

Once these conditions are met, the invisible rank can be used as a placebo check to ensure that our other controls adequately capture any remaining relationship between ability and later outcomes. If we use appropriate controls, we expect to find a small coefficient on the invisible rank. As any omitted variable, and particularly ability, would not exhibit the asymmetry in saliency, a large and significant coefficient on the visible rank will be evidence that this rank is operating through a channel that changes students’ perceptions and that is orthogonal to ability.

\subsection{Main Specification}
\begin{figure}[H]
\centering
\begin{subfigure}[b]{0.45\textwidth}
\centering
\caption{Visible Rank}\label{subfig:abilityrank_salient_ts}
\includegraphics[width=\textwidth]{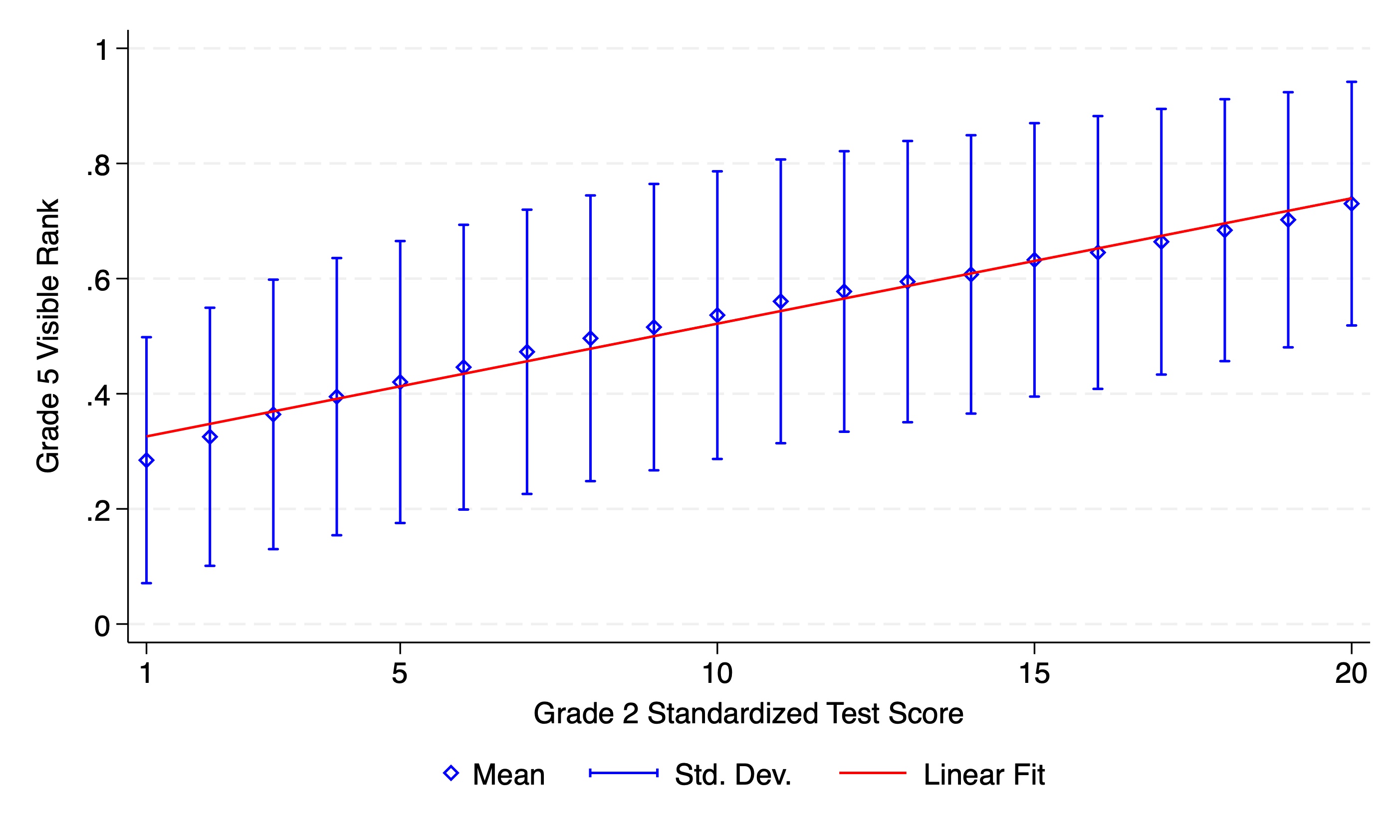}
\end{subfigure}
\begin{subfigure}[b]{0.45\textwidth} 
\centering
\caption{Invisible Rank}\label{subfig:abilityrank_undisclosed_ts}
\includegraphics[width=\textwidth]{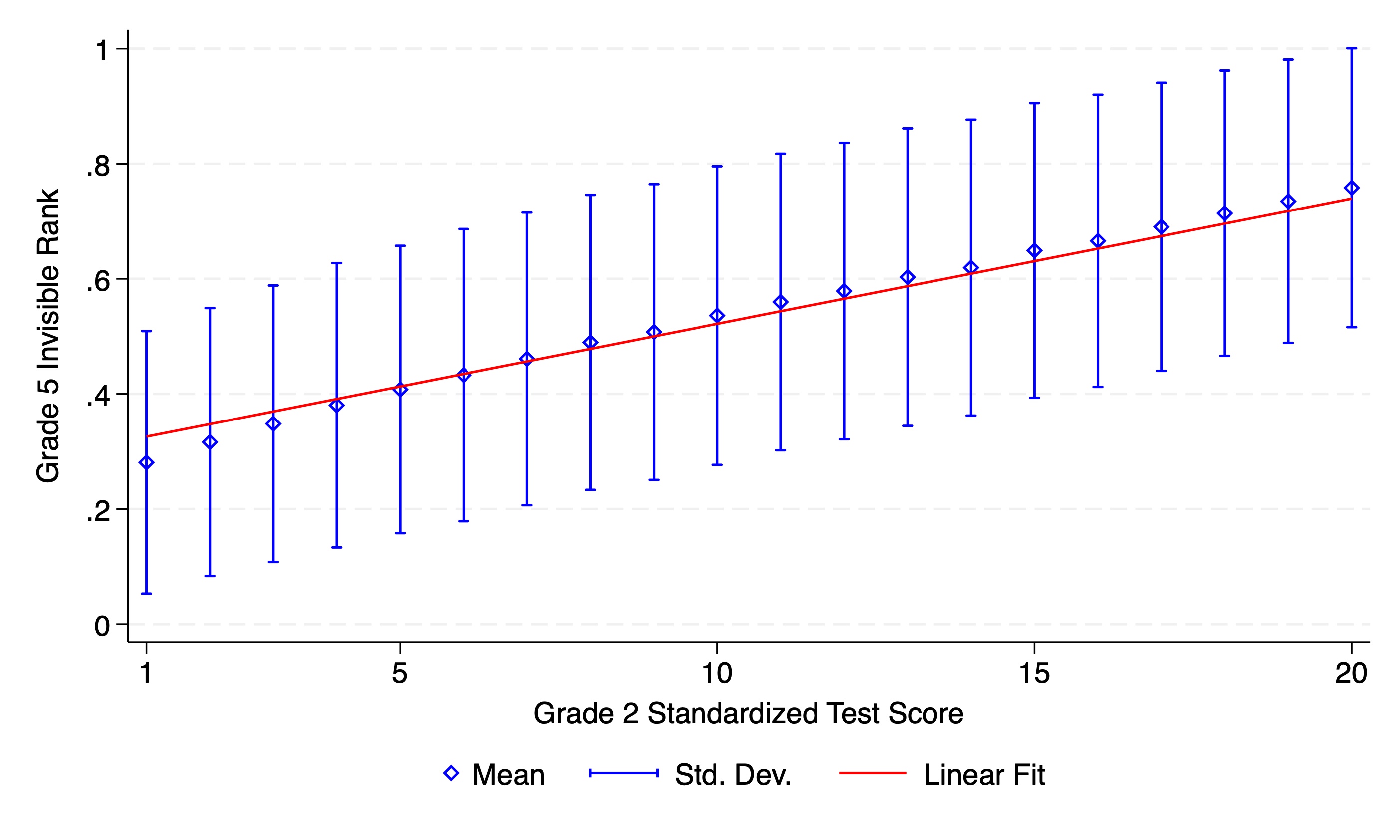}
\end{subfigure}
\footnotesize \begin{tabular}{p{14cm}}
\caption{\textbf{Correlation Between Grade 2 Test Scores and Ranks }}\label{fig:abilityrank_corr_g2g5_undisclosed}
\setstretch{1} \textbf{Note}: This Figure displays a binscatter plot of the correlation between ventiles in grade 2 standardized test scores and grade 5 visible rank (\ref{subfig:abilityrank_salient_ts}) and invisible rank (\ref{subfig:abilityrank_undisclosed_ts}). See Section \ref{sec:empirical_strategy} for details.
\end{tabular}
\end{figure}

The first necessary condition to estimate a credible rank effect is that our specification passes the ability placebo check for both ranks. Figure \ref{subfig:abilityrank_salient_ts} shows that the visible rank exhibits a very strong and positive relationship with grade 2 test scores, indicating the visible rank is heavily correlated with underlying ability too. 

We start from the simplest specification, very close to \eqref{eq:mw}, but using both the visible and the invisible ranks as our main variables of interest:
\begin{equation}\label{eq:simple_specification}
    T^g_{ics} = \alpha + \beta R^V_{ics}+ \delta R^I_{ics} + \sum_{t=1}^{100} 1[T^5_{ics} = t] + \sum_{k=1}^{10}  1[C_{ics} = k] + \theta_{cs}  + \varepsilon_{ics} 
\end{equation}
 We control for dummies in percentiles of grade 5 standardized test scores $T^5_{ics}$ and dummies in rounded grade 5 class grades $C_{ics}$ to make it as flexible as possible. Class grades are an additional control for ability and non-cognitive skills as well as a way to account for the absolute effect of class grades on students (e.g., holding the rank constant, getting a 9 rather than a 7 may, in and of itself, affect students' perceptions).

 Table \ref{robcheck_g2g8_class_cohort} displays the results of the ability placebo check, i.e., using the companion dataset. Panel A reports the estimates of Equation \eqref{eq:both_ranks_uncond}, which show that both ranks have comparable impact on past or future past or future test scores when we do not control for ability: in both grades 2 (Column 1) and 8 (Column 2), the two coefficients are of similar magnitude. That of the invisible rank is slightly larger, thus ruling out concerns that the visible rank is inherently more correlated to ability. 

Column 1 of Panel B demonstrates that Equation \eqref{eq:simple_specification} fails the ability placebo check: despite our controls, both ranks remain strongly correlated with previous measures of ability. Besides, Column 2 of Panel B shows that the impact of the invisible rank on future test score is very large and significant, further proving that our controls do not suffice to fully account for ability. 

In the same spirit as \citetalias{dmw21}, we have to tweak Equation \eqref{eq:simple_specification} to pass the ability placebo check: we add a student fixed effect and we interact standardized test scores and class grades with indicators for groups defined from classes' means and variances in standardized test scores and class grades. These interaction variables are a way to account for a part of the ability distribution of the class that might generate omitted variable bias. 

The need for student fixed effects suggests that scores and grades, which are subject-specific controls for ability, fail to adequately account for underlying across-subject ability, which appears to be determinant. Student fixed effects also control for unobservables, such as grit, competitiveness or parental support, which seem to also play a role in influencing the rank.

Our main specification thus becomes:
\begin{align}\label{eq:main_specification}
\begin{split}
    T^g_{ics} & = \alpha + \beta R^V_{ics} + \delta R^I_{ics} + \theta_{cs} + \gamma_i + \sum_{D=1}^{25} \sum_{t=1}^{100} 1[T^5_{ics} = t] 1[d_c = D] \\
    &  + \sum_{G=1}^{50}  \sum_{k=1}^{10}  1[C_{ics} = k] 1[g_c = G]  + \sum_{E=1}^{1875}  \sum_{k=1}^{10}  1[C_{ics} = k] 1[e_c = E] + \varepsilon_{ics} \\
    & \\
    & = \alpha + \beta R^V_{ics} + \delta R^I_{ics} + \Gamma_{ics} + \varepsilon_{ics}
\end{split}
\end{align}
where $\gamma_i$ is a student fixed effect. We also interact test scores with classes' 25-quantile in the distribution of classes' means in standardized test scores $d_c$ and class grades with 50-quantile in the distribution of class' means in class grades $g_c$. Additionally, we construct 25 groups of classes based on their means in standardized test scores and 75 groups of classes defined by their variances in standardized test scores. Together, they constitute 1,875 groups $e_c$ that we interact with class grades. For the sake of notational simplicity, we henceforth designate our controls by $\Gamma_{ics}$.

Columns 1 and 2 of Table \ref{robcheck_g2g8_class_cohort} - Panel C shows that this specification ensures that both ranks are uncorrelated with previous measures of ability. This is evidence that our controls credibly account for ability and that our claim to identification of a perception effect is warranted, as this effect can logically impact only later outcomes and not past ones. Importantly, we chose the specification so that both coefficients are small and uncorrelated with measures of ability in grade 2. Column 3 of Panel C shows that such a specification implies that the invisible rank has a small, non-significant effect on performances in middle school, thereby testifying to the relevance of the ability placebo check.

However, there is no free lunch, as we have to include student fixed effects to effectively control for ability. The rank effect is therefore estimated off of within-student variations. This should not be a problem for estimation: The correlation between the visible ranks in Italian and math is 0.76 in our sample and an analysis of the residuals of the regression of one on the other reveals that, holding the rank in Italian constant, there is a variation of at least one (four) math rank for 70\% (20\%) of our sample.

Although this precludes us from investigating whether the effect of the rank is different by subjects, this is the price to pay to avoid spuriously capturing any ability component.

\section{Main Results}\label{sec:main_results}

\subsection{Effect of the Rank on Subsequent Academic Outcomes}

Our main results are estimated from our main dataset and are presented in Table \ref{main_spec_class_cohort}. Panel A reports the results of the estimation of Equation \eqref{eq:both_ranks_uncond}. Again, we observe that, unconditionally, the visible and the invisible ranks comparably affect later outcomes. 

Panel B reports the results of our main specification Equation \eqref{eq:main_specification}. In both grades 8 and 10, the coefficient of the undisclosed rank is close to zero and non-significant. The effect of the rank measured from class grades is large: \textit{ceteris paribus}, ranking at the top of class compared to the bottom in the last year of primary school leads to a gain of 8.5 percentiles in the national standardized grade distribution in middle school and 7.7 in high school.\footnote{The estimated impact of the rank on performances in middle school is not significantly different from the one estimated with our companion dataset and reported in Panel C - Column 2 of Table \ref{robcheck_g2g8_class_cohort} ($\Delta$ = 0.72, p-value = 0.50).} This is equivalent to moving up by about 37,000 ranks nationally.

We find a stronger rank effect than when using Equations \eqref{eq:mw} or \eqref{eq:dmw}. Columns 1 and 3 of Table \ref{tab:comparison_ourresults_vs_mwdmw} show that coefficients from these specifications and ours are significantly different at conventional levels. Columns 2 and 4 show that, in grade 8, the effect is 43\% smaller than ours when estimated through \eqref{eq:mw} and 76\% smaller when estimated through \eqref{eq:dmw}. In grade 10, the respective underestimations are 19\% and 69\%.

How can we account for this substantial discrepancy? The previous methods suffer from omitted variable bias by construction. However, its sign is a priori unclear, even though Table \ref{replication_mw_dmw_ability_placebo} suggests that it is likely positive. However, this upward bias may be more than offset by the use of a proxy only loosely related to the rank actually perceived by students, which could lead to an attenuation bias. How the two effects interact is unclear, although it seems to yield an overall downward bias in our case. 

\subsection{Heterogeneity}\label{subsec:results_nonlinear_rankeffect}

\paragraph{Non-Linear Rank Effects.} To explore potential non-linear rank effects, we run the following specification:
\begin{align}\label{eq:main_specification_ventiles}
\begin{split}
    T^g_{ics} = \alpha & + \sum_{v \in \left\{ V,I\right\}} \left( \beta^v_{top} \mathbf{1}[R^v_{ics} = 1] + \beta^v_{bot} \mathbf{1}[R^v_{ics} = 0] + \sum\limits_{\substack{1\leq k \leq 20 \\ k\neq 10}} \beta^v_k \mathbf{1}[d^v_{ics} = k]  \right)  + \Gamma_{ics} + \varepsilon_{ics} 
\end{split}
\end{align}
where $d^V_{ics}$ and $d^I_{ics}$ are the respective ventiles of student's $i$ visible and invisible ranks in class $c$ and subject $s$.

The reference group is the 10th ventile. Figure \ref{fig:nonlinear_rank_effect_tworanks} plots $\left\{(\hat{\beta}^V_k,\hat{\beta}^I_k)\right\}_{k=1}^{20}$. As expected, the effect of the invisible rank is close to zero and imprecisely estimated at all ventiles and for both grades. 

In grade 8, the effect of the visible rank is mostly non-negative: students ranking below the 10th percentile do not seem to be negatively impacted, but students ranking above benefit greatly from it, with a linear increase in each ventile from the 11th. In contrast, in grade 10, the rank effect is linear across the distribution. In both cases, coefficients around the median are imprecisely estimated, suggesting that the rank may be less salient to students who stand in the middle of the class distribution.

\begin{figure}[H]
\centering
\begin{subfigure}[b]{0.45\textwidth}
\centering
\caption{Grade 8 - Visible Rank}
\label{subfig:rankeffect_gender_g8_salient}
\includegraphics[width=\textwidth]{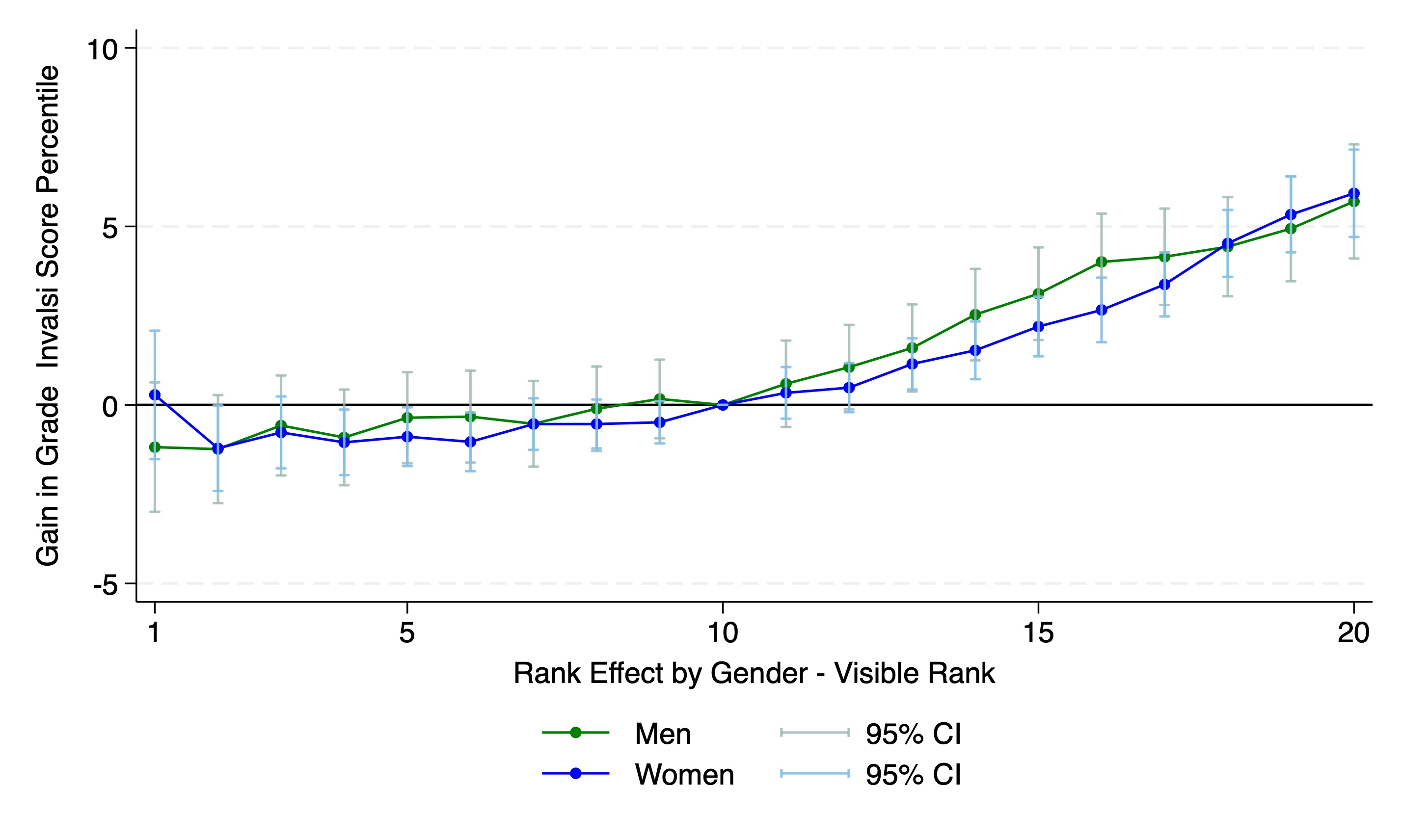}
\end{subfigure}
\begin{subfigure}[b]{0.45\textwidth} 
\centering
\caption{Grade 8 - Invisible Rank}
\label{subfig:rankeffect_gender_g8_undisclosed}
\includegraphics[width=\textwidth]{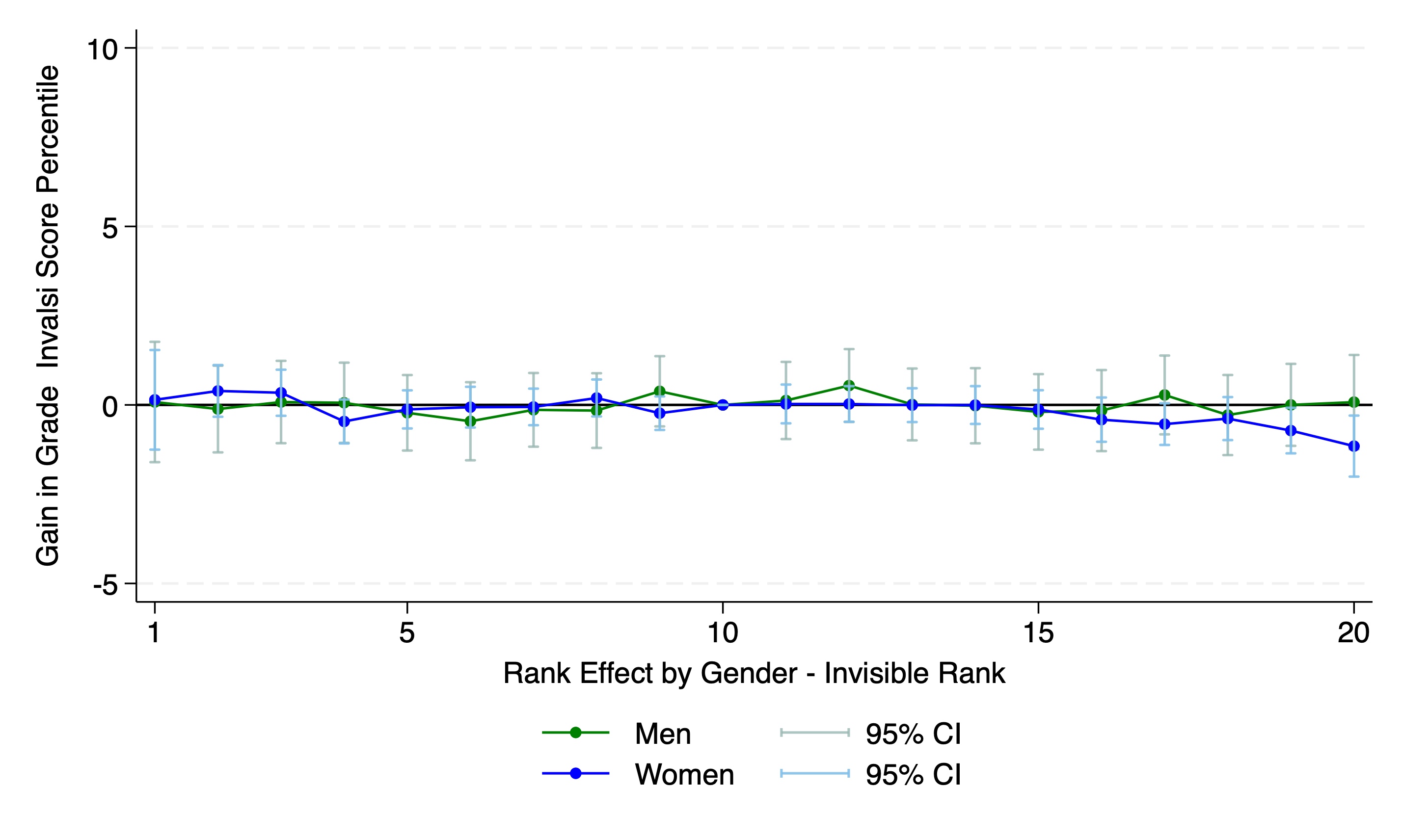}
\end{subfigure}
\begin{subfigure}[b]{0.45\textwidth}
\centering
\caption{Grade 10 - Visible Rank}
\label{subfig:rankeffect_gender_g10_salient}
\includegraphics[width=\textwidth]{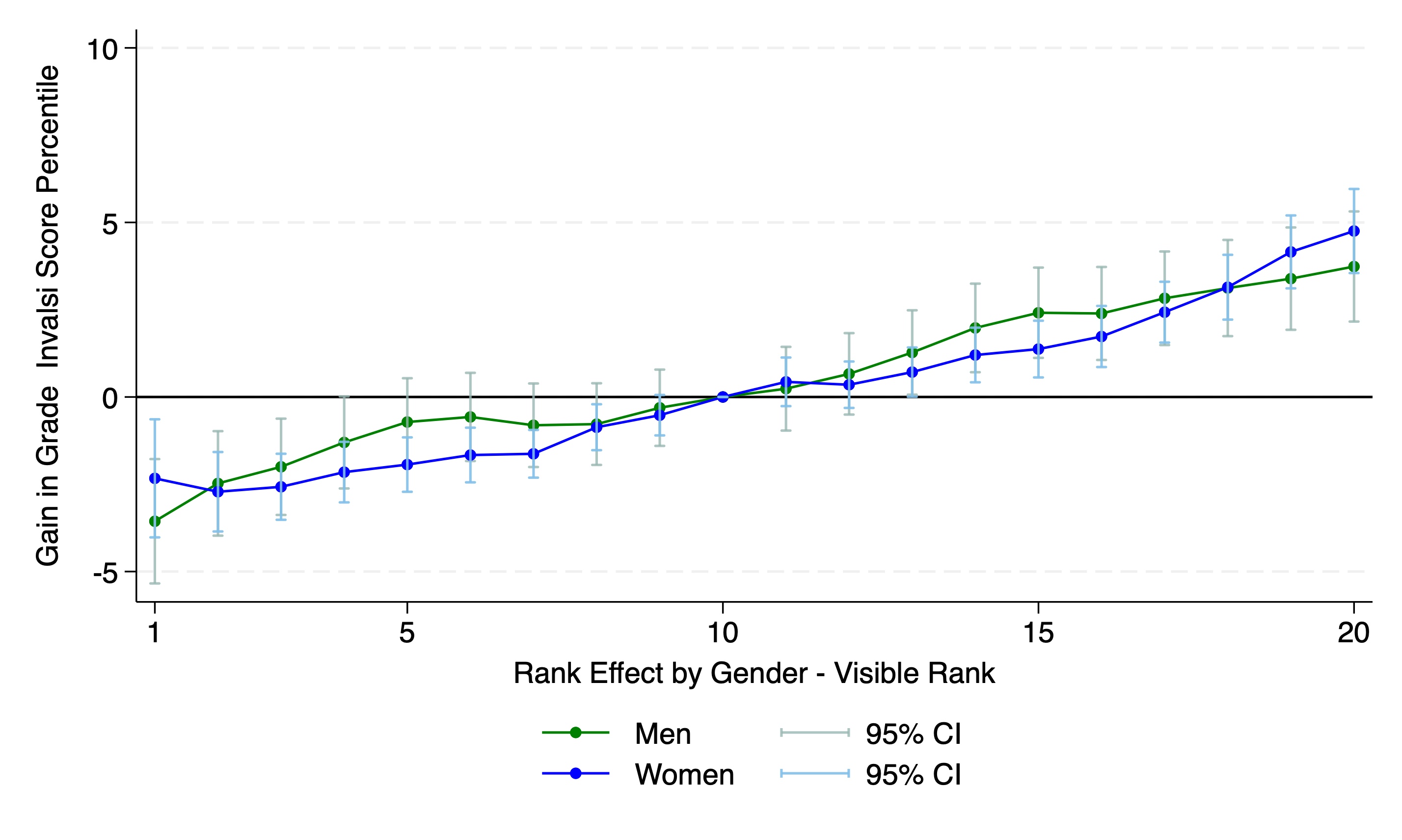}
\end{subfigure}
\begin{subfigure}[b]{0.45\textwidth} 
\centering
\caption{Grade 10 - Invisible Rank}
\label{subfig:rankeffect_gender_g10_undisclosed}
\includegraphics[width=\textwidth]{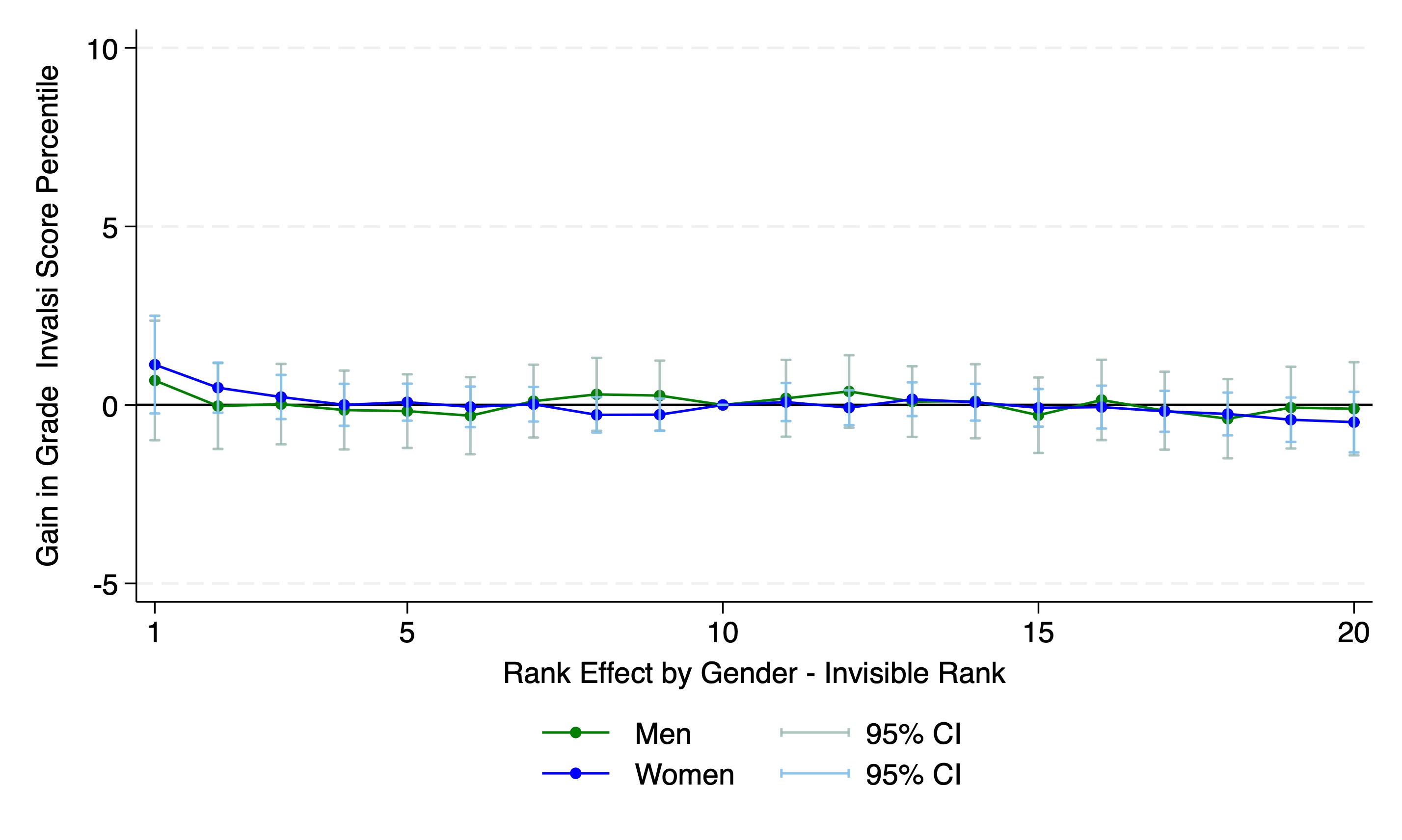}
\end{subfigure}
\caption{Non-Linear Rank Effects - Gender}
\label{fig:nonlinear_rank_effect_gender}
\vspace{.2cm}
\footnotesize \begin{tabular}{p{14cm}}
\setstretch{1} \textbf{Note}: This Figures compare the effect of the visible (from class scores) and invisible (from standardized test scores) ranks by ventiles and gender. See Section \ref{subsec:results_nonlinear_rankeffect} for details.
\end{tabular}
\end{figure}

\begin{figure}[H]
\centering
\begin{subfigure}[b]{0.45\textwidth}
\centering
\caption{Grade 8 - Visible Rank}
\label{subfig:rankeffect_immigrant_g8_salient}
\includegraphics[width=\textwidth]{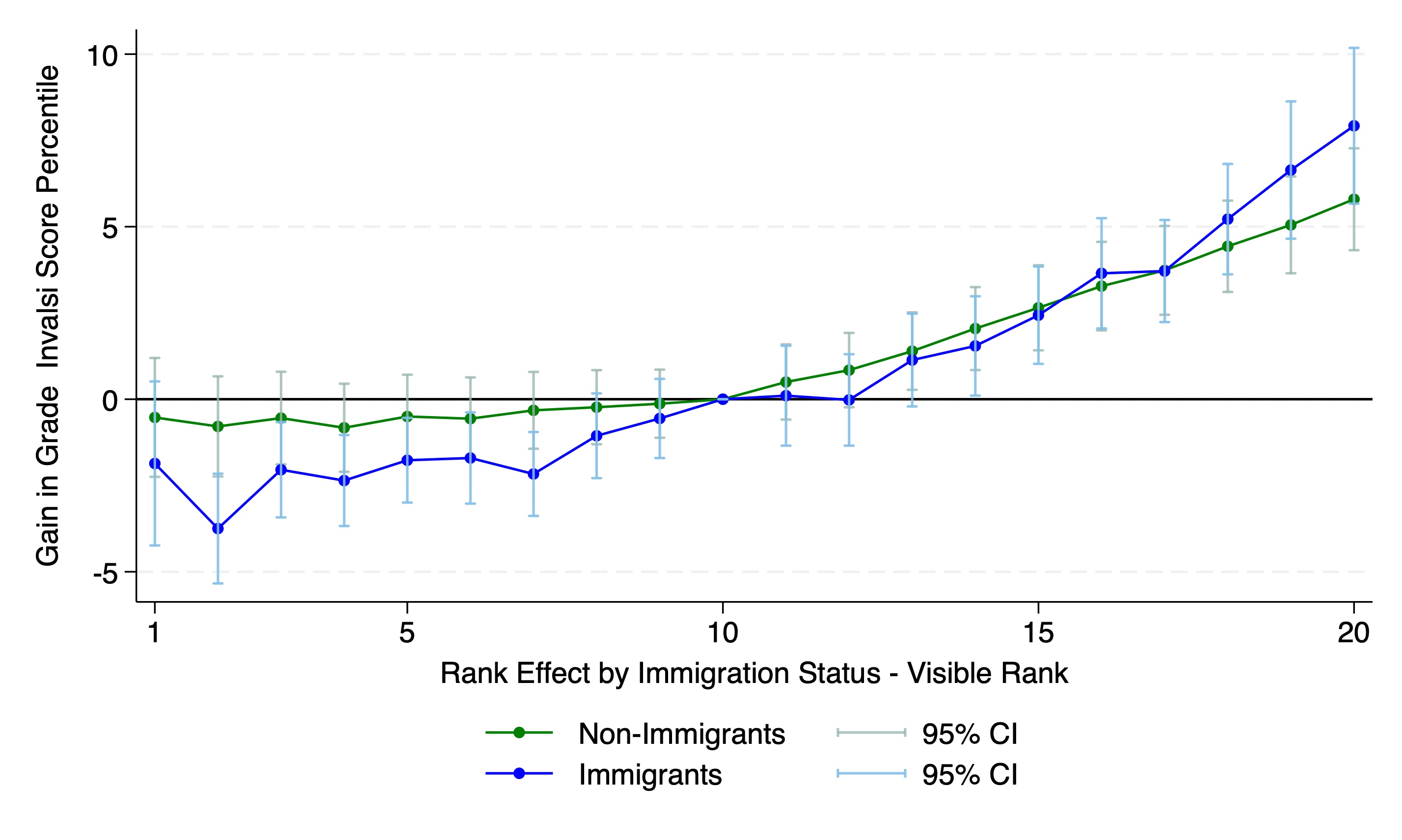}
\end{subfigure}
\begin{subfigure}[b]{0.45\textwidth} 
\centering
\caption{Grade 8 - Invisible Rank}
\label{subfig:rankeffect_immigrant_g8_undisclosed}
\includegraphics[width=\textwidth]{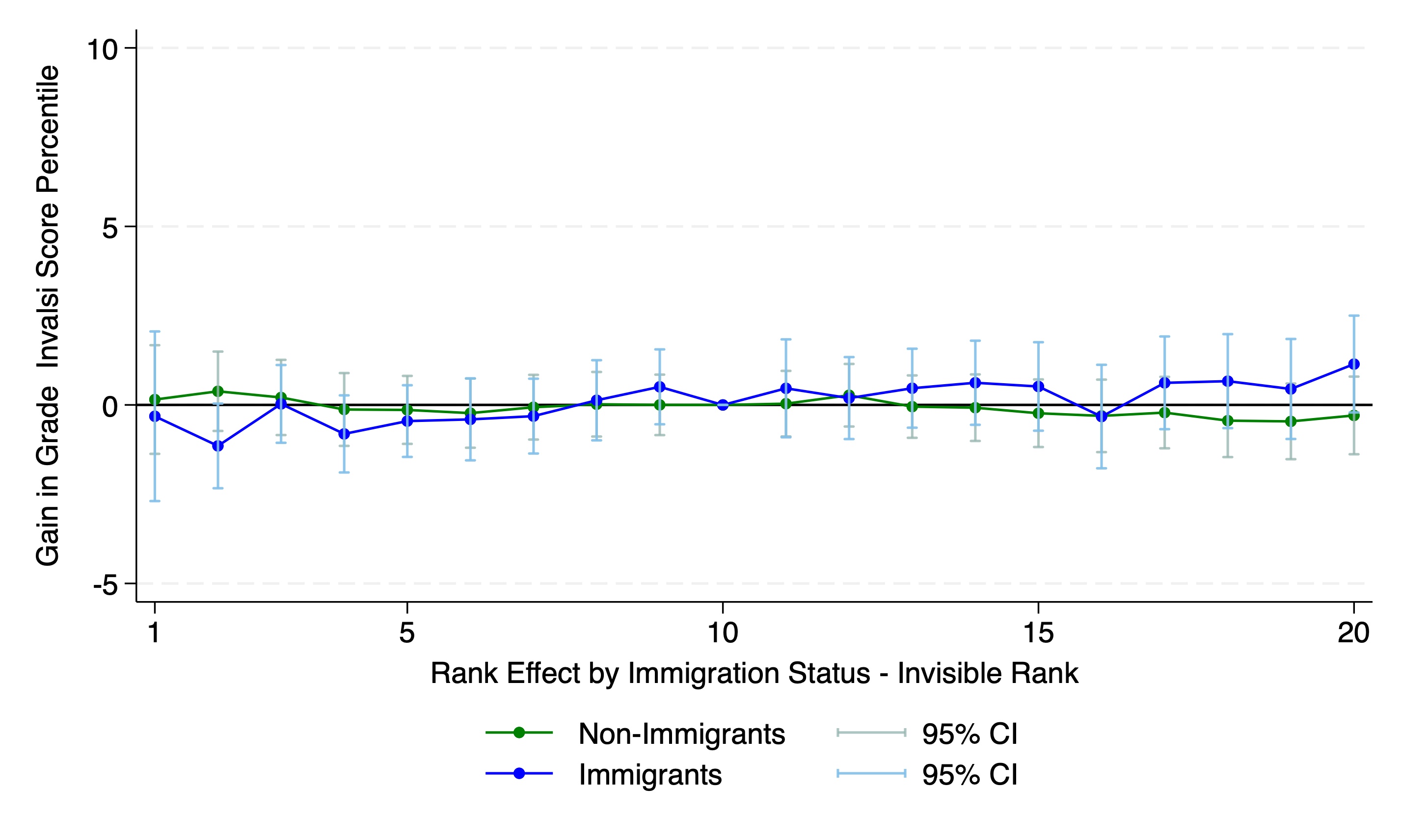}
\end{subfigure}
\begin{subfigure}[b]{0.45\textwidth}
\centering
\caption{Grade 10 - Visible Rank}
\label{subfig:rankeffect_immigrant_g10_salient}
\includegraphics[width=\textwidth]{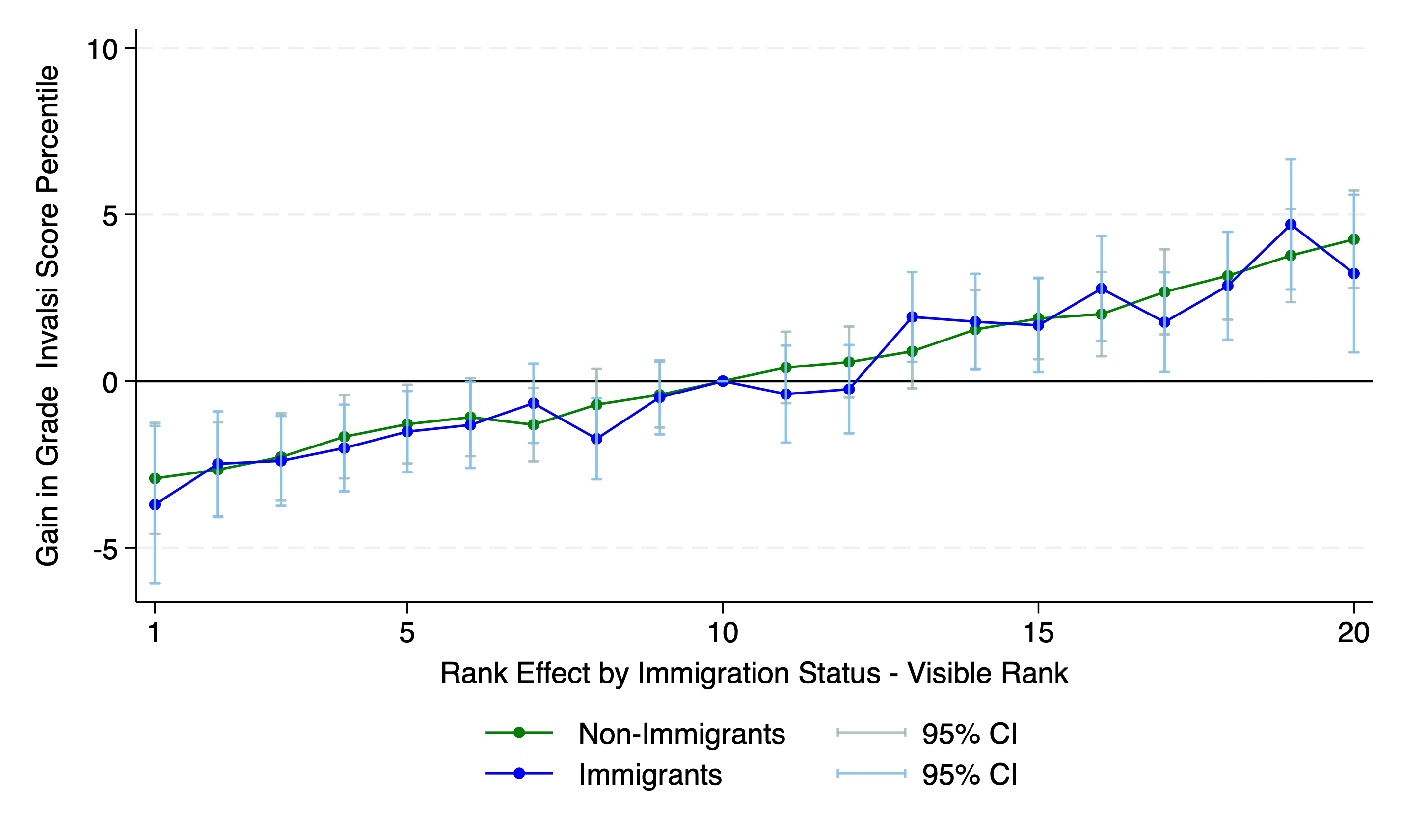}
\end{subfigure}
\begin{subfigure}[b]{0.45\textwidth} 
\centering
\caption{Grade 10 - Invisible Rank}
\label{subfig:rankeffect_immigrant_g10_undisclosed}
\includegraphics[width=\textwidth]{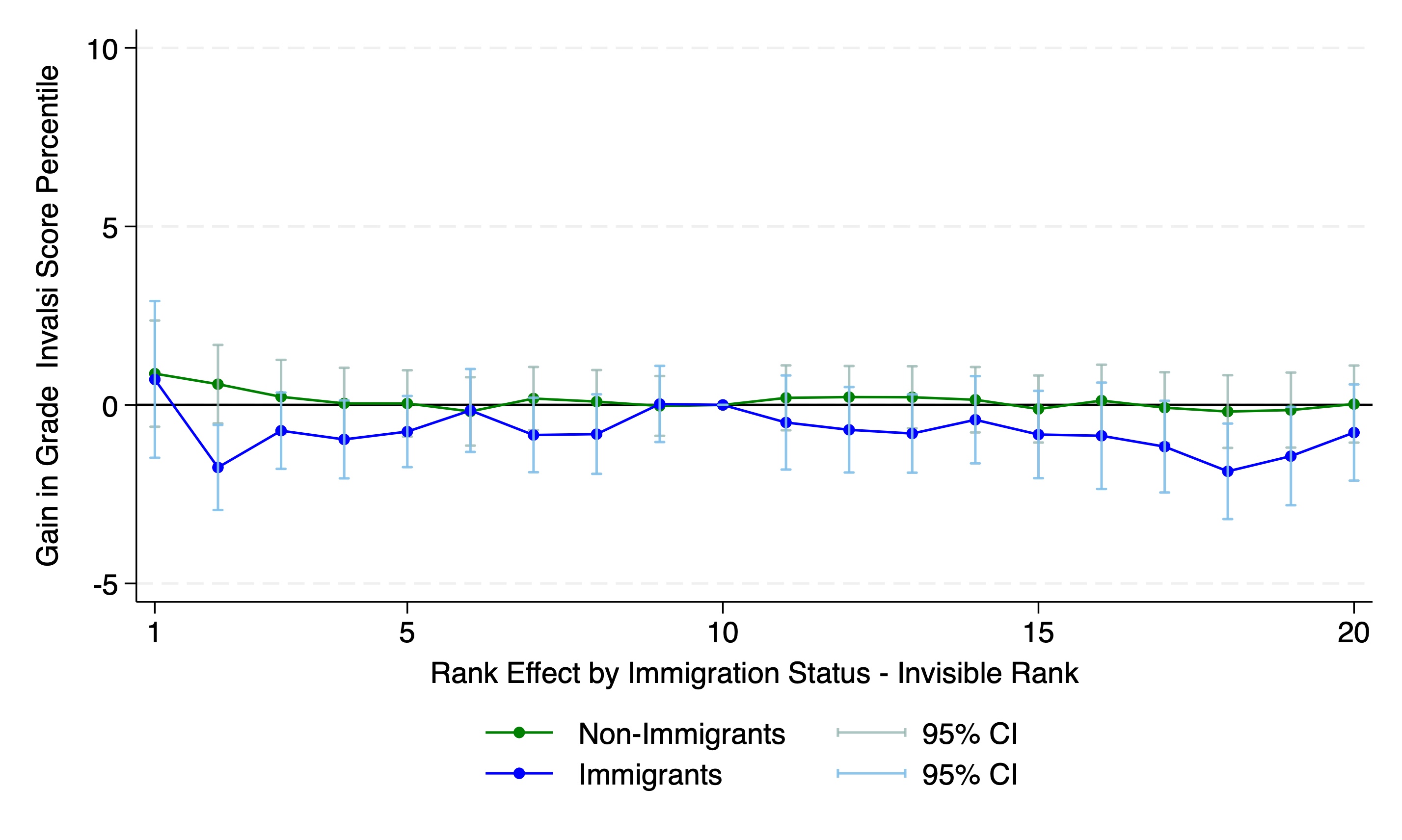}
\end{subfigure}
\caption{Non-Linear Rank Effects - Immigrant}
\label{fig:nonlinear_rank_effect_immigrant}
\vspace{.2cm}
\footnotesize \begin{tabular}{p{14cm}}
\setstretch{1} \textbf{Note}: This Figures compare the effect of the visible (from class scores) and invisible (from standardized test scores) ranks by ventiles and immigrant status. See Section \ref{subsec:results_nonlinear_rankeffect} for details.
\end{tabular}
\end{figure}

\begin{figure}[H]
\centering
\begin{subfigure}[b]{0.45\textwidth}
\centering
\caption{Grade 8 - Visible Rank}
\label{subfig:rankeffect_lowses_g8_salient}
\includegraphics[width=\textwidth]{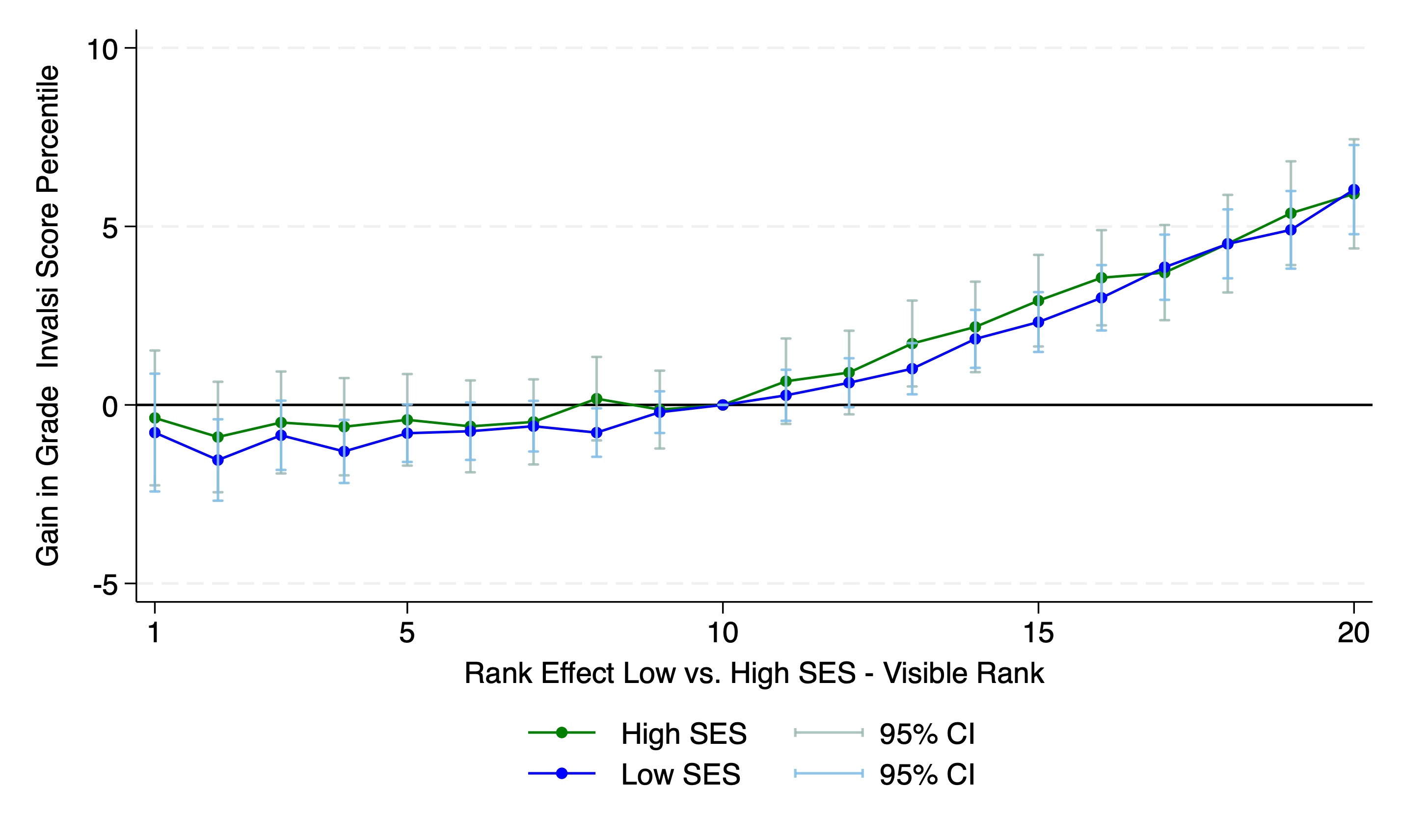}
\end{subfigure}
\begin{subfigure}[b]{0.45\textwidth} 
\centering
\caption{Grade 8 - Invisible Rank}
\label{subfig:rankeffect_lowses_g8_undisclosed}
\includegraphics[width=\textwidth]{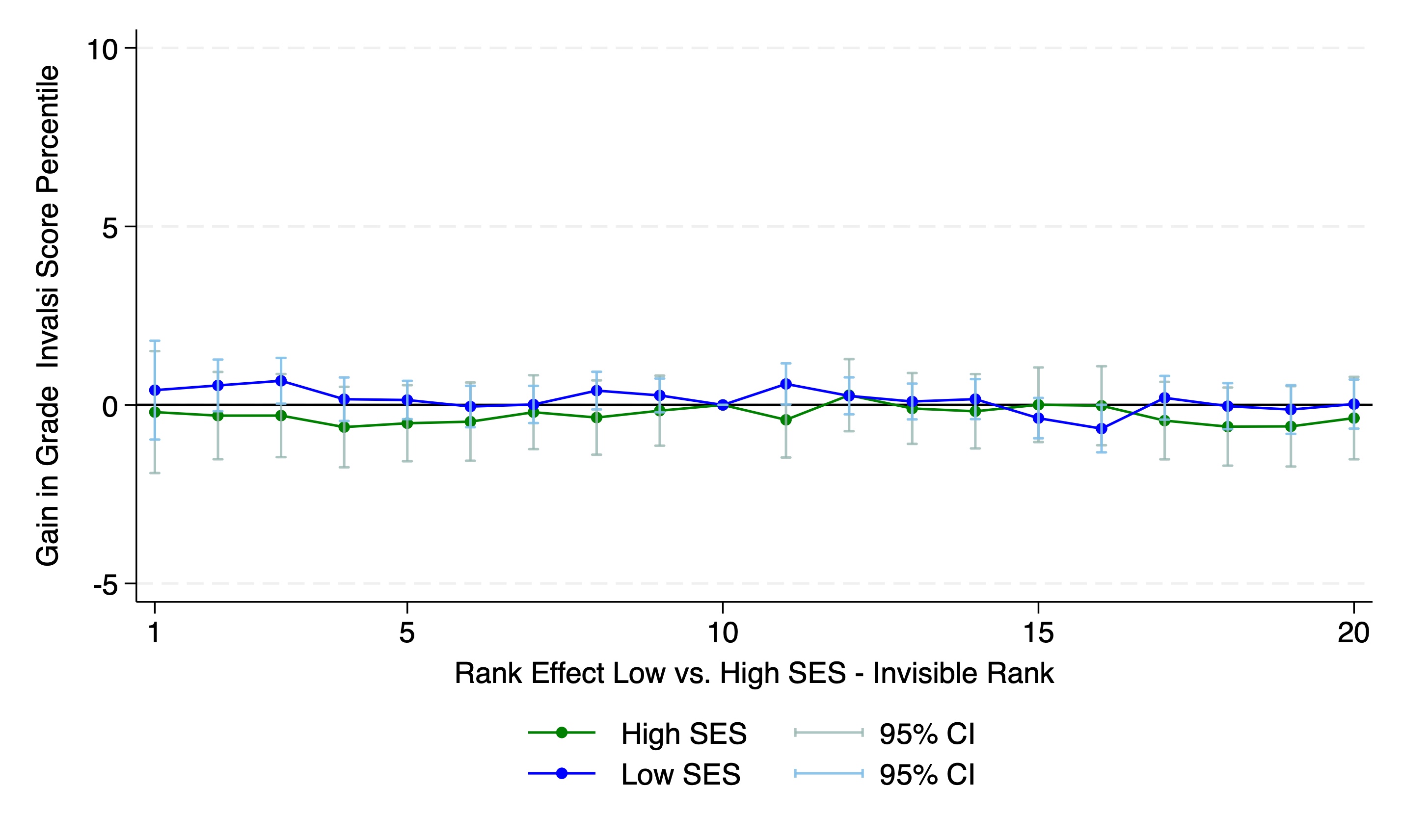}
\end{subfigure}
\begin{subfigure}[b]{0.45\textwidth}
\centering
\caption{Grade 10 - Visible Rank}
\label{subfig:rankeffect_lowses_g10_salient}
\includegraphics[width=\textwidth]{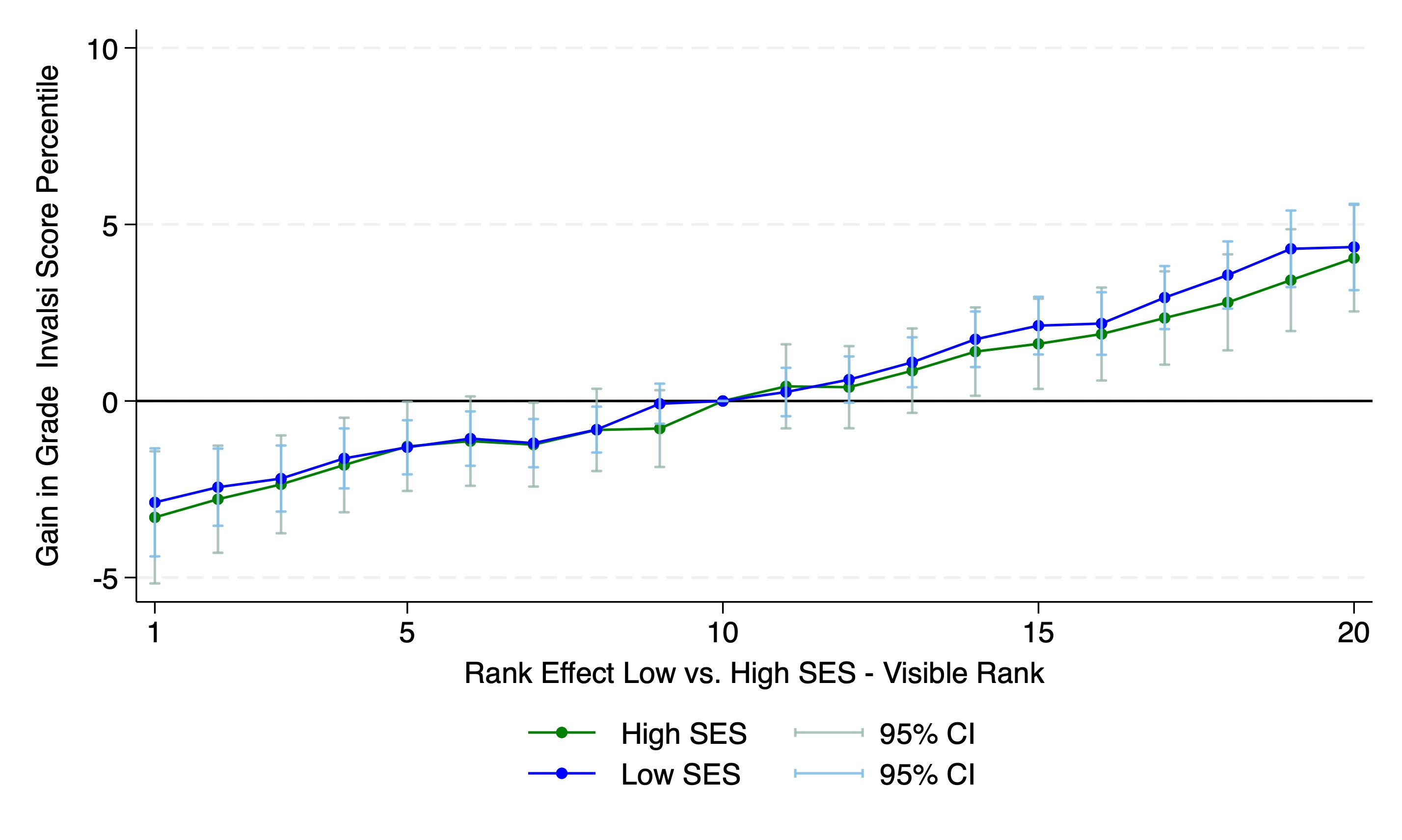}
\end{subfigure}
\begin{subfigure}[b]{0.45\textwidth} 
\centering
\caption{Grade 10 - Invisible Rank}
\label{subfig:rankeffect_lowses_g10_undisclosed}
\includegraphics[width=\textwidth]{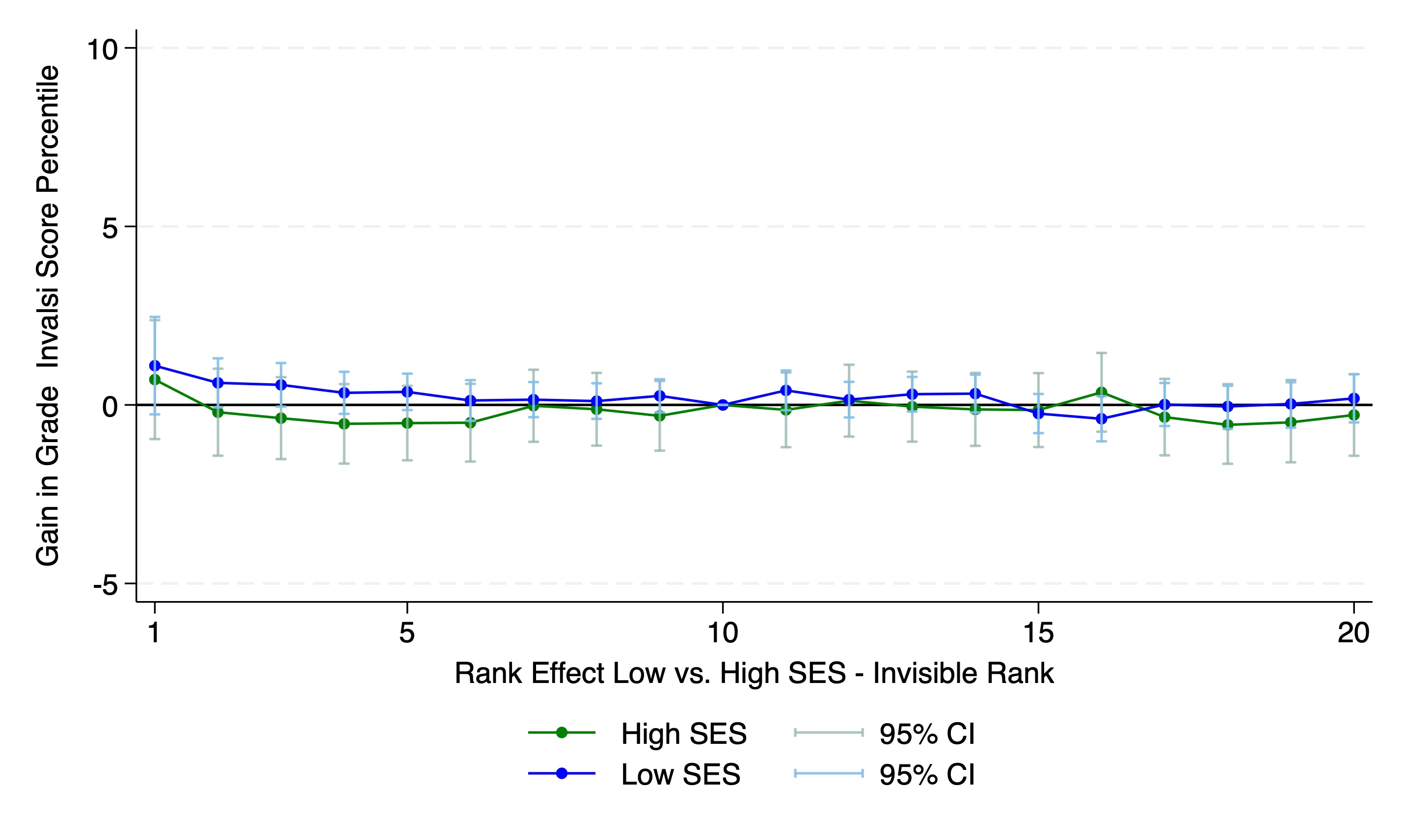}
\end{subfigure}
\caption{Non-Linear Rank Effects - Low vs. High SES}
\label{fig:nonlinear_rank_effect_lowses}
\vspace{.2cm}
\footnotesize \begin{tabular}{p{14cm}}
\setstretch{1} \textbf{Note}: This Figures compare the effect of the visible (from class scores) and invisible (from standardized test scores) ranks by ventiles and low/high SES. See Section \ref{subsec:results_nonlinear_rankeffect} for details.
\end{tabular}
\end{figure}

\paragraph{Heterogeneity by Demographics.} Recent literature has shown that peer effects might vary by gender \citep{zarate2023uncovering, lee2020causal, bargagli2020heterogeneous}. To investigate whether the rank affects students differently based on their demographics, we estimate Equation \eqref{eq:main_specification_ventiles} where we interact ventile coefficients with a dummy for female, immigrant and low SES. Results are plotted in Figures \ref{fig:nonlinear_rank_effect_gender}, \ref{fig:nonlinear_rank_effect_immigrant} and \ref{fig:nonlinear_rank_effect_lowses} respectively. Again, the estimates of the invisible rank ventiles are close to zero and imprecisely estimated. However, no significant difference across groups is detected on the visible rank ventiles.

\paragraph{Effect of Class Size.}\label{subsec:class_size}
We now explore if the rank effect differs by class size. The effect of class size per se on student achievement is unclear \citep{hoxby_2000,angrist1999using}. A priori, it is also the case regarding how class size affects the rank effect. On the one hand, ranking high among a large student group might might bolster self-confidence and lead to a stronger rank effect. On the other hand, students in the middle of the distribution might find it harder to know where they stand when comparing to a large number of peers. To investigate this question, we interact the two ranks with class size in grade 5. We restrict the analysis to class whose size lies between 5 and 26 students. We discard the classes whose number of students is outside that range (representing 1.4\% of our sample), as there are too few observations by each class size to yield meaningful coefficients. We thus estimate the following regression:
\begin{align}\label{eq:main_specification_class_size}
\begin{split}
    T^g_{ics} = \alpha + \sum_{v \in \left\{ V,I\right\}}\sum\limits_{\substack{5\leq k \leq 25}} \left( \beta^v_k  R^v_{ics} \times \mathbf{1}[Size_{c} = k]  \right) + \Gamma_{ics} + \varepsilon_{ics} 
\end{split}
\end{align}
where $\mathbf{1}[Size_{c} = k]$ is an indicator equal to one if there are $k$ students in class $c$. Notice that the class fixed effect in $\Gamma_{ics}$ controls for class size already. 

\begin{figure}[H]
\centering
\begin{subfigure}[b]{0.45\textwidth}
\centering
\caption{Grade 8}
\label{subfig:rankeffect_class_size_g8}
\includegraphics[width=\textwidth]{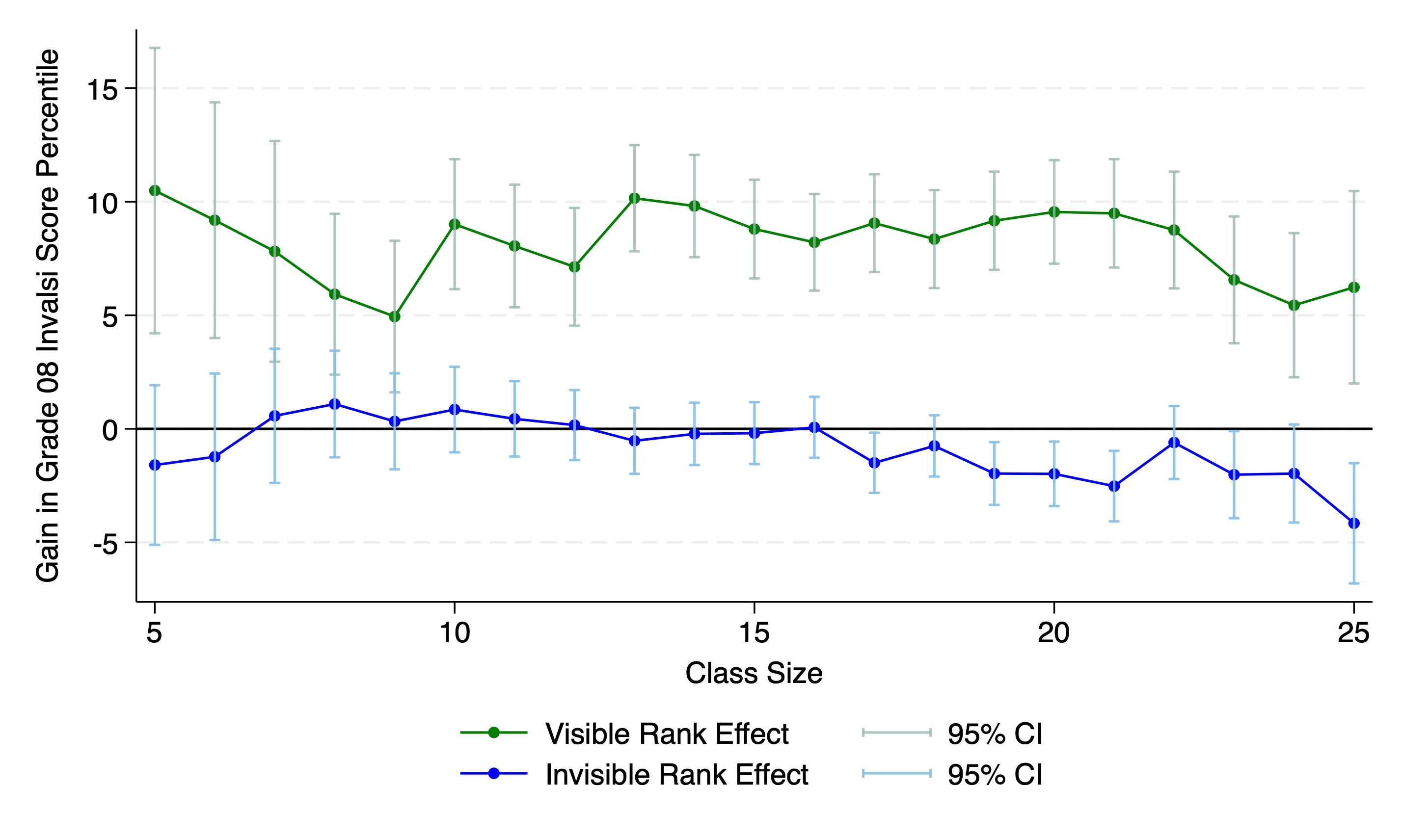}
\end{subfigure}
\begin{subfigure}[b]{0.45\textwidth} 
\centering
\caption{Grade 10}
\label{subfig:rankeffect_class_size_g10}
\includegraphics[width=\textwidth]{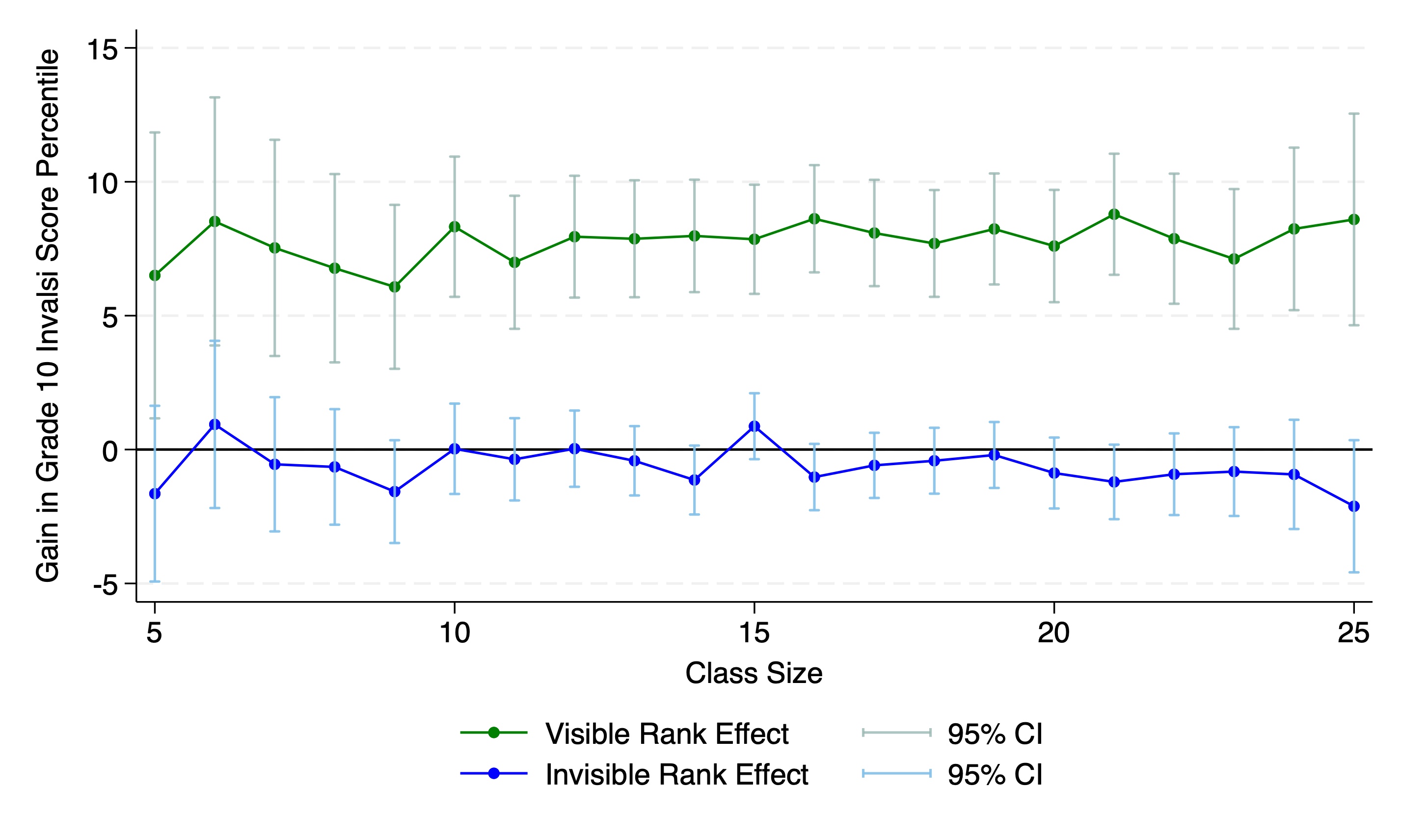}
\end{subfigure}
\caption{Rank Effect by Class Size}
\label{fig:rankeffect_class_size}
\vspace{.2cm}
\footnotesize \begin{tabular}{p{14cm}}
\setstretch{1} \textbf{Note}: This Figures compare the effect of the visible (from class scores) and invisible (from standardized test scores) ranks by class size, as estimated through Equation \ref{eq:main_specification_class_size}. See Section \ref{subsec:class_size} for details.
\end{tabular}
\end{figure}

Figures \ref{subfig:rankeffect_class_size_g8} and \ref{subfig:rankeffect_class_size_g10} plot $\left\{(\hat{\beta}^V_k,\hat{\beta}^I_k)\right\}_{k=5}^{25}$. Results suggest that the size of the reference group does not meaningfully influence the rank effect. 

In grade 8, coefficients indicate that the visible rank matters less in very small classes (with fewer than nine students) and in very large classes (with more than 23 students). The invisible rank is very small and non-significant at almost every size. In grade 10, the visible rank effect seems to be the same across sizes, although the effect in classes of 24 students appears to be much stronger - but this may be due to the few observations we have for classes of these sizes. As expected, the coefficients on the invisible rank is close to zero and imprecisely estimated at all sizes.

\paragraph{Effect of Peers' and Individual Ability.}\label{subsec:peerquality}
The importance of peers' behaviors and abilities on student outcomes has been documented \citep[see, e.g.,][]{carrell2018long, bargagli2022heterogeneous}. We thus conclude this section with an exploration of how peers' and individual ability impact the rank effect. In both cases, we use standardized test scores to proxy for ability, as there are unbiased and comparable across classes. 

Being first among higher-achieving peers might be more rewarding than among lower-quality peers. To explore this, we proxy peer quality by the leave-out mean of average standardized test scores in subject $s$ in grade 5. For each subject, we then group classes by ventile of peer quality $d_{ics}$ and estimate the following regression:
\begin{align}\label{eq:main_specification_peerquality}
\begin{split}
    T^g_{ics} = \alpha + \beta R^V_{ics} + \sum\limits_{\substack{1\leq k \leq 20 \\ k\neq 10}} \left( \beta^V_k R^V_{ics} \times \mathbf{1}[d^V_{ics} = k]  \right)  + \sum\limits_{1\leq k \leq 20} \left( \beta^I_k R^I_{ics} \times \mathbf{1}[d^I_{ics} = k]  \right) + \Gamma_{ics} + \varepsilon_{ics} 
\end{split}
\end{align}

\begin{figure}[H]
\centering
\begin{subfigure}[b]{0.45\textwidth}
\centering
\caption{Grade 8}
\label{subfig:rankeffect_peerquality_g8}
\includegraphics[width=\textwidth]{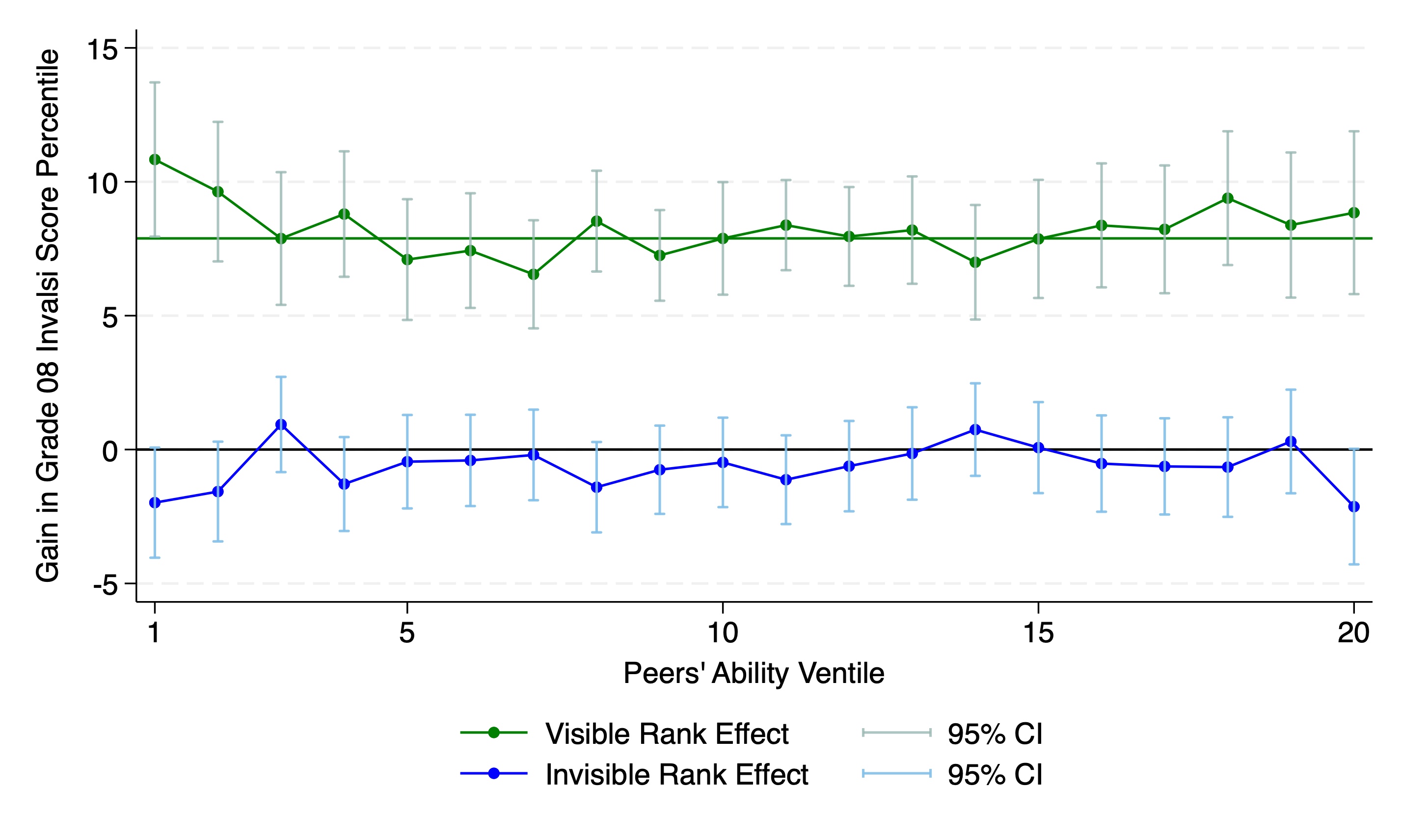}
\end{subfigure}
\begin{subfigure}[b]{0.45\textwidth} 
\centering
\caption{Grade 10}
\label{subfig:rankeffect_peerquality_g10}
\includegraphics[width=\textwidth]{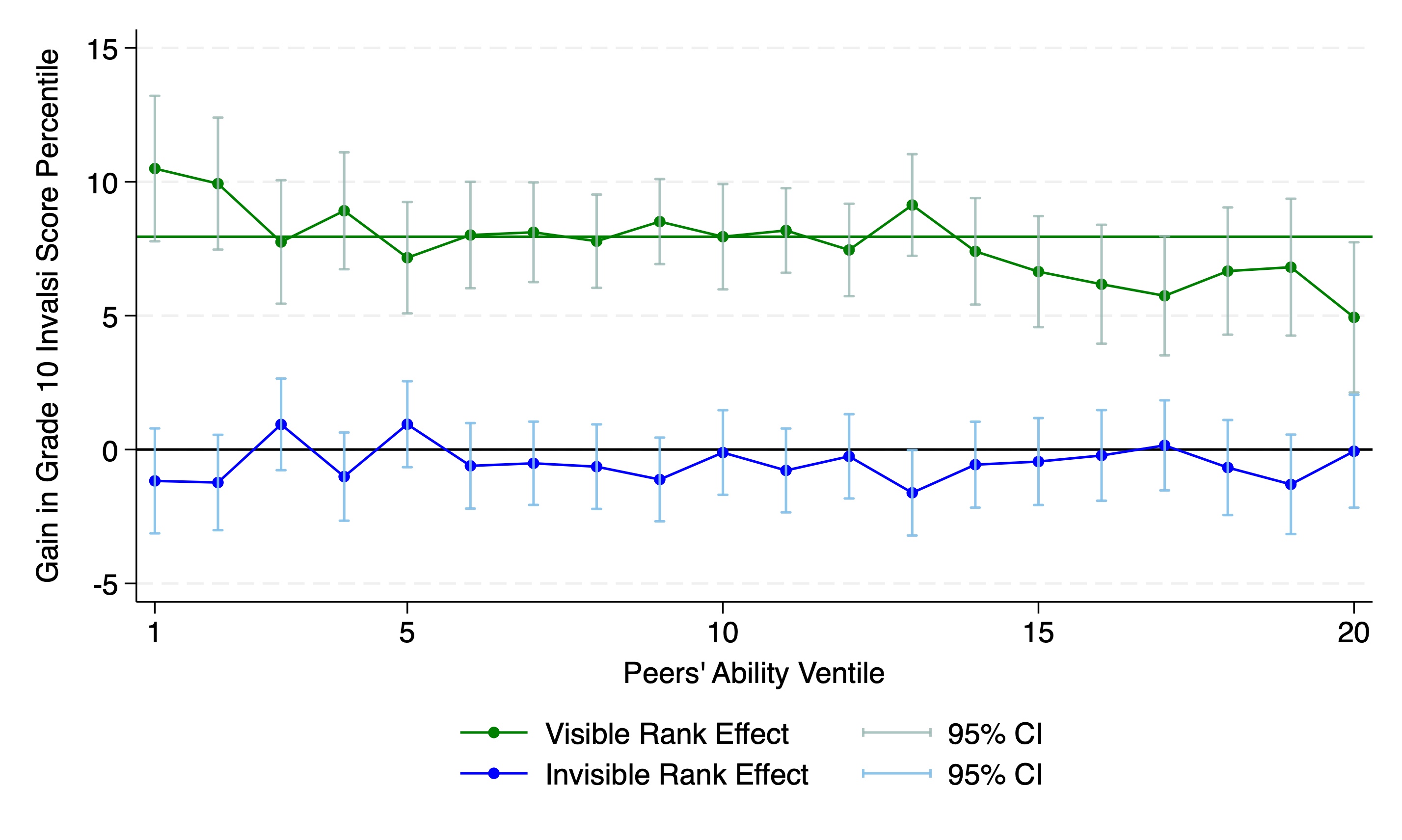}
\end{subfigure}
\caption{Rank Effect by Peer Quality}
\label{fig:rankeffect_peerquality}
\vspace{.2cm}
\footnotesize \begin{tabular}{p{14cm}}
\setstretch{1} \textbf{Note}: This Figures compare the effect of the visible (from class scores) and invisible (from standardized test scores) ranks by class size, as estimated through Equation \ref{eq:main_specification_peerquality}. The colored lines are drawn at the value of the 10th ventile for each rank type. For the visible rank, confidence intervals for every ventile except the 10th have to be compared to the green line. See Section \ref{subsec:peerquality} for details.
\end{tabular}
\end{figure}

Figures \ref{subfig:rankeffect_peerquality_g8}
and \ref{subfig:rankeffect_peerquality_g10} plots plot $\left\{(\hat{\beta}^V_k,\hat{\beta}^I_k)\right\}_{k=1}^{20}$. For the visible rank, we also report the coefficient of the 10th ventile ($\hat{\beta}^V$) and the corresponding 95\% confidence interval. Therefore, confidence intervals for the visible rank that include the green line are not significantly different from the 10th ventile. Invisible rank ventile coefficients are compared to zero. 

As expected, the effect on the invisible rank is always small and not significantly different from zero. Regarding the visible rank, we do not detect any pattern: the rank effect is pretty much the same across every ventile.

\begin{figure}[H]
\centering
\begin{subfigure}[b]{0.45\textwidth}
\centering
\caption{Grade 8}
\label{subfig:rankeffect_peerquality_indivg8}
\includegraphics[width=\textwidth]{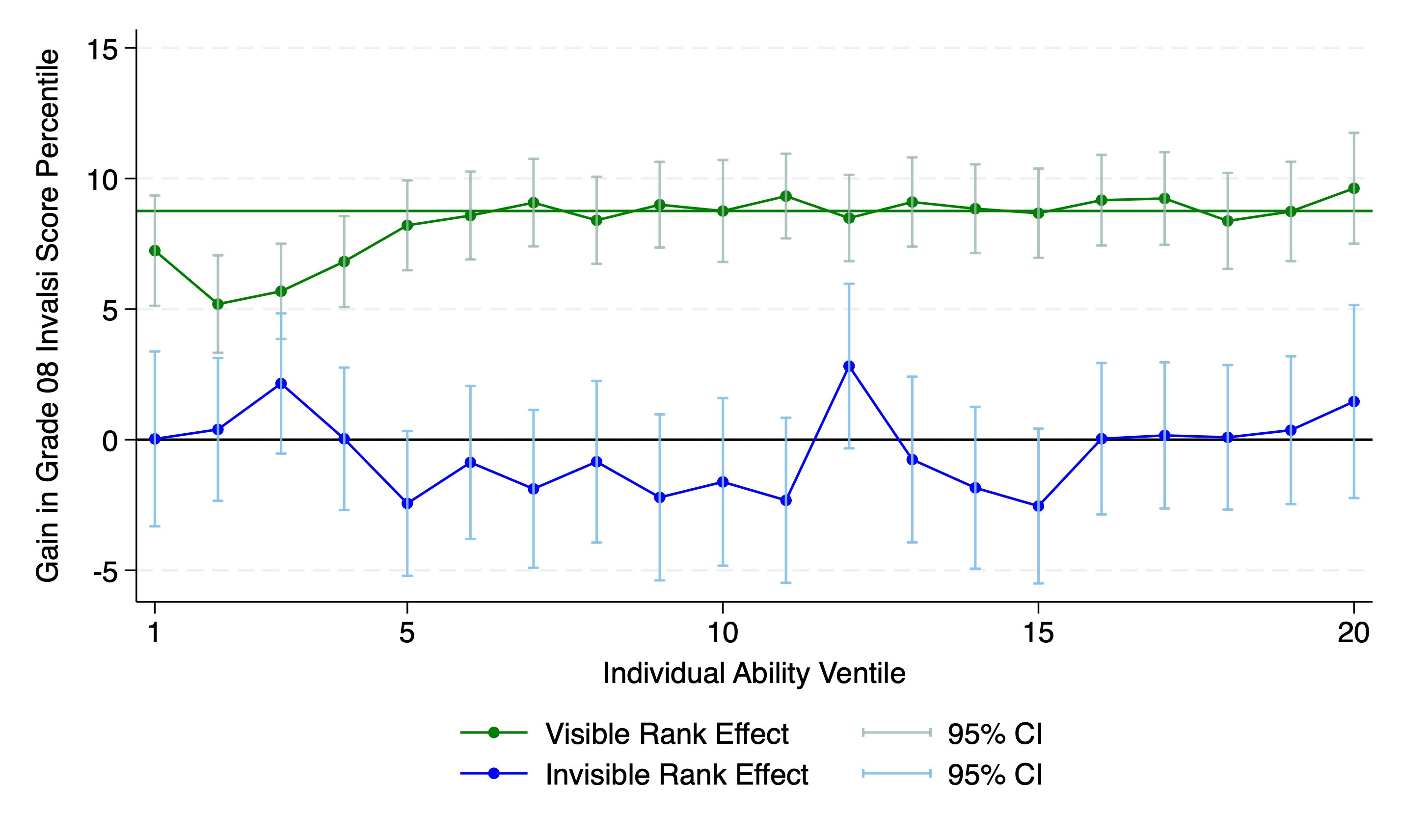}
\end{subfigure}
\begin{subfigure}[b]{0.45\textwidth} 
\centering
\caption{Grade 10}
\label{subfig:rankeffect_peerquality_indivg10}
\includegraphics[width=\textwidth]{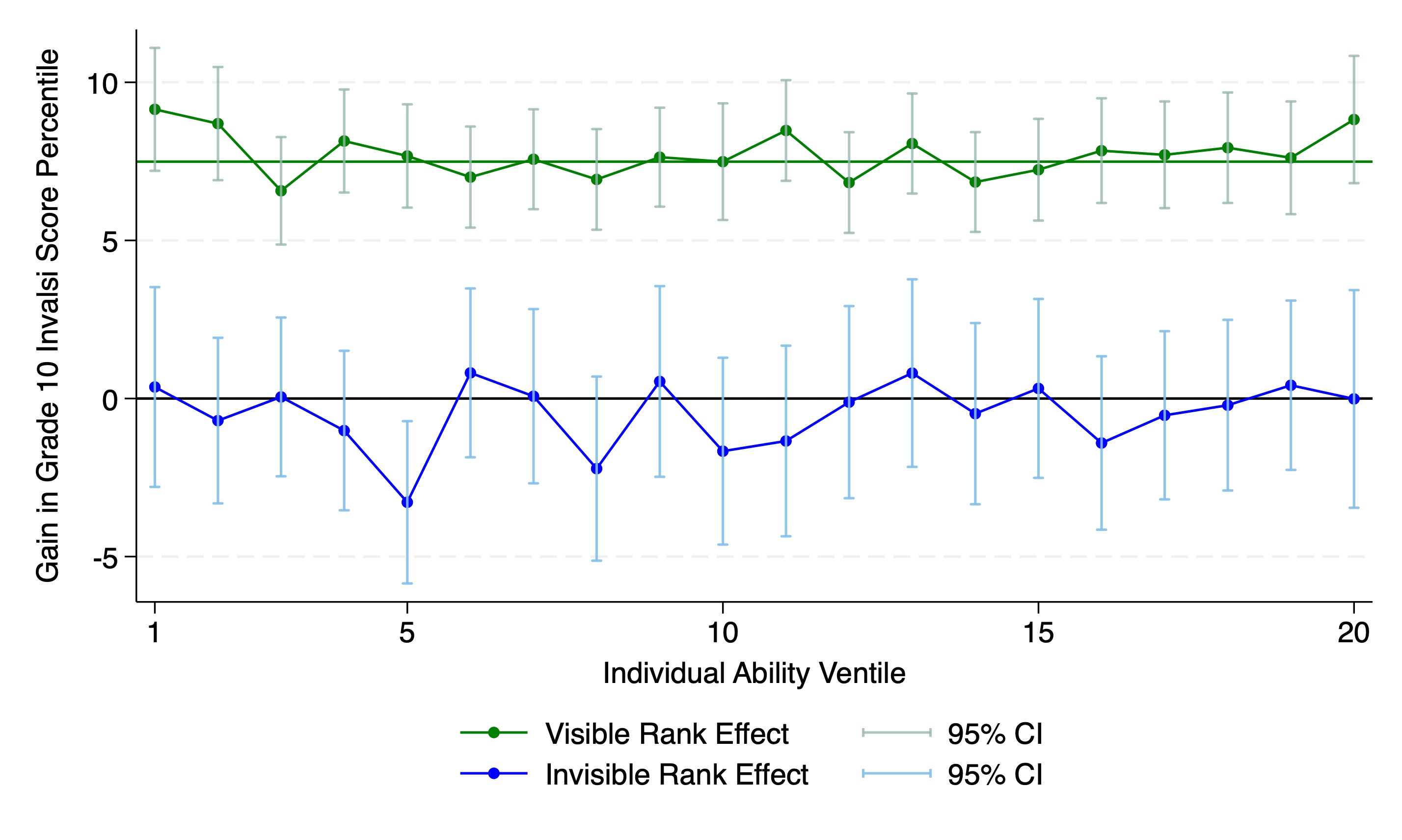}
\end{subfigure}
\caption{Rank Effect by Individual Ability}
\label{fig:rankeffect_peerquality_indiv}
\vspace{.2cm}
\footnotesize \begin{tabular}{p{14cm}}
\setstretch{1} \textbf{Note}: This Figures compare the effect of the visible (from class scores) and invisible (from standardized test scores) ranks by class size, as estimated through Equation \ref{eq:main_specification_peerquality}. The colored lines are drawn at the value of the 10th ventile for each rank type. For the visible rank, confidence intervals for every ventile except the 10th have to be compared to the green line. See Section \ref{subsec:peerquality} for details.
\end{tabular}
\end{figure}

We perform the same exercise but interacting the rank with ventiles in individual standardized test scores. Figures \ref{subfig:rankeffect_peerquality_indivg8} and \ref{subfig:rankeffect_peerquality_indivg10} reports the coefficients. As before, no significant different is detected in the rank effect across ventiles.

\section{Rank Effects and Class Effects}\label{sec:rankvsclass}
In this section, we seek to quantify the importance of the rank effect relative to the class effect. To extend our metaphor, we have estimated that, conditional on being in a village or in Rome, it is better to be ranked higher. How does this effect compare to that of being in a village compared to being in Rome, conditional on rank? 

\textbf{Measuring Class Quality.} In the same vein as \cite{chettyetal_2011} and \citetalias{dmw21}, we estimate class quality as class value-added from the class-by-subject fixed effects in Equation \eqref{eq:main_specification}. Thus defined, class quality is a measure of all class inputs (e.g., teacher and peer effects, class size, and amenities) affecting subsequent academic performances.

We find that increasing class quality by one standard deviation is associated with a gain of 13.0 national percentiles in the standardized test score distribution in middle school and 13.3 in high school. By contrast, an increase in class rank by one standard deviation results in a gain of 2.2 national percentiles in the standardized test score distribution in middle school and 1.9 in high school. Therefore, the class effect is about six times larger than the rank effect: to compensate for a decrease in one standard deviation in class quality, the class rank would need to increase by six standard deviations.

\textbf{What drives primary school class quality?} To offer a tentative answer, we look at the correlation between class quality and the following observable characteristics at the class level in primary school: proportion of women and immigrants, size, average SES, and average standardized test score. Table \ref{reg_class_quality_cohort} reports the results, after standardizing all variables. We observe that class quality is strongly and positively correlated with the average SES and the average standardized test score in the primary school class. The proportion of women appears to be slightly positively correlated to class quality. The proportion of immigrants in primary school classes has an ambiguous effect: it is negatively correlated with the effect of class quality on performances in middle school but positively correlated with the effect of class quality on performances in high school. Lastly, we observe that class quality is also positively correlated to class size.

To highlight potential trade-offs more concretely, we examine how a student's class rank in primary school relates to (i) the class mean SES and (ii) the class mean achievement as measured by the average standardized test score. Column 1 of Table \ref{correlation_outcomes_scoreses} reports the correlation between class rank and the class mean SES and achievement, controlling for demographics, and a fixed effect in individual grade 5 test scores. An increase of one standard deviation in the class mean SES is associated with a 0.02 decrease in student's visible rank, which suggests a mild ``relative deprivation effect'' \citep{mayer_jencks_1989,meyer_1970}. Similarly, an increase of one standard deviation in the class mean achievement is associated with a 0.06 decrease in student's class rank.\footnote{Therefore, through the rank channel, an increase of one standard deviation in the class mean achievement leads to a loss of $0.06\times 8.4 = 0.5$ percentiles in the national distribution of standardized test scores in middle school.} 

We next estimate the impact of the class mean SES and standardized test score on students' future academic performances, controlling for both ranks in Grade 5, gender, immigration status and SES. Columns 2 and 3 of Table \ref{correlation_outcomes_scoreses} show that, holding a student's ranks constant in the distribution of both scores, an increase of one standard deviation in the class mean SES leads to a gain of 2.9 percentiles in the national distribution of standardized test scores in middle school and 4.2 in high school. An increase in one standard deviation in the class mean achievement generates a gain of 3.8 national percentiles in middle school and 3.4 in high school.

We can conclude about the net effect on performance of increasing the class mean SES or the class mean achievement by one standard deviation. The former yields a net gain of 2.7 national percentiles in middle school and 4 in high school; the latter yields a net gain of 3.3 national percentiles in middle school and 2.9 in high school. Therefore, it is worth trading a lower rank for the possibility of interacting with students with higher SES and higher levels of academic achievement.

This is at odds with some recent findings in the education literature which have shown that exposure to higher-achieving peers bears little effect on students' later outcomes \citep{abdulkadirouglu2014elite,dobbie2014impact}. Interestingly, both papers employ a regression discontinuity design. At the cutoff, the baseline peer test score mean  jumps from 0.17 to 0.5 std. Our estimation shows that the net effect should be positive but quite small. It is also possible that the RD leads to comparing students who rank first (at the left of the cutoff) with students who rank last (at the right). In that case, the rank effect could be large enough to offset any gain from exposure to higher-achieving peers.

\section{Mechanisms}\label{sec:mechanisms}
\subsection{Self-Sorting in Middle School and High School}\label{subsec:mecha_school_sorting}
We first set out to analyze the influence of class rank in primary school on choices of middle and high schools. Since 2014, and so in the period we study, schools have been legally mandated to make their average standardized test scores public, and this has become a way to attract students and increase enrollment \citep{lucifora_2020}. This makes the exploration of this channel relevant, as these scores can be seen by students, who can in turn take them into account when deciding which school to attend.  

An important feature of the Italian education system is that students self-select into middle and high schools. Nonetheless, middle-school enrollment is constrained by residency criteria and students can only choose among a few different local schools. They have much more leeway in selecting their high schools, by type (academic, vocational, or technical) and location. High schools are also specialized: for instance, academic high schools can be focused on either humanities or science. High schools usually do not discriminate based on grades, unless there is an oversupply of students.

In this section, we measure school quality by the average level of student achievement in standardized tests. We use an additional dataset, which closely resembles that used for our main analysis. It tracks students from an earlier cohort, which was in grade 8 in 2014, grade 10 in 2016, and grade 13 in 2019. Our measure of school quality is constructed using standardized test scores in 2014 for middle schools and 2016 for high schools. We compute the average by school and subject and convert the resulting measure to a percentile so that our school quality metric lies between 1 and 100.\footnote{Following Invalsi guidelines in computing the scores that are then made public, we exclude classes with a probability of cheating larger than 50\% from the computation at the school level. These schools are excluded from the analysis as they are not allowed to publicly release their standardized test results. This results in 40,902 students missing from our main sample.\label{footnote_schoolsorting}}

We estimate Equation \eqref{eq:main_specification} with school quality by subject as the dependent variable. Results are displayed in Table \ref{school_sorting}. As expected, neither the visible nor the invisible rank impacts middle school quality (Column 1). In contrast, the rank has a sizable effect on that of high school: going from last to first in one's primary school class leads to attending a high school whose average achievement in standardized tests is 4.5 national percentiles higher (Column 2). The coefficient on the invisible rank is significant but of negligible magnitude.

\begin{figure}[H]
\centering
\begin{subfigure}[b]{0.45\textwidth}
\centering
(a) Grade 8 \\
\includegraphics[width=\textwidth]{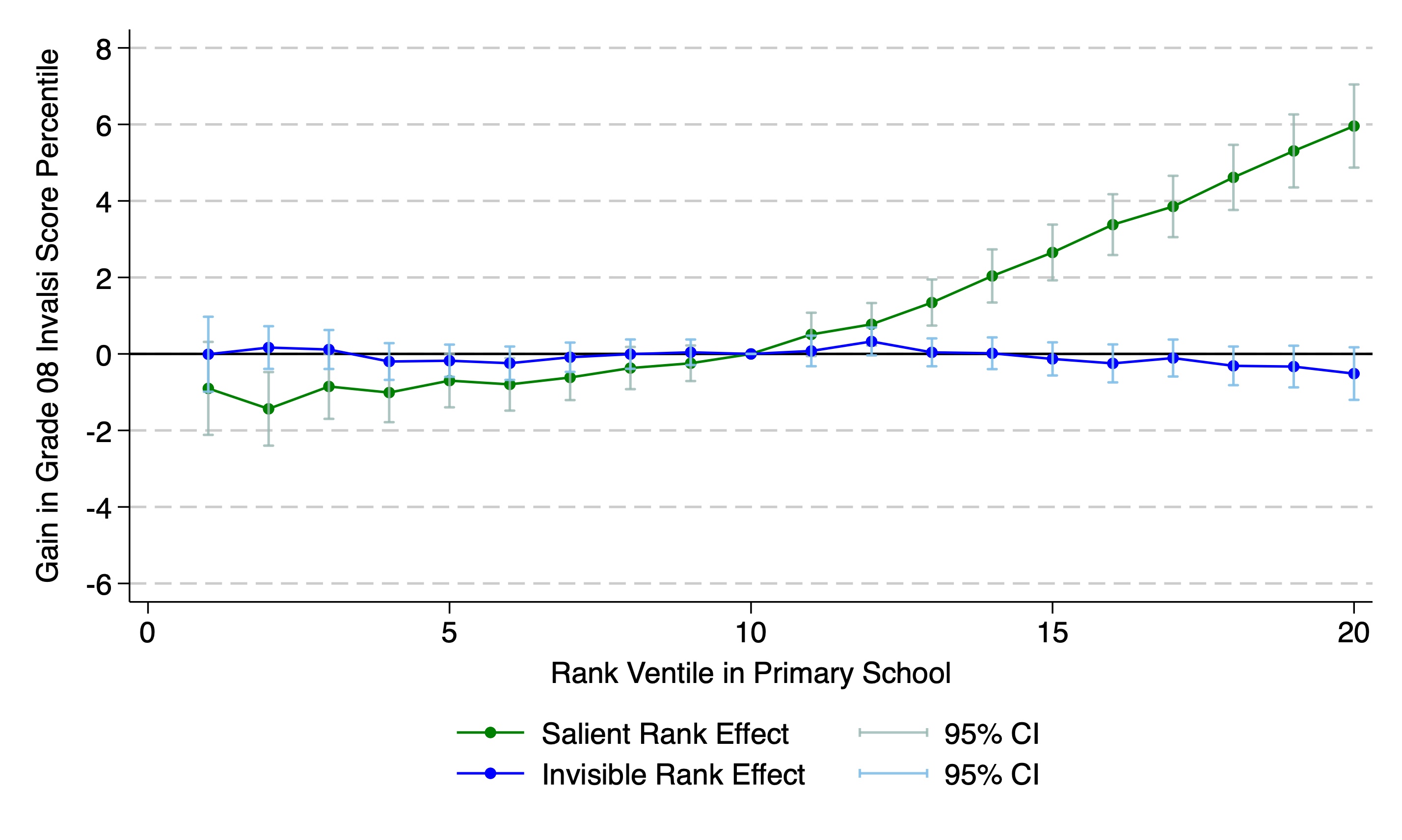}
\end{subfigure}
\begin{subfigure}[b]{0.45\textwidth} 
\centering
(b) Grade 10 \\
\includegraphics[width=\textwidth]{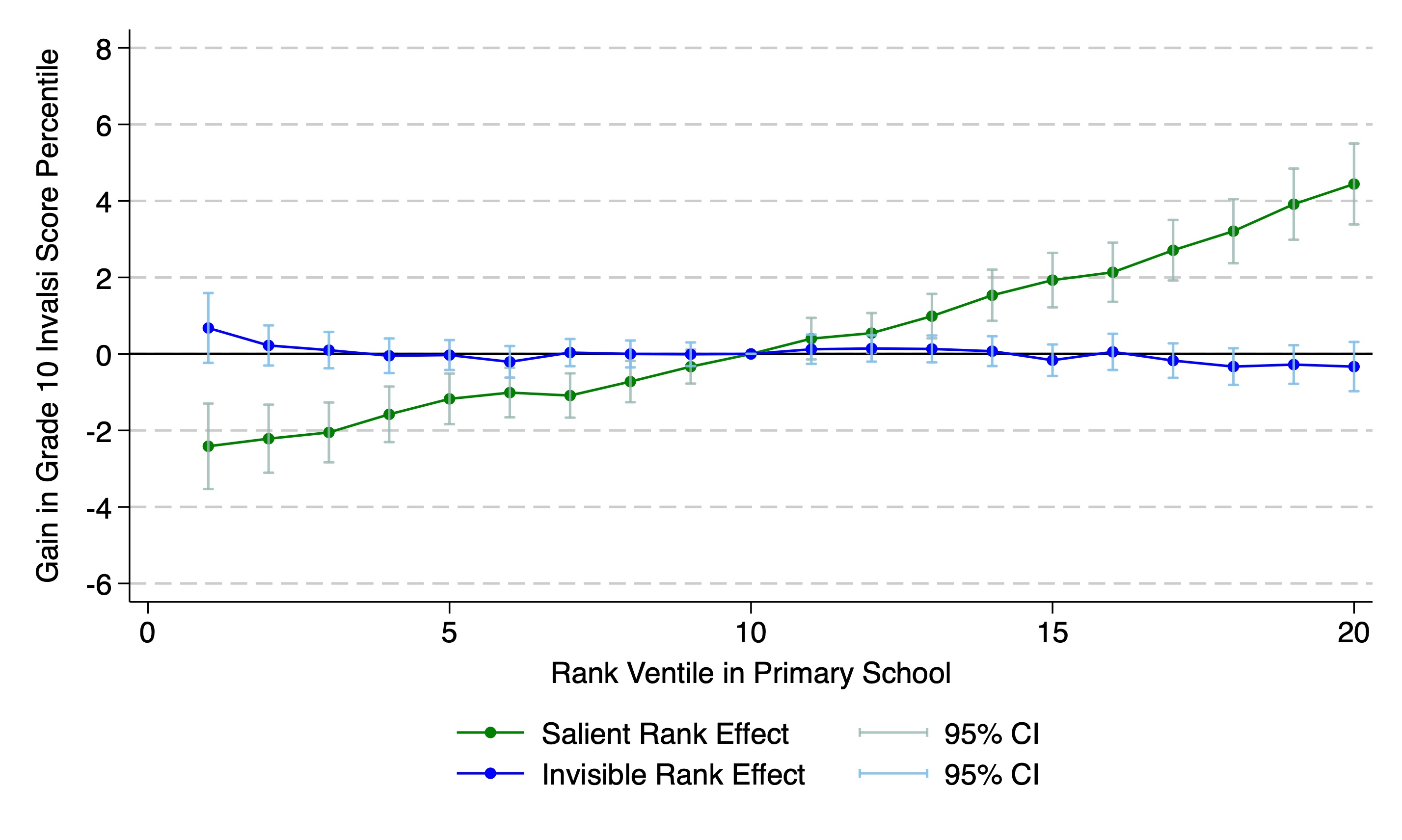}
\end{subfigure}
\caption{Non-Linear Rank Effects}
\label{fig:nonlinear_rank_effect_tworanks}
\vspace{.2cm}
\footnotesize \begin{tabular}{p{14cm}}
\setstretch{1} \textbf{Note}: This Figures compare the effect of the visible (from class scores) and invisible (from standardized test scores) ranks by ventiles, as estimated through Equation \ref{eq:main_specification_ventiles}. See Section \ref{subsec:results_nonlinear_rankeffect} for details.
\end{tabular}
\end{figure}

This could help explain the differences in the heterogeneous effect of the rank we observe in Figure \ref{fig:nonlinear_rank_effect_tworanks}. Absent any self-selection in middle school, the rank effect is mostly positive, i.e., low-ranking students do not perform significantly worse because of their rank. By contrast, the negative impact of low rank on performance in high school may be mediated by self-selection into significantly worse high schools.

\subsection{Motivation}\label{subsec:mecha_motivation}
To further explore mechanisms of motivation, we exploit students' answers to the \textit{Questionario di Contesto} (see Section \ref{sec:institutional_context}). This questionnaire covers several topics: e.g., how students perceive school, how interested they are in Italian and math, or the amount of support they receive from their parents. On each main topic, students are asked several questions to which they respond using a six-point Likert scale. We average the scores by topic and convert them to national percentiles prior to removing any students.

Importantly, only questions about subject interest are subject-specific, allowing us to use Equation \eqref{eq:main_specification} with student fixed effects. For the other channels, we will use a version of the main specification without student fixed effects, investigating the effect of rank in Italian and math separately on each topic: 
\begin{align}\label{eq:motivation_spec}
\begin{split}
    M^{10}_{ict}   = \alpha + \beta R^V_{ics} + \delta R^I_{ics}  + x'_i \gamma + \Gamma_{ics}  + \varepsilon_{ics} 
\end{split}
\end{align}
where for student $i$ in class $c$ studying subject $s$, $M^{10}_{ict}$ is the percentile average score to questions on topic $t$ in grade $10$ 
and $x_i'$ are students' observables. Even if we cannot use our main specification, the coefficient on the invisible rank will continue to serve as a placebo check. 

The questionnaire was administered only in 2018 due to the uproar it stirred in the Italian media and the negative press coverage it received. Therefore, we restrict our analysis of motivation to the 2018 cohort.\footnote{We apply the same criteria of sample restriction as described in Section \ref{subsec:sample_selection}, with two modifications: we now keep students even if we do not observe grade 8 standardized scores and we drop those for whom we do not observe scores for each topic of the questionnaire.}

\paragraph{Subject Confidence.}\label{subsec:mecha_subject_confidence} We first explore how the rank in each subject affects students' interest and confidence in Italian and math. As questions are similar for both subjects, we can continue using Equation \eqref{eq:main_specification}. We also compare the results with those obtained using Equation \eqref{eq:motivation_spec} to gauge whether we can infer anything meaningful from regressions that do not include student fixed effects. 

Results are reported in Table \ref{mecha_subject_interest}. We first observe that not including student fixed effects in the specification leads us to estimate a lower bound to the rank effect. Importantly, the coefficient on the invisible rank is small and not significant in both cases. We interpret this as the fact that students fixed effects are not needed to adequately capture the effect of cognitive ability on this particular outcome. 

The rank has a very strong effect on the confidence of students in the subject studied. Going from first to last of one's primary school class leads to a 4.0 gain in percentiles in the national distribution of subject confidence scores. Interestingly, being an immigrant or a woman seems to be associated with substantially higher subject confidence. The coefficient on SES is as expected. 

\paragraph{Parental Support.}\label{subsec:mecha_parental_support} We now turn to explore how primary school ordinal rank affects parental support. A priori, the effect is not obvious. One may expect that students with higher ranks would prompt their parents to provide them with more support, due to higher returns. It could also be that parents are less involved with higher-ranked students because they feel they do not need to be encouraged. Parents may also be induced to support their children more strongly if they perceive them as struggling in school. Columns 1 and 2 of Table \ref{mecha_allchannels} show that a higher rank appears to be negatively correlated with parental support, although the coefficient is imprecisely estimated. Women and students from higher socio-economic backgrounds appear to receive more parental support while immigrants receive significantly less.\footnote{These results prompt us to explore non-linearities in the rank effect on parental support, but we fail to detect any pattern.}

\paragraph{Self-Esteem.} One channel through which class rank in primary school may have a lasting impact on later outcomes is confidence about one's intellectual abilities. Indeed, rank at an early age can shape one's conceptions about one's intellectual worth. Table \ref{mecha_allchannels} (Columns 3 and 4) shows that the visible rank in both Italian and math significantly impact self-esteem. However, the coefficient of the invisible rank in math is large relatively to that of the visible rank and significant at the 10\% level, suggesting that our specification might not be ideal. We are thus more confident in the effect of the visible rank in Italian. Furthermore, being an immigrant does not significantly affect self-esteem but being a woman is associated with lower levels of confidence. Students with a higher SES also display significantly higher levels of self-confidence.

\paragraph{Peer Recognition.} As class rank is known among students, it may be a metric used to compare to each other beyond school achievements. The standing associated with one's class rank may thus affect a student's relationship with her peers. Table \ref{mecha_allchannels} (Columns 5 and 6) shows that this is the case both in Italian and math, albeit the effect of the Italian is stronger. Again, we are warier of interpreting the effect of the rank in math due to a coefficient of the invisible rank relatively large and significant at the 10\% level. We also observe that being a woman or an immigrant leads to reporting significantly lower levels of peer recognition. In contrast, higher SES is correlated with higher peer recognition. 

\paragraph{Career Expectations.} Columns 7 and 8 of Table \ref{mecha_allchannels} display the effect of rank on career expectations. Italian and math ranks do not seem to have any effects on how confident students are about their future careers. Interestingly, stated expectations are higher for women and immigrants, as well as for high SES.

\paragraph{Perception of School.} The last mechanism we explore is the perception that students have of school. Questions related to this topic have two different scales: some questions are negative, with the highest score indicating that students have a hard time at school, and some are positive, with the highest score indicating that students enjoy their time at school. For this reason, we group them into two categories: ``bad perception'' and ``good perception''. Table \ref{mecha_schoolperception} shows that a high visible rank in primary school reduces bad perception of school. Interestingly, the visible rank in Italian seems to also play a role in improving the perception of school, while the visible rank in math does not.

\section{Concluding Remarks}\label{sec:conclusion}
In various fields in economics, recent research has shown that peer comparison could have substantial impacts on individuals’ behaviors and perceptions. In this paper, we propose a new strategy to identify the effect of relative position using the Italian public school system as a case study. 

We demonstrate that student class rank in primary school has a strong effect on later academic achievements in middle school and high school. We find significant non-linearities: the rank effect is detrimental to performances in high-school for low-ranking students but beneficial for high-ranking students. Interestingly, we find another pattern in middle-school whereby students ranking in the low-half of the distribution are not negatively affected while high-ranking students benefit from a strong effect. We find no evidence that the rank effect changes according to the size of the peer group, or peers' and individual ability. We also find that higher-ranked students are more likely to sort into high schools whose average student achievement is significantly higher. This is not observed in middle schools, whose enrollment is constrained by residency-based criteria. This might explain why the performance of lower-ranking students is not affected in middle school. Finally, we show that a higher visible rank enhances students' interests in the subjects studied and improves their perception of school. It also appears to positively affect their self-esteem and perceived recognition by their peers. The rank in Italian seems to have a larger impact than the rank in math on these dimensions.  

Our first key contribution pertains to the identification strategy. This paper is the first attempt at using two scores, one visible and the other invisible to students, to estimate how the rank impacts future performances by affecting students' perceptions. This is a major departure from the current literature, which has been constrained by its use of a unique score both to proxy for cognitive ability and to construct students' ranks. Indeed, in this setting, the identification strategy requires exploiting variations in ability peer effects, which leads to omitted variable bias. Replicating the common specifications of the literature allows us to formally establish that there is indeed omitted variable bias due to unobserved ability. Strikingly, we find that even ensuring that the rank is uncorrelated to students' pretermined characteristics, such as gender or immigration status, is no guarantee that it is as good as random.

In contrast, using both a visible and an invisible ranks yields two main benefits: First, it makes it theoretically possible to identify the portion of the rank effect that affects students' perceptions, since they know their ranking. Second, as long as both scores are similarly correlated with students' latent abilities, we can use the invisible rank as a placebo test to assure that our controls effectively capture any remaining relationship between cognitive ability and later performances, and our data allows us to conduct additional checks to shore up this claim. We can thus identify a rank effect under substantially relaxed assumptions.

Our second main contribution is to implement our strategy by harnessing comprehensive and detailed data, including a unique survey covering all Italian students in Grade 10 in 2018 and analyzed here for the first time. To our knowledge, we are the first in the rank literature to leverage data covering all students in a country at the class level. This allows us to construct an accurate rank measure, to precisely measure class quality, and to have greater confidence in the external validity of the psychological channels to the rank effect that we pin down.  

Our findings reveal that specifications prevalent in the literature lead to consistently underestimate the rank effect. We see two main reasons to explain this discrepancy. First, other estimates from identification strategies are inherently biased. Second, the use a proxy only loosely related to the rank perceived by students may conduce to attenuate the effect measured. 

We show that the effect of increasing class quality, measured as class value-added, by one std is six-fold larger than increasing the rank by one std. Furthermore, we find that class quality is strongly correlated with the students' average test score and socio-economic status. To illustrate a concrete trade-off, we look at the net effect on subsequent academic achievement of attending a class whose average achievement is higher by one standard deviation. The gain in performance due to having higher-performing peers is found to be also five times larger than the loss in performance due to a lower rank. 

The main objective of this study is to suggest a way to credibly estimate the effect of relative position. As such, our paper is less about exploring policy recommendations, which would require future work. 
However, our results point to possible interventions. For instance, given the linearity exhibited by the rank effect, one might think of disclosing ranks only for top students, to retain the positive effect of ranking high while not discouraging weaker students. Furthermore, given the advantages of being with higher-achieving peers outweigh the downsides of a lower rank, our findings suggest avoiding grouping students in ability-based classes.


\vspace{-0.2cm}
\doublespacing
\bibliography{biblio.bib}

\newpage
\newgeometry{top=1in, bottom=1in, left=1in, right=1in}  
\singlespacing
\section{Tables}\label{sec:tables}


\subsection{Data and Descriptive Statistics}

\begin{table}[H]\centering
\begin{threeparttable}[b]
\caption{Standardized Test Attendance}
\footnotesize
\label{tab:invalsi_attendance}
\vspace*{0.4cm}
\begin{tabular}{lM{10em}M{10em}M{10em}}
\toprule
\toprule
& \multicolumn{3}{c}{\textbf{Panel A: 2018 Cohort}} \\
\cmidrule(lr){2-4}
    & Grade 5 (2013) & Grade 8 (2015)  & Grade 10 (2018) \\
\midrule
Actual Number of Students & 413,596 & 442,971 & 529,039\\
Observed Number of Students & 371,574 & 407,211 & 496,331 \\
Fraction Missing & 10.2\% & 8.1\% & 6.2\% \\
& & & \\
& \multicolumn{3}{c}{\textbf{Panel B: 2019 Cohort}} \\
\cmidrule(lr){2-4}
    & Grade 5 (2014)& Grade 8 (2016)  & Grade 10 (2019) \\
\midrule
Actual Number of Students & 424,668 & 452,153 & 535,132\\
Observed Number of Students & 372,668 & 419,239 & 501,224 \\
Fraction Missing & 12.2\% & 7.3\% & 6.3\% \\
\bottomrule 
\bottomrule 
\end{tabular} 
\footnotesize
\begin{tablenotes} \setstretch{1} \textbf{Note:} The actual number of students is the number of students that could have been observed if all Italian students had taken the test and none of them had repeated or skipped a year. It is compared with the number of students observed every year in our dataset. The increase in the number of students over time is due to repetition: as the likelihood of repeating a year is higher in middle and high school, the number of students in a given year mechanically increases in higher grades. See Section \ref{sec:institutional_context} for details.
\end{tablenotes}
\end{threeparttable}
\end{table}

\begin{table}[H] \centering \begin{threeparttable}[b]  \footnotesize
\caption{Student Characteristics} \label{tab:cohort_students_chara}
\renewcommand{\arraystretch}{0.9}
\begin{tabular}{lM{10em}M{10em}M{10em}}
\toprule
\toprule
&  \multicolumn{3}{c}{\textbf{Panel A: Restricted Sample}} \\
\cmidrule(lr){2-4}
& Mean & Std. Dev. & Obs. \\ 

\midrule
\hspace{0.4cm} \% of Women &    0.51  &    0.50 &   387,614 \\
\hspace{0.4cm} Socio-Economic Status &    0.12  &    0.97 &      376,562 \\
\hspace{0.4cm} \% of Immigrants &    0.11 & 0.31   &    387,614 \\
\vspace{0.05cm}\\
&  \multicolumn{3}{c}{\textbf{Panel B: All Sample}} \\
\cmidrule(lr){2-4}
& Mean & Std. Dev. & Obs. \\ 
\midrule
\hspace{0.4cm} \% of Women &    0.51  &    0.50 &    744,242 \\
\hspace{0.4cm} Socio-Economic Status &    0.10  &    0.98 &  705,858 \\
\hspace{0.4cm} \% of Immigrants &    0.14  &    0.35 &      744,242 \\

\bottomrule
\bottomrule 
\end{tabular}
\begin{tablenotes} \setstretch{1}
\textbf{Note:} This Table compares students included the restricted sample (Panel A) with students in the original sample (Panel B) on different dimensions. SES data is missing for 11,052 students of our sample (2.9\%). See Section \ref{subsec:sample_selection} for details. 
\end{tablenotes}
\end{threeparttable}
 \end{table}

\subsection{Results}

\subsubsection{Replications of MW and DMW}

\begin{table}[H]%
\caption{Replication of MW and DMW}%
 \label{replication_mw_dmw_balance}%
 \scriptsize \centering %
  \resizebox{\textwidth}{!}{ %
 \begin{threeparttable} %
\begin{tabular}{lM{6.8em}M{6.8em}M{6.8em}M{6.8em}M{6.8em}M{6.8em}M{6.8em}}%
\vspace{0mm}\\%
\toprule
\toprule
 & \multicolumn{2}{c}{Equation \eqref{eq:mw}} & \multicolumn{5}{c}{Equation \eqref{eq:dmw}} \\
\cmidrule(lr){2-3}  \cmidrule(lr){4-8} 
 & \multicolumn{2}{c}{Test Score in} & \multicolumn{2}{c}{Test Score in} & \multicolumn{3}{c}{Balance Checks} \\
 \cmidrule(lr){2-3} \cmidrule(lr){4-5}  \cmidrule(lr){6-8} 
&Grade 8&Grade 10&Grade 8&Grade 10&Female&Immigrant& SES\\%
&(1)&(2)&(3)&(4)&(5)&(6)&(7)\\%
 \midrule%
Invisible Rank &4.834***&6.236***&2.044***&2.417***&0.014&{-}0.005&0.136\\%
&(0.285)&(0.265)&(0.416)&(0.401)&(0.010)&(0.005)&(0.499)\\%
Female&2.114***&0.520***&2.244***&0.574***&&&\\%
&(0.065)&(0.087)&(0.066)&(0.085)&&&\\%
SES&2.415***&2.606***&2.421***&2.599***&&&\\%
&(0.036)&(0.042)&(0.036)&(0.042)&&&\\%
Immigrant&{-}2.336***&{-}4.457***&{-}2.328***&{-}4.499***&&&\\%
&(0.112)&(0.126)&(0.114)&(0.127)&&&\\%
&\\
Observations&752,193&752,193&752,193&752,193&752,193&752,193&752,193\\%
\bottomrule
\bottomrule%
\end{tabular}%
\begin{tablenotes} \regfootnotesrepmwdmw \end{tablenotes}%
\end{threeparttable}%
}%
\end{table}

\begin{table}[H]%
\caption{Ability Placebo Check with MW and DMW}%
 \label{replication_mw_dmw_ability_placebo}%
 \scriptsize \centering %
  \resizebox{\textwidth}{!}{ %
 \begin{threeparttable} %
\begin{tabular}{lM{6.8em}M{6.8em}M{6.8em}M{6.8em}M{6.8em}}%
\vspace{0mm}\\%
\toprule
\toprule
 & \multicolumn{1}{c}{Equation \eqref{eq:mw}} & \multicolumn{4}{c}{Equation \eqref{eq:dmw}} \\
\cmidrule(lr){2-2}  \cmidrule(lr){3-6}
 & \multicolumn{1}{c}{Test Score in} &  \multicolumn{1}{c}{Test Score in} & \multicolumn{3}{c}{Balance Checks} \\
 \cmidrule(lr){2-2} \cmidrule(lr){3-3} \cmidrule(lr){4-6} 
& Grade 2  & Grade 2 &Female&Immigrant& SES\\%
&(1)&(2)&(3)&(4)&(5)\\%
\midrule
Invisible Rank&3.447***&1.256**&0.019&0.007&0.575\\%
&(0.391)&(0.579)&(0.014)&(0.007)&(0.696)\\%
Female&0.445***&0.703***&&&\\%
&(0.070)&(0.072)&&&\\%
SES&2.163***&2.156***&&&\\%
&(0.045)&(0.046)&&&\\%
Immigrant&{-}4.042***&{-}4.007***&&&\\%
&(0.138)&(0.140)&&&\\%
& \\
Observations&500,782&500,782&500,782&500,782&500,782\\%

\bottomrule%
\bottomrule
\end{tabular}%
\begin{tablenotes} \regfootnotesrepmwdmwplacebo \end{tablenotes}%
\end{threeparttable}%
}%
\end{table}

\subsubsection{Our Specification}

\begin{table}[H]%
\caption{Ability Placebo Check}%
 \label{robcheck_g2g8_class_cohort} %
 \footnotesize \centering %
 \begin{threeparttable} %
\begin{tabular}{lM{12em}M{12em}}%
\vspace{-.7cm}\\%
 \toprule%
  \toprule
  & \multicolumn{2}{c}{Test Score in} \\
    \cmidrule(lr){2-3}  
  & Grade 2 & Grade 8 \\
&(1)&(2) \\%
\midrule
&\multicolumn{2}{c}{ \textbf{Panel A: No Control for Ability (Eq. \ref{eq:both_ranks_uncond}) }} \\ %
\midrule%
Visible Rank&27.834***&35.785***\\%
&(0.183)&(0.183)\\%
Invisible Rank&32.402***&38.481***\\%
&(0.208)&(0.180)\\%
& \\
 &\multicolumn{2}{c}{ \textbf{Panel B: Simple Specification (Eq. \ref{eq:simple_specification})  }} \\ %
\midrule%
Visible Rank&11.196***&15.390***\\%
&(0.572)&(0.503)\\%
Invisible Rank&{-}1.429***&5.727***\\%
&(0.370)&(0.334)\\%
& \\
 &\multicolumn{2}{c}{ \textbf{Panel C: Main Specification (Eq. \ref{eq:main_specification})  }} \\ %
\midrule%
Visible Rank&0.030&7.675***\\%
&(1.101)&(0.896)\\%
Invisible Rank&0.896&0.679\\%
&(0.574)&(0.486)\\%
& \\
Observations&502,806&502,806\\%
\bottomrule%
\bottomrule
\end{tabular}%
\begin{tablenotes} \regfootnotesrobcheckabilitycorr \end{tablenotes}%
\end{threeparttable}%
\end{table}

\begin{table}[H]%
\caption{Main Results}%
 \label{main_spec_class_cohort} %
 \footnotesize \centering %
 \begin{threeparttable} %
\begin{tabular}{lM{12em}M{12em}}%
\vspace{-.75cm}\\%
\toprule
\toprule
& \multicolumn{2}{c}{Test Score in} \\
\cmidrule(lr){2-3}
&Grade 8&Grade 10\\%
&(1)&(2)\\%
 \midrule%
 &\multicolumn{2}{c}{ \textbf{Panel A: No Control for Ability (Eq. \ref{eq:both_ranks_uncond}) }} \\ %
\midrule%
Visible Rank&34.355***&31.102***\\%
&(0.148)&(0.173)\\%
Invisible Rank&36.035***&33.803***\\%
&(0.172)&(0.138)\\%
& \\
 &\multicolumn{2}{c}{ \textbf{Panel B: Main Specification (Eq. \ref{eq:main_specification})  }} \\ %
\midrule%
Visible Rank&8.468***&7.721***\\%
&(0.773)&(0.714)\\%
Invisible Rank&{-}0.605&{-}0.480\\%
&(0.395)&(0.367)\\%
& \\
Observations&775,228&775,228\\%
\bottomrule%
\bottomrule%
\end{tabular}%
\begin{tablenotes} \regfootnotesmainspec \end{tablenotes}%
\end{threeparttable}%
\end{table}

 \begin{table}[H]\centering
\begin{threeparttable}[b]
\caption{Effect of the Rank in our Specification vs. \ref{eq:mw} and \ref{eq:dmw} }\label{tab:comparison_ourresults_vs_mwdmw}
\footnotesize
\label{table}
\begin{tabular}{lM{6em}M{6em} M{6em}M{6em}}
\toprule
\toprule

& \multicolumn{2}{c}{Main Specification vs. \eqref{eq:mw}} & \multicolumn{2}{c}{Main Specification vs. \eqref{eq:dmw}} \\

\cmidrule(lr){2-3} \cmidrule(lr){4-5}
 & Diff.  & Ratio  & Diff.  & Ratio  \\
 & (1) & (2) & (3) & (4) \\
\midrule

Rank Effect on Grade 8 Perf. & 3.61*** & 57\% & 6.42*** & 24\% \\ 
& (0.88) & & (0.90) & \\

Rank Effect on Grade 10 Perf. & 1.49* & 81\% &5.30*** & 31\%  \\ 
& (0.76) & & (0.82) &\\
\bottomrule 
\bottomrule 
\end{tabular} 
\begin{tablenotes}
\setstretch{1} \textbf{Note:} Standard errors in parentheses. *** $p<0.01$, ** $p<0.05$, * $p<0.1$. This Table compares the estimation of the rank effect with our Specification \eqref{eq:main_specification} and \eqref{eq:mw} and \eqref{eq:dmw}. Columns 1 and 3 reports the difference and its significance. Columns 3 and 4 reports the ratio between the coefficients of \eqref{eq:mw} and \eqref{eq:dmw} and the corresponding coefficients from our specification. 
\end{tablenotes}
\end{threeparttable}
\end{table}

\subsection{Rank Effect vs. Class Effects}

\begin{table}[H]%
\caption{Correlation Between Class Value-Added and Observables}%
 \label{reg_class_quality_cohort} %
 \scriptsize \centering %
 \begin{threeparttable} %
\begin{tabular}{lM{5.8em}M{5.8em}}%
\vspace{0mm}\\%
 \toprule%
  \toprule%
 & \multicolumn{2}{c}{Class Value-Added as measured by class effects on Test Score in} \\
 \cmidrule(lr){2-3}
&Grade 8&Grade 10\\%
&(1)&(2)\\%
 \toprule%
Mean SES&0.257***&0.379***\\%
&(0.006)&(0.006)\\%
Fraction of Immigrants&{-}0.061***&0.039***\\%
&(0.005)&(0.005)\\%
Fraction of Women&0.036***&0.050***\\%
&(0.005)&(0.005)\\%
Class Size&0.066***&0.052***\\%
&(0.006)&(0.006)\\%
Mean Test Score&0.304***&0.290***\\%
&(0.006)&(0.006)\\%
& \\
Observations&60,564&60,564\\%

\bottomrule%
\bottomrule%
\end{tabular}%
\begin{tablenotes} \regfootnotesrankvsclasseffects \end{tablenotes}%
\end{threeparttable}%
\end{table}

\begin{table}[H]%
\caption{Impact of Class Mean SES and Class Mean Achievement}%
 \label{correlation_outcomes_scoreses} %
 \scriptsize \centering %
 \begin{threeparttable} %
\begin{tabular}{lM{6.8em}M{6.8em}M{6.8em}}%
\vspace{0mm}\\%
 \toprule%
 \toprule%
& Class Rank in & \multicolumn{2}{c}{Test Score in} \\
\cmidrule(lr){2-2} \cmidrule(lr){3-4}
& Grade 5 & Grade 8&Grade 10\\%
&(1)&(2)&(3)\\%
 \toprule%
Mean SES&{-}0.017***&3.161***&4.419***\\%
&(0.000)&(0.082)&(0.078)\\%
Mean Test Score&{-}0.059***&3.761***&3.465***\\%
&(0.001)&(0.078)&(0.064)\\%
Observations&753,124&753,124&753,124\\%
\bottomrule%
\bottomrule%
  
\end{tabular}%
\begin{tablenotes} \regfootnotescorrperfscore \end{tablenotes}%
\end{threeparttable}%
\end{table}

\subsection{Mechanisms}

\begin{table}[H]%
\caption{School Sorting}%
 \label{school_sorting} %
 \scriptsize \centering %
 \begin{threeparttable} %
\begin{tabular}{lM{6.8em}M{6.8em}}%
\vspace{0mm}\\%
\toprule
\toprule
 & \multicolumn{2}{c}{School Quality in}\\
 \cmidrule(lr){2-3}
&Grade 8&Grade 10\\%
&(1)&(2)\\%
\midrule
Visible Rank&0.085&4.496***\\%
&(0.412)&(0.489)\\%
Invisible Rank&0.233&{-}0.936***\\%
&(0.198)&(0.241)\\%
\midrule%
Observations&693,424&693,424\\%
\bottomrule%
\bottomrule%
\end{tabular}%
\begin{tablenotes} \regfootnotesmechaschoolsorting \end{tablenotes}%
\end{threeparttable}%
\end{table}

\begin{table}[H]%
\caption{Impact of Class Rank on Subject Interest}%
 \label{mecha_subject_interest} %
 \scriptsize \centering %
 \begin{threeparttable} %
\begin{tabular}{lM{6.8em}M{6.8em}}%
\vspace{0mm}\\%
 \toprule%
  \toprule%
&\multicolumn{2}{c}{Subject Confidence}\\%
&(1)&(2)\\%
 \midrule
Visible Rank&3.536***&6.603***\\%
&(0.969)&(1.450)\\%
Invisible Rank&{-}0.052&0.873\\%
&(0.609)&(0.776)\\%
& \\
Female&4.242***&\\%
&(0.125)&\\%
SES&2.309***&\\%
&(0.063)&\\%
Immigrant&1.035***&\\%
&(0.224)&\\%
& \\
Student FE&No&Yes\\%
Observations&397,868&397,868\\%
\bottomrule%
\bottomrule%
\end{tabular}%
\begin{tablenotes} \regfootnotesmechasubjectinterest \end{tablenotes}%
\end{threeparttable}%
\end{table}

\begin{table}[H]%
\caption{Other Channels}%
 \label{mecha_allchannels} %
 \scriptsize \centering 
  \resizebox{\textwidth}{!}{ %
 \begin{threeparttable} %
\begin{tabular}{lM{5.8em}M{5.8em}M{5.8em}M{5.8em}M{5.8em}M{5.8em}M{5.8em}M{5.8em}}%
\vspace{0mm}\\%
 \toprule%
  \toprule%
&\multicolumn{2}{c}{Parental Support}&\multicolumn{2}{c}{Career Expectations}&\multicolumn{2}{c}{Self{-}Esteem}&\multicolumn{2}{c}{Peer Recognition}\\%
\cmidrule(lr){2-3} \cmidrule(lr){4-5} \cmidrule(lr){6-7} \cmidrule(lr){8-9}
&(1)&(2)&(3)&(4)&(5)&(6)&(7)&(8)\\%
\midrule%
Visible Math Rank&{-}0.669&&2.078&&3.200**&&3.138**&\\%
&(1.409)&&(1.397)&&(1.405)&&(1.423)&\\%
Invisible Math Rank&0.614&&0.835&&1.639*&&1.453*&\\%
&(0.867)&&(0.878)&&(0.872)&&(0.877)&\\%
Visible Italian Rank&&{-}0.131&&2.270&&4.822***&&4.882***\\%
&&(1.413)&&(1.454)&&(1.401)&&(1.394)\\%
Invisible Italian Rank&&0.608&&{-}0.863&&{-}0.463&&0.224\\%
&&(0.915)&&(0.922)&&(0.887)&&(0.920)\\%
& \\
Female&1.900***&2.794***&1.771***&1.641***&{-}2.893***&{-}5.022***&{-}5.238***&{-}5.669***\\%
&(0.144)&(0.143)&(0.155)&(0.151)&(0.143)&(0.142)&(0.153)&(0.153)\\%
SES&5.660***&5.472***&4.791***&4.733***&3.722***&3.584***&2.544***&2.464***\\%
&(0.077)&(0.078)&(0.080)&(0.079)&(0.079)&(0.080)&(0.080)&(0.080)\\%
Immigrant&{-}6.122***&{-}5.987***&0.913***&1.009***&{-}0.408&0.101&{-}4.682***&{-}4.329***\\%
&(0.279)&(0.282)&(0.284)&(0.287)&(0.283)&(0.288)&(0.288)&(0.291)\\%
& \\
Observations&198,643&198,643&198,643&198,643&198,643&198,643&198,643&198,643\\%
\bottomrule%
\bottomrule%
\end{tabular}%
\begin{tablenotes} \regfootnotesmechaunstacked \end{tablenotes}%
\end{threeparttable}%
}%
\end{table}

\begin{table}[H]%
\caption{Perception of School}%
 \label{mecha_schoolperception} %
 \scriptsize \centering 
 \begin{threeparttable} %
\begin{tabular}{lM{6.8em}M{6.8em}M{6.8em}M{6.8em}}%
\vspace{0mm}\\%
 \toprule%
  \toprule%
&Bad Perception&Good Perception&Bad Perception&Good Perception\\%
&(1)&(2)&(3)&(4)\\%
\midrule%
Visible Math Rank&{-}4.317***&1.212&&\\%
&(1.405)&(1.442)&&\\%
Invisible Math Rank&{-}1.553*&{-}0.532&&\\%
&(0.876)&(0.895)&&\\%
Visible Italian Rank&&&{-}4.624***&2.556*\\%
&&&(1.376)&(1.414)\\%
Invisible Italian Rank&&&0.349&1.075\\%
&&&(0.919)&(0.899)\\%
& \\
Female&{-}8.933***&5.293***&{-}8.972***&5.157***\\%
&(0.142)&(0.151)&(0.147)&(0.150)\\%
SES&{-}2.357***&2.299***&{-}2.264***&2.216***\\%
&(0.079)&(0.080)&(0.080)&(0.080)\\%
Immigrant&{-}0.582**&{-}1.651***&{-}0.808***&{-}1.615***\\%
&(0.277)&(0.285)&(0.303)&(0.291)\\%
& \\
Observations&198,269&198,269&198,269&198,269\\%
\bottomrule%
\bottomrule%
\end{tabular}%
\begin{tablenotes} \regfootnotesmechaunstackedschoolperception \end{tablenotes}%
\end{threeparttable}%
\end{table}

\newgeometry{top=.5in, bottom=.5in, left=1in, right=1in}  
\section{Figures}\label{sec:figures}

\subsection{Empirical Strategy}

\subsection{Results}

\subsubsection{Non-Linear Effects}

\subsubsection{Does Class Size Matter?}

\subsubsection{Do Peer or Individual Abilities Impact the Rank Effect?}

\newpage
\newgeometry{top=1in, bottom=1in, left=1in, right=1in}  
\begin{center} \LARGE{\textbf{FOR ONLINE PUBLICATION}}
\end{center}
\onehalfspacing
\setcounter{table}{0}
\renewcommand{\thetable}{A\arabic{table}}
\setcounter{figure}{0}
\renewcommand{\thefigure}{A\arabic{figure}}
\begin{appendix}
\newgeometry{top=1in, bottom=1in, left=1in, right=1in}  
\section{Additional Information about the Analysis Sample}\label{sec:app_data}

\subsection{Students Characteristics by High School Type}

\subsubsection{All Sample}

\begin{table}[H] \begin{threeparttable}[b] \centering \footnotesize
\caption{Students' Characteristics by High School Type}\label{tab:vocational_vs_academic_all}
\renewcommand{\arraystretch}{0.9}
\begin{tabular}{l M{12em}M{12em}M{12em} }
\toprule
&   Academic &  Technical  &  Vocational \\ 
 \toprule 
\multicolumn{4}{l}{\textbf{Demographics}} \\
\hspace{0.4cm} Socio-Economic Status &    0.32  &   -0.20 &   -0.58  \\
 & [0.95] &  [0.92] &  [0.94]\\
\hspace{0.4cm} \% of Immigrants &  0.14  &    0.23 &    0.35   \\
 & [0.34] &  [0.42] &  [0.48]\\
\hspace{0.4cm} \% of Women&    0.60  &    0.31 &    0.43 \\
 & [0.49] &  [0.46] &  [0.50]\\
 \midrule
 \multicolumn{4}{l}{\textbf{Observations}} \\
\hspace{0.4cm}  Number of Students & 526,747 & 299,809 & 163,631  \\ 
\hspace{0.4cm}  Proportion & 53.2\% & 30.3\%
& 16.5\% \\ 
\bottomrule

\end{tabular}
\begin{tablenotes}  \tabfnhighschooltypeall
\end{tablenotes}
\end{threeparttable}
 \end{table}

\subsubsection{Restricted Sample}

\begin{table}[H] \begin{threeparttable}[b] \centering \footnotesize 
\caption{Students' Characteristics by High School Type}\label{tab:vocational_vs_academic_restricted}
\renewcommand{\arraystretch}{0.9}
\begin{tabular}{l M{10em}M{10em}M{10em}}
\toprule 
&   Academic &  Technical  &  Vocational \\ 
 \toprule
\multicolumn{4}{l}{\textbf{Demographics}} \\
\hspace{0.4cm} Socio-Economic Status &    0.36  &   -0.16 &   -0.54\\
 & [0.93] &  [0.90] &  [0.91]\\
\hspace{0.4cm} \% of Immigrants &  0.08  &    0.13 &    0.19 \\
 & [0.27] &  [0.34] &  [0.40] \\
\hspace{0.4cm} \% of Women&    0.61  &    0.32 &    0.48  \\
 & [0.49] &  [0.47] &  [0.50]\\
  \midrule
 \multicolumn{4}{l}{\textbf{Observations}} \\
\hspace{0.4cm}  Number of Students & 234,407 & 110,930 & 42,277  \\
\hspace{0.4cm}  Proportion & 60.5\% & 28.6\% & 10.9\% \\
\bottomrule
\end{tabular}
\begin{tablenotes} 
\tabfnhighschooltyperestricted
\end{tablenotes}
\end{threeparttable}
 \end{table}

\subsection{Missing Students and Retention Rate}\label{subsec:app_retention_rate}

In our dataset, there are two reasons for which a given student is not observed in primary and/or middle school. We do not observe repeating students: a student who repeated a class will have taken six years to go from grade 5 to grade 10 instead of five years. If she is in the 2018 Cohort for instance, this implies that she was not in grade 8 in 2015 but in grade 7 and thus not observed in previous years, and this reasoning stands for any students that repeated or skipped one or more years. The remaining missing students have simply not shown up for the test.

The actual number of students showed in Table \ref{tab:invalsi_attendance} gives us an indication about the retention rate, as only grade repetition or skipping can explain it. We cannot observe if a student skipped a grade but it is quite rare in Italy. Thus, we can see that the retention rate is approximately 15.9\% between grade 8 and grade 10 and 21.2\% between grade 5 and grade 10, which is in line with \cite{salza_2022}. Among students we observe in our sample, the retention rate is smaller: 12.5\% between grades 8 and 10, and 15.7\% between grades 5 and 10. This suggests that those who did not take the test in grade 10 are on average more likely to have repeated a year and can be considered as weaker academically.

\subsection{Class Size Distribution and Coverage}\label{app:subsec_class_size_distribution}

By law, class size should range from 9 to 28 students but exceptions are possible: Classes in schools located in ``disadvantaged areas'' are not subject to these limits and class size limits are to be satisfied at the beginning of each section: therefore, if a student leaves her school, she will not be replaced, and class size may fall below the legal requirement. The distribution is shown in Table \ref{tab:cohort_classroom_chara_all_sample}.

\begin{table}[H] \centering \begin{threeparttable}[b]  \footnotesize
\caption{Class and School Size Distribution}\label{tab:cohort_classroom_chara_all_sample}
\renewcommand{\arraystretch}{0.9}
\begin{tabular}{lccccc}
\toprule
& Mean & Std. Dev.  & 25th Perc. & Median & 75th Perc. \\ 
 \midrule 
\multicolumn{6}{l}{\textbf{Primary School (Grade 5)}} \\
\hspace{0.4cm} Actual Class Size	&	14.6	&	5.0	&	11	&	15	&	18	\\
\hspace{0.4cm} Observed Class Size	&	12.9	&	4.8	&	10	&	13	&	16	\\
\hspace{0.4cm} Actual School Size	&	60.0	&	34.2	&	32	&	58	&	83	\\
\hspace{0.4cm} Observed School Size	&	53.3	&	30.7	&	29	&	51	&	73	\\
\hspace{0.4cm} Number of Classes Per School	&	4.1	&	2.1	&	2	&	4	&	6	\\

\vspace{0.05cm}\\
\multicolumn{6}{l}{\textbf{Middle School (Grade 08)}} \\
\hspace{0.4cm} Actual Class Size	&	16.6	&	4.7	&	14	&	17	&	20	\\
\hspace{0.4cm} Observed Class Size	&	15.3	&	4.8	&	12	&	15	&	19	\\
\hspace{0.4cm} Actual School Size	&	77.2	&	47.6	&	46	&	69	&	98	\\
\hspace{0.4cm} Observed School Size	&	71.3	&	45.0	&	41	&	63	&	91	\\
\hspace{0.4cm} Number of Classes Per School	&	4.6	&	2.4	&	3	&	4	&	6	\\

\vspace{0.05cm}\\
\multicolumn{6}{l}{\textbf{High School (Grade 10)}} \\
\hspace{0.4cm} Actual Class Size	&	20.5	&	5.1	&	18	&	21	&	24	\\
\hspace{0.4cm} Observed Class Size	&	19.2	&	5.4	&	16	&	20	&	23	\\
\hspace{0.4cm} Actual School Size	&	134.2	&	99.1	&	24	&	139	&	204	\\
\hspace{0.4cm} Observed School Size	&	125.8	&	94.7	&	23	&	128	&	192	\\
\hspace{0.4cm} Number of Classes Per School	&	6.6	&	4.4	&	1	&	7	&	10	\\

\bottomrule
\end{tabular}
\begin{tablenotes} \setstretch{1}
\textbf{Note:} This Table reports the distribution of class and school sizes per Grade. See Section \ref{sec:institutional_context} for details.
\end{tablenotes}
\end{threeparttable}
 \end{table}

Notice that the distribution of the number of classes by high school is skewed. Indeed, as school is compulsory until age 16, municipalities must ensure that the different types of high schools are accessible to local students. As middle school track is common to all students and there are three different types of high schools with many sub-tracks within each type, it is no surprise that there are a fair number of small high schools with few classes within each.

\begin{table}[H] \centering \begin{threeparttable}[b]  \footnotesize
\caption{Class Coverage}\label{tab:coverage_class}
\renewcommand{\arraystretch}{0.9}
\begin{tabular}{lcccccccc}
\toprule
& \textbf{Mean} & \textbf{Std. Dev.} & \textbf{Min} & \textbf{p25} & \textbf{Median} & \textbf{p75} & \textbf{Max} & \textbf{N}  \\ 
 \toprule
 \multicolumn{9}{c}{\textbf{Panel A: 2018 Cohort}} \\
\midrule
Grade 5 & 91.3\% &   8.7 &  5\% &  87\% &   93.3\% &  100\% & 100\% & 28,825 \\
Grade 8 &    92.8\% & 7.5 &    27.3\% &  89.5\% &   95\% &  100\% &   100\% &  26,748 \\
Grade 10 &  93.4\%  & 9.7 & 4.2\% &   90.5\% &   95.8\% &  100\% &   100\% &   25,723\\
\\
 \multicolumn{9}{c}{\textbf{Panel B: 2019 Cohort}} \\
\midrule
Grade 5 & 90.1\% &   9.8 &  4.2\% &  84\% &   90.9\% &  95.7\% & 100\% & 28,649 \\
Grade 8 &    93.5\% & 6.8 &    25\% &  90.5\% &   95.2\% &  100\% &   100\% &  27,130 \\
Grade 10 &  93.2\%  & 9.7 & 3.6\% &   90\% &   95.7\% &  100\% &   100\% &  26,226 \\
\bottomrule
\end{tabular}
\begin{tablenotes} \setstretch{1}
\textbf{Note:} This Table reports the distribution of the fraction of students we observe by class. For instance, in the 2018 Cohort we observe 93.3\% of the students or more in 50\% of the classes. See Section \ref{subsec:sample_selection} for details.
\end{tablenotes}
\end{threeparttable}
 \end{table}

\subsection{Cheating}\label{subsec:app_cheating}
To discourage cheating, Invalsi devised an algorithm yielding a Cheating Propensity Indicator $CPI_{cgs}$, by class $c$, grade $g$, and subject $s$ \citep{lucifora_2020}. If the CPI is above 50\%, scores are not returned to classes and are excluded from the computation of the school average. If the CPI is less than 50\%, scores are ``corrected''---i.e., deflated by $(1-CPI_{cgs})$ at the class level. Importantly, this ``correction'' is intended as a sanction to discourage cheating and is not a way to recover what students' performances would have been in the absence of cheating. In this paper, we include classes with a cheating probability below 50\%---i.e., those which are more likely to have played by the rules than not---and we also use non-corrected scores. 

In the paper, and in line with the literature, we turn raw Invalsi scores into percentiles, after excluding classes with a cheating probability larger than 50\%. Finally, note that the proportion of students in classes with a high likelihood of cheating at baseline is minuscule: 90\% of students are in classes with a cheating probability lower than 7\% in grade 5 and 8\% in grade 10.

\subsection{Characteristics of the Selected Sample}

\begin{table}[H] \centering \footnotesize\begin{threeparttable}[b]  
\caption{Characteristics of the Selected Sample}\label{tab:selected_sample}
\renewcommand{\arraystretch}{0.9}
\begin{tabular}{lccc}
\toprule
 & Number of Students & Number of Classes & Number of Schools \\
 \midrule
&  \multicolumn{3}{c}{\textbf{Panel A: 2018 Cohort}} \\

\midrule
Grade 5	&	205,123	&	16,393	&	5,984	\\
Grade 8	&	205,123	&	23,990	&	5,570	\\
Grade 10	&	205,123	&	24,621	&	3,635	\\

& \multicolumn{3}{c}{\textbf{Panel B: 2019 Cohort}}\\
\midrule

Grade 5	&	182,491	&	13,893	&	5,584	\\
Grade 8	&	182,491	&	23,238	&	5,499	\\
Grade 10	&	182,491	&	24,995	&	3,659	\\

\bottomrule
\end{tabular}
\begin{tablenotes} \setstretch{1}
\textbf{Note:} This Table reports the number of students, classes and schools observed in the sample after selection. See Section \ref{subsec:sample_selection} for details.
\end{tablenotes}
\end{threeparttable}
 \end{table}

\subsection{Distribution of Scores}

\begin{table}[H] \centering \begin{threeparttable}[b]  \footnotesize 
\caption{Class Grade Distribution}\label{tab:desc_stat_cohort_class_score}
\renewcommand{\arraystretch}{0.9}
\begin{tabular}{p{0.2\textwidth}cccccccc}
\toprule
& Mean & Std. Dev. & Min & p25 & p50 & p75 & Max  \\ 
 \toprule \\ 
\multicolumn{8}{l}{\textbf{Panel A: Grade 5}} \\
\hspace{0.4cm} Italian &    7.97  &    1.00 &    1.00 &    7.00 &    8.00 &    9.00 &   10.00 \\
\hspace{0.4cm} Math &    8.05  &    1.04 &    1.00 &    7.00 &    8.00 &    9.00 &   10.00 \\
\multicolumn{8}{l}{\textbf{Panel B: Grade 8}} \\
\hspace{0.4cm} Italian &    7.29  &    1.14 &    1.00 &    6.00 &    7.00 &    8.00 &   10.00 \\
\hspace{0.4cm} Math &    7.15  &    1.34 &    1.00 &    6.00 &    7.00 &    8.00 &   10.00 \\
\multicolumn{8}{l}{\textbf{Panel C: Grade 10}} \\
\hspace{0.4cm} Italian &    6.48  &    1.03 &    1.00 &    6.00 &    6.50 &    7.00 &   10.00 \\
\hspace{0.4cm} Math &    6.15  &    1.41 &    1.00 &    5.00 &    6.00 &    7.00 &   10.00 \\
\bottomrule
\end{tabular}
\begin{tablenotes} \setstretch{1} 
\textbf{Note:} This Table shows the distribution of class scores in each grade. See Section \ref{subsec:scores} for details.
\end{tablenotes}
\end{threeparttable}
 \end{table}

\begin{table}[H] \centering \begin{threeparttable}[b]  \footnotesize 
\caption{Percentile Standardized Test Score Distribution}\label{tab:cohort_ipscore_chara}
\renewcommand{\arraystretch}{0.9}
\begin{tabular}{p{0.2\textwidth}M{5em}M{5em}M{5em}M{5em}M{5em}}
\toprule 
& Mean & Std. Dev. & p25 & p50 & p75 \\ 
 \midrule \\ 
\multicolumn{6}{l}{\textbf{Panel A: Grade 5 Standardized Test Grades}} \\
\hspace{0.4cm} Italian &   50.43  &   28.78 &   26.00 &   50.00 &   75.00 \\
\hspace{0.4cm} Math &   50.55  &   28.78  &   26.00 &   50.00 &   75.00 \\
\multicolumn{6}{l}{\textbf{Panel B: Grade 8 Standardized Test Grades}} \\
\hspace{0.4cm} Italian &   52.50  &   28.57 &     28.00 &   53.00 &   77.00 \\
\hspace{0.4cm} Math &   52.62  &   28.87 &     28.00 &   53.00 &   77.00 \\
\multicolumn{6}{l}{\textbf{Panel C: Grade 10 Standardized Test Grades}} \\
\hspace{0.4cm} Italian &   54.81  &   28.04 &      32.00 &   56.00 &   79.00  \\
\hspace{0.4cm} Math &   54.41  &   28.34 &     31.00 &   56.00 &   79.00 \\
\bottomrule
\end{tabular}
\begin{tablenotes} \setstretch{1}
\textbf{Note:} This Table shows the distribution of percentile standardized test grades in each grade in the restricted sample. The distribution is slightly skewed for Panels B and C because our selection leaves out the students who are likely to be academically weaker. See Sections \ref{subsec:overview_dataset} and \ref{subsec:scores} for details.
\end{tablenotes}
\end{threeparttable}
 \end{table}

\subsection{Rank Computation}\label{subsec:app_qqplot}
\begin{figure}[H]
\centering
\begin{subfigure}[b]{0.4\textwidth}
\centering
(a) Italian \\
\includegraphics[width=\textwidth]{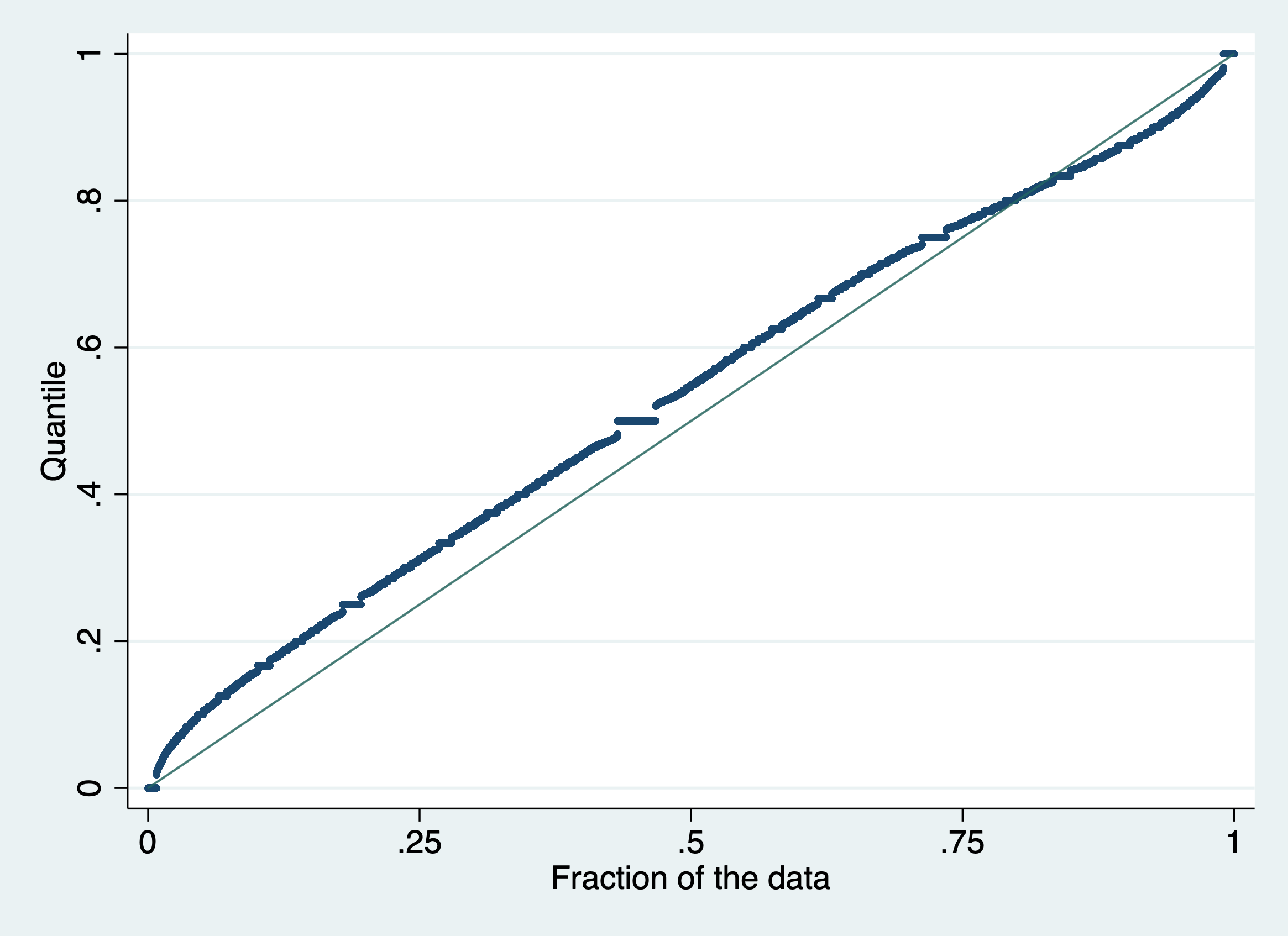}
\end{subfigure}
\begin{subfigure}[b]{0.4\textwidth} 
\centering
(b) Math \\
\includegraphics[width=\textwidth]{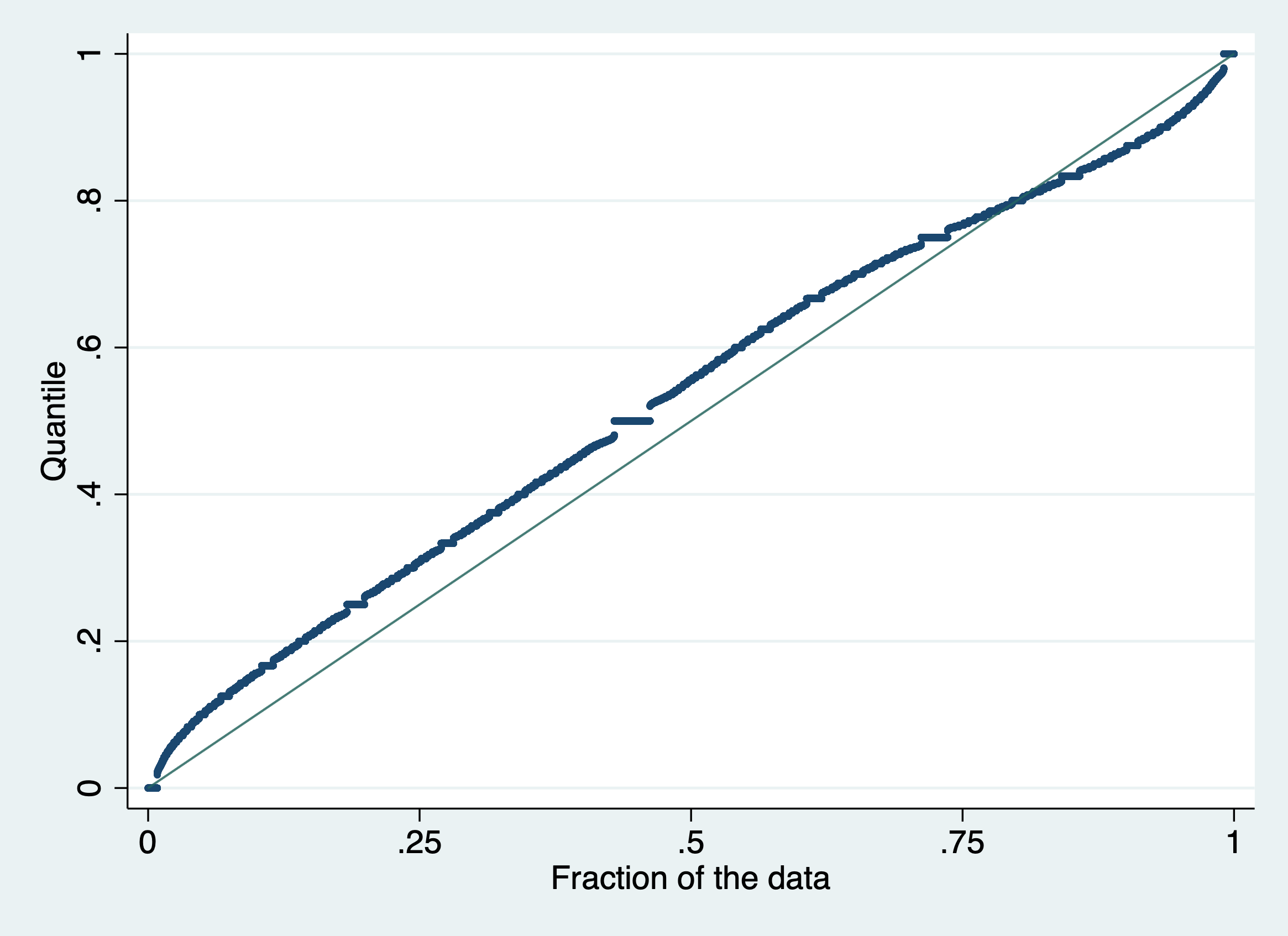}
\end{subfigure}
\caption{\textbf{QQ-plot of Class Rank Quantiles based on Class Grades}}
\label{fig:qqplot_prank}
\vspace{.2cm}
\footnotesize \begin{tabular}{p{14cm}}
\setstretch{1} \textbf{Note}: This Figure displays the QQ-plot of the quantiles of the distribution of ranks based on class grades. See Section \ref{subsec:rank_computation} for details.
\end{tabular}
\end{figure}

\begin{figure}[H]
\centering
\begin{subfigure}[b]{0.4\textwidth}
\centering
(a) Italian \\
\includegraphics[width=\textwidth]{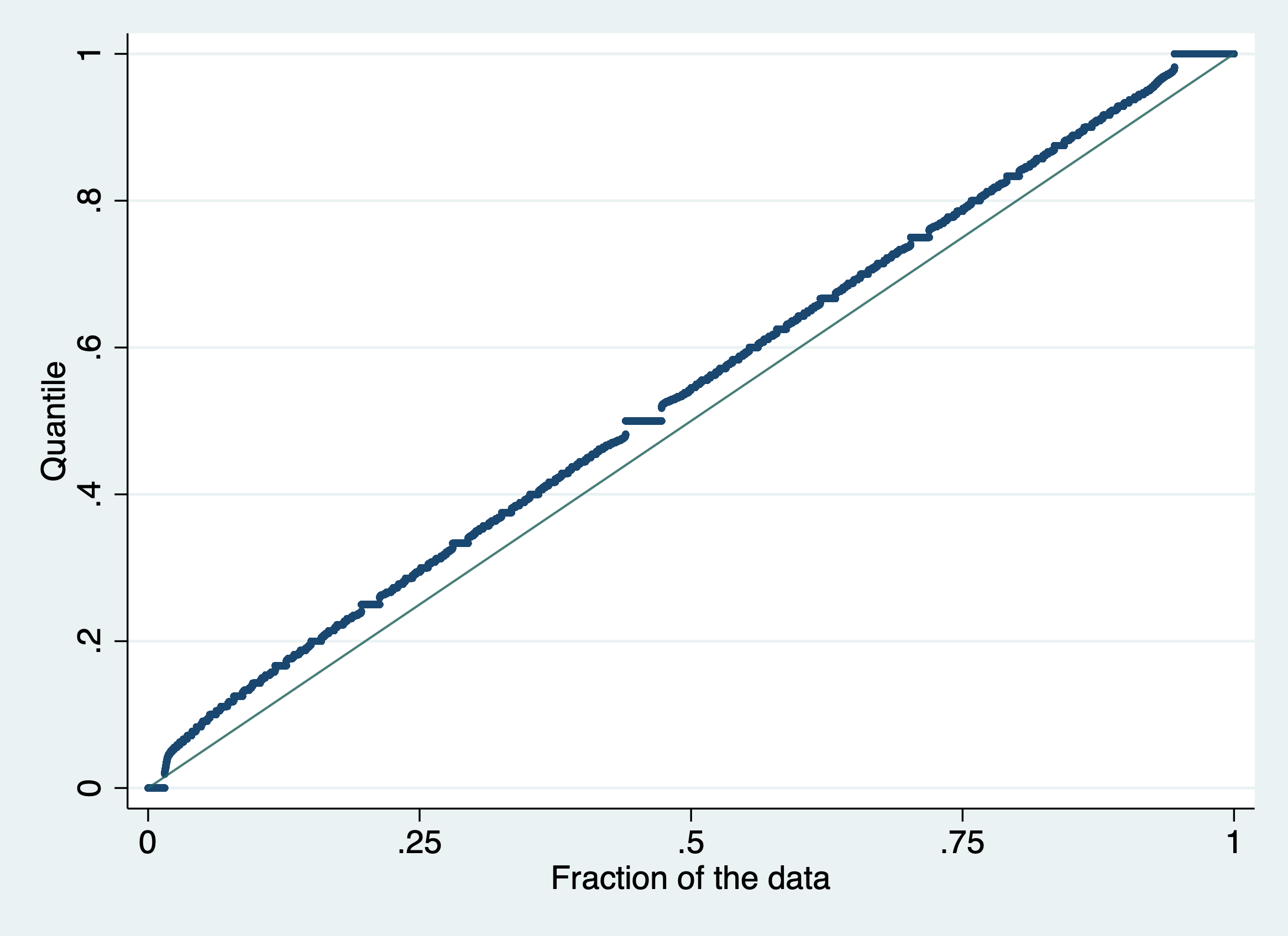}
\end{subfigure}
\begin{subfigure}[b]{0.4\textwidth} 
\centering
(b) Math \\
\includegraphics[width=\textwidth]{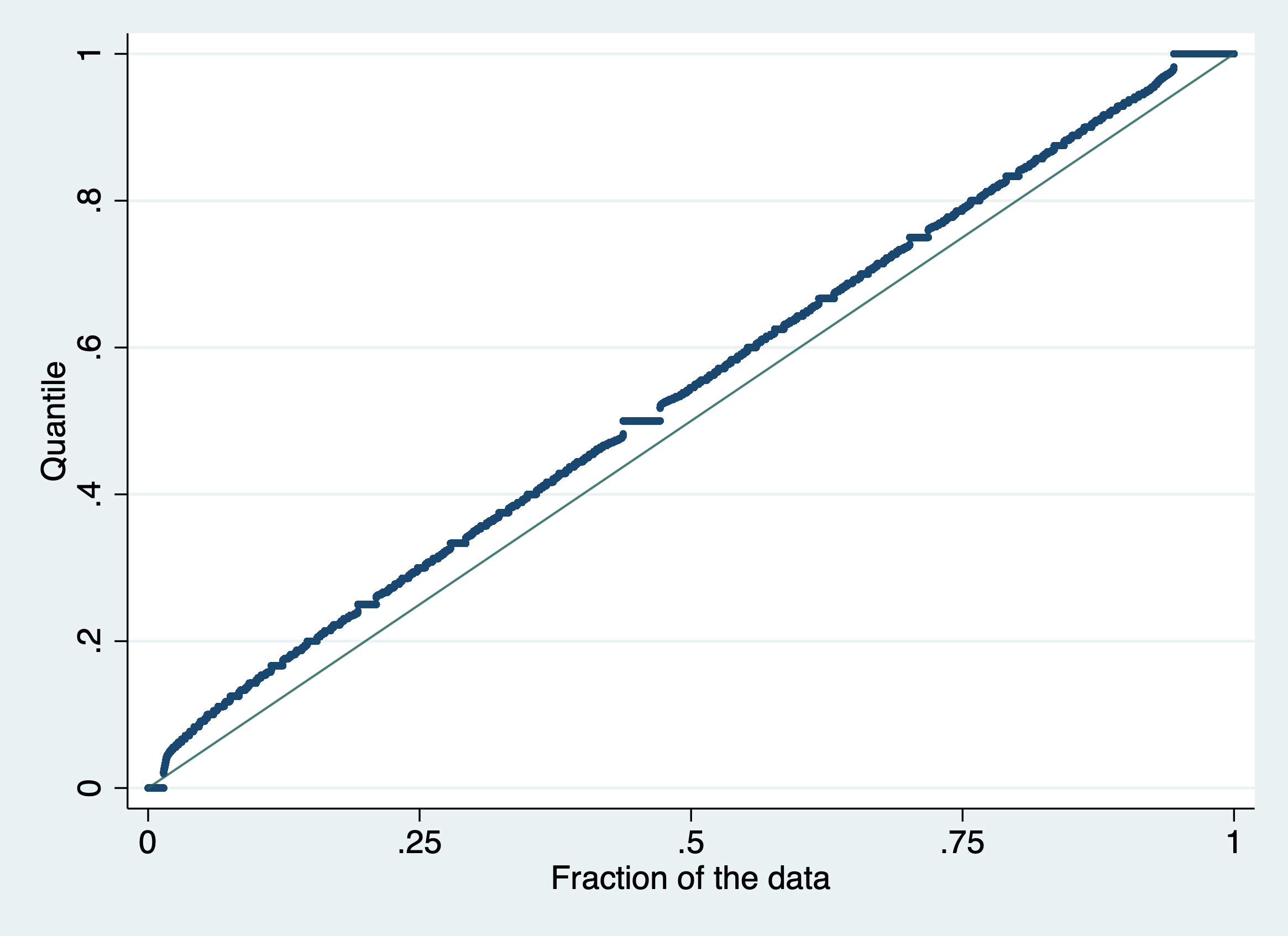}
\end{subfigure}
\caption{\textbf{QQ-plot of Class Rank Quantiles based on Standardized Scores}}
\label{fig:qqplot_iprank}
\vspace{.2cm}
\footnotesize \begin{tabular}{p{14cm}}
\setstretch{1} \textbf{Note}: This Figure displays the QQ-plot of the quantiles of the distribution of ranks based on standardized test scores. See Section \ref{subsec:rank_computation} for details.
\end{tabular}
\end{figure}

Figure \ref{fig:qqplot_prank} shows distribution of our rank measure based on class grades for Italian and math. Figure
\ref{fig:qqplot_iprank} shows a similar distribution for the rank measure based on standardized grades. 

Notice that, if we observed every student and there were no ties, the way we construct our rank variable would make it uniformly distributed with support $(0,1)$. In both cases, the actual distribution is very close to a uniform distribution but is slightly skewed, as expected for we assume that unobserved students would have ranked at the bottom (see Section \ref{subsec:sample_selection}).

\subsection{Empirical Strategy: Additional Figures}

\begin{figure}[H]
\centering
\includegraphics[width=.6\textwidth]{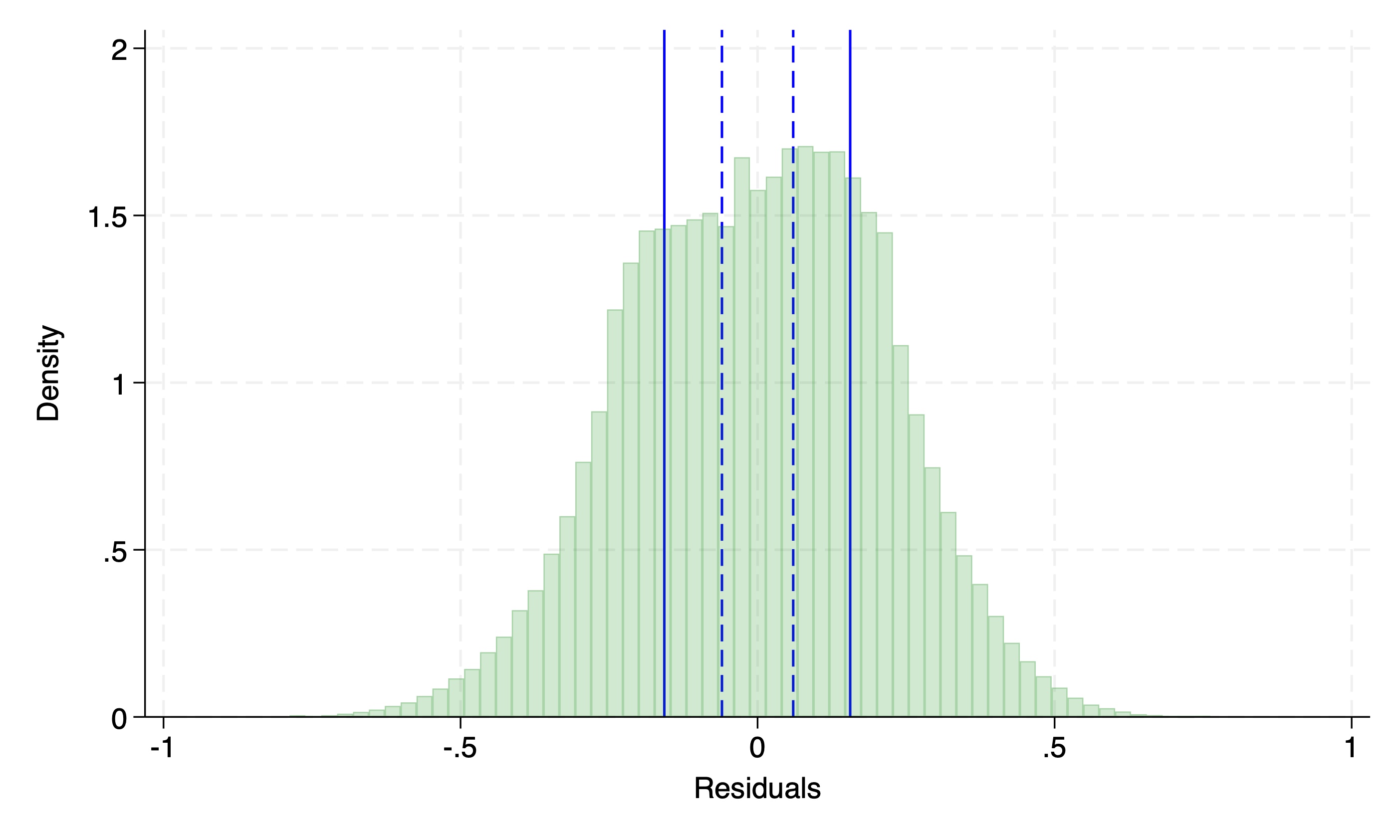}
\caption{\textbf{Distribution of Residuals}}
\label{fig:hist_rank_res}
\vspace{.2cm}
\footnotesize \begin{tabular}{p{15cm}}
\setstretch{1} \textbf{Note}: This Figure plots the residuals of the regression of the visible rank on the invisible rank. The blue lines indicate the 25th and 75th percentiles. The blue dashed lines are at -0.06 and 0.06 respectively, which is equivalent of one rank in a class of size 15. See Section \ref{subsec:strategy_motivation} for details. 
\end{tabular}
\end{figure}

\section{Rank and Ties}\label{sec:app_rankandties}
In this Section, we check that our tie-breaking approach does not drive our results.

As explained in Section \ref{sec:data_descstat}, we followed the literature and assigned the mean rank to ties. For instance, as shown in Table \ref{tab:rank_ties_explanation}, take the case of a toy class with just 6 students where two students got a grade of 10, one got a grade of 9, and three students got a grade of 8. In our analysis, the first two students are assigned rank 1.5, the third is assigned rank 3 while the other three students are assigned rank 5. This allows us to normalize the sum of the ranks by class, so that classes of the same size have the same weight.\footnote{If there were no ties in this class, the sum of the ranks would be 21, which is also the case with the Mean tie-breaking rule. The Min. rule underestimates it while the Max. rule overestimates it.} However, it is legitimate to wonder what would be the result of treating ties differently, especially assigning to each student in a tied group the minimum or maximum rank of that group. Table \ref{tab:rank_ties_explanation} shows how ranks would be attributed according to each tie-breaking rule.

\begin{table}[H]\centering
\begin{threeparttable}[b]
\caption{Tie-Breaking Rules}
\small
\label{tab:rank_ties_explanation}
\vspace*{0.4cm}
\begin{tabular}{cccc}
\toprule
\multirow{2}{*}{Grade} & \multicolumn{3}{c}{\textbf{Tie-Breaking Rule}} \\
\cline{2-4}
 & Mean Rank & Min. Rank & Max. Rank \\
\midrule
10 & 1.5 & 1 & 2 \\
10 & 1.5 & 1 & 2 \\
9 & 3 & 3 & 3\\
8 & 5 & 4 & 6 \\
8 & 5 & 4 & 6 \\
8 & 5 & 4 & 6 \\
\bottomrule 
\end{tabular} 
\begin{tablenotes}\textbf{Note:} This Table shows how ranks are imputed depending on the tie-breaking rule.
\end{tablenotes}
\end{threeparttable}
\end{table}

We are wary of using these two alternative tie-breaking rules as they would generate an artificially high number of top or bottom students, giving some classes disproportionate weight.\footnote{If, say, all students in a class of twenty got the same grade, we would end up with either twenty top or twenty bottom students while they are arguably no better than the class average and are not distinguishable from each other on this dimension.} We expect these tie-breaking rules to yield much lower coefficients, as they essentially defuse the rank effect by polarizing ranks artificially. 

However, we want to make sure that the rank effect is not overestimated due to the presence of ties. To that end, we compare the rank effect estimated from the main specification (Equation \ref{eq:main_specification}) with the effect measured from a specification in which we account for ties:
\begin{align}\label{eq:dummytie}
\begin{split}
    T^g_{ics} & = \alpha + \beta R^S_{ics}+  \delta R^U_{ics} + \sum_k \beta_k  R_{ics} \times \mathbf{1}[t_{ics}=k] + \mathbf{1}[t_{ics}=k]  + \Gamma_{ics} + \varepsilon_{ics}
\end{split}
\end{align}
where $\mathbf{1}[t_{ics}=k]$ is a dummy equal to one if student $i$ in class $c$ and subject $s$ is tied with $k$ other students. 

Results are reported in Table \ref{tab:dummytie_breaking}. As expected, the impact of the rank when using the minimum or maximum rank to break ties is much diminished. However, the use of the Mean tie-breaking rule is vindicated: the rank effect remains virtually the same once ties are accounted for. As shown by Columns (1) and (4) of Table \ref{tab:dummytie_breaking_pval}, the p-value of the difference is consistently well above 10\% in every case.

By contrast, with the other two methods, the rank effect doubles once we control for ties, becoming close to our baseline estimates. Columns (2), (3), (5), and (6) of Table \ref{tab:dummytie_breaking_pval} show that the p-value of the difference hovers around 0.3 in both Grades for both tie-breaking rules. This assuages concerns that results are driven by ties.

\begin{table}[H]\centering
\begin{threeparttable}[b]
\caption{Impact of Tie-Breaking Rules on the Rank Effect Estimate}
\footnotesize
\label{tab:dummytie_breaking}
\vspace*{0.4cm}
\begin{tabular}{lcccccc}
\toprule
 & \multicolumn{3}{c}{\textbf{Grade 8}} & \multicolumn{3}{c}{\textbf{Grade 10}}  \\
 \cmidrule(lr){2-4}  \cmidrule(lr){5-7}
 &  Mean & Min. & Max.  & Mean & Min. & Max.  \\
 & (1) & (2) & (3) & (4) & (5) & (6)  \\
\midrule

\multirow{2}{*}{Main Specification
(Eq. \ref{eq:main_specification})} &
8.468***	&	3.031***	&	3.582***	&	7.721***	&	3.082***	&	2.894***	\\
& (0.773)	&	(0.463)	&	(0.491)	&	(0.714)	&	(0.425)	&	(0.460)	\\
&\\
\multirow{2}{*}{Accounting for Ties
(Eq. \ref{eq:dummytie})} &
	
8.379***	&	6.750***	&	7.000***	&	7.564***	&	6.152***	&	6.247***	\\
& (0.820)	&	(0.761)	&	(0.747)	&	(0.762)	&	(0.702)	&	(0.702)	\\
\bottomrule 
\end{tabular} 
\begin{tablenotes} \textbf{Note:} Robust standard errors in parentheses. *** $p<0.01$, ** $p<0.05$, * $p<0.1$. Clustering at the middle school (Grade 8) or high school (Grade 10) level. This Table reports the coefficient on the visible rank $\hat{\beta}$ from Equations \ref{eq:main_specification} and \ref{eq:dummytie} using three different tie-breaking rules: assigning the Mean, Minimum or Maximum rank.
\end{tablenotes}
\end{threeparttable}
\end{table}

\begin{table}[H]\centering
\begin{threeparttable}[b]
\caption{Difference with Baseline Estimates - T-Tests}
\small
\label{tab:dummytie_breaking_pval}
\vspace*{0.4cm}
\begin{tabular}{lM{4em}M{4em}M{4em}M{4em}M{4em}M{4em}}
\toprule
 & \multicolumn{3}{c}{\textbf{Grade 8}} & \multicolumn{3}{c}{\textbf{Grade 10}}  \\
 \cmidrule(lr){2-4}  \cmidrule(lr){5-7}
 &  Mean & Min. & Max.  & Mean & Min. & Max.  \\
  & (1) & (2) & (3) &(4) &(5) &(6) \\
\midrule
Difference &
0.09	&	1.72	&	1.47	&	0.16	&	1.57	&	1.47	\\
P-value & 0.94	&	0.11	&	0.17	&	0.88	&	0.12	&	0.14	\\

\bottomrule 
\end{tabular} 
\begin{tablenotes} \textbf{Note:} The first row of this Table reports the difference between our baseline estimate $\hat{\beta}$ (i.e., from the standard specification \ref{eq:main_specification} using the Mean tie-breaking rule) and $\hat{\beta}$ from Equation \ref{eq:dummytie} using the three different tie-breaking rules. The second row reports the p-value of the difference.
\end{tablenotes}
\end{threeparttable}
\end{table}

\section{Testing the Weak Student Assumption}\label{sec:app_other_subset_classes}

In this Section, we look at how our assumption that missing students are the weakest affect our results (see Section \ref{sec:data_descstat}). Our main sample is comprised of classes whose coverage exceeds 90\%. The mean coverage in our main sample is 95.9\% and the median is 95.5\%, which means that we miss fewer than one student on average.  

We start by restricting the sample of classes to those with 100\% coverage so that we know that all students are observed and the rank is well-defined. 

\begin{table}[H]%
\caption{Rank Effect on Fully Covered Classes}%
 \label{tab:robcheck_fullcoverage} %
 \scriptsize \centering %
 \begin{threeparttable} %
\begin{tabular}{lM{6.8em}M{6.8em}}%
\vspace{0mm}\\%
 \toprule%
 \toprule
 & \multicolumn{2}{c}{Test Score in} \\
 \cmidrule(lr){2-3}
&Grade 8&Grade 10\\%
&(1)&(2)\\%
\midrule
Visible Rank&8.881***&6.452***\\%
&(1.302)&(1.187)\\%
Invisible Rank &{-}0.354&{-}1.385**\\%
&(0.627)&(0.581)\\%
& \\
Observations&269,784&269,784\\%
\bottomrule%
\bottomrule%
\end{tabular}%
\begin{tablenotes} \regfootnotesrobcheckcoveragefull \end{tablenotes}%
\end{threeparttable}%
\end{table}

Results, displayed in Table \ref{tab:robcheck_fullcoverage}, show that estimates are extremely close to those from our main sample (the difference between the two is not significant at the 10\% level).

We then test the assumption that students for whom we lack information are the weakest in their class. To see how previous results are affected, we start from the fully covered sample and we drop the lowest-performing students according to the size of the class: 0 for a size below 9, 1 for a size between 10 and 19, 2 for a size between 20 and 29 and 3 for a size of 30. Results are displayed in Table \ref{tab:robcheck_coverage_drop}.

\begin{table}[H]%
\caption{Testing the Weak Student Assumption}%
 \label{tab:robcheck_coverage_drop} %
 \scriptsize \centering %
 \begin{threeparttable} %
\begin{tabular}{lM{6.8em}M{6.8em}}%
\vspace{0mm}\\%
  \toprule%
 \toprule
 & \multicolumn{2}{c}{Performance in} \\
 \cmidrule(lr){2-3}
&Grade 8&Grade 10\\%
&(1)&(2)\\%
\midrule
Visible Rank&10.678***&5.941***\\%
&(1.502)&(1.381)\\%
Invisible Rank&{-}0.298&{-}1.185*\\%
&(0.654)&(0.609)\\%
& \\
Observations&241,410&241,410\\%
\bottomrule%
\end{tabular}%
\begin{tablenotes} \regfootnotesrobcheckcoveragedrop \end{tablenotes}%
\end{threeparttable}%
\end{table}

We see that both estimates are close to those from Table \ref{tab:robcheck_fullcoverage}. For neither grades are estimates statistically different at the 10\% level. Furthermore, the magnitude on the invisible rank are exceedingly close. Importantly, the invisible rank has a significant impact in grade 10, albeit of negligible magnitude, before and after dropping the weakest students. We can thus conclude that dropping the weakest students does not significantly alter the results that we would obtain had we observed everyone. 

\section{Why Does Exploiting Variations in the Ability Distribution of Peers Lead to Omitted Variable Bias?}\label{sec:app_toyexample}

\begin{figure}[H]
\centering
\begin{subfigure}[b]{0.45\textwidth}
\centering
(a) Above-the-Mean Abilities \\
\includegraphics[width=\textwidth]{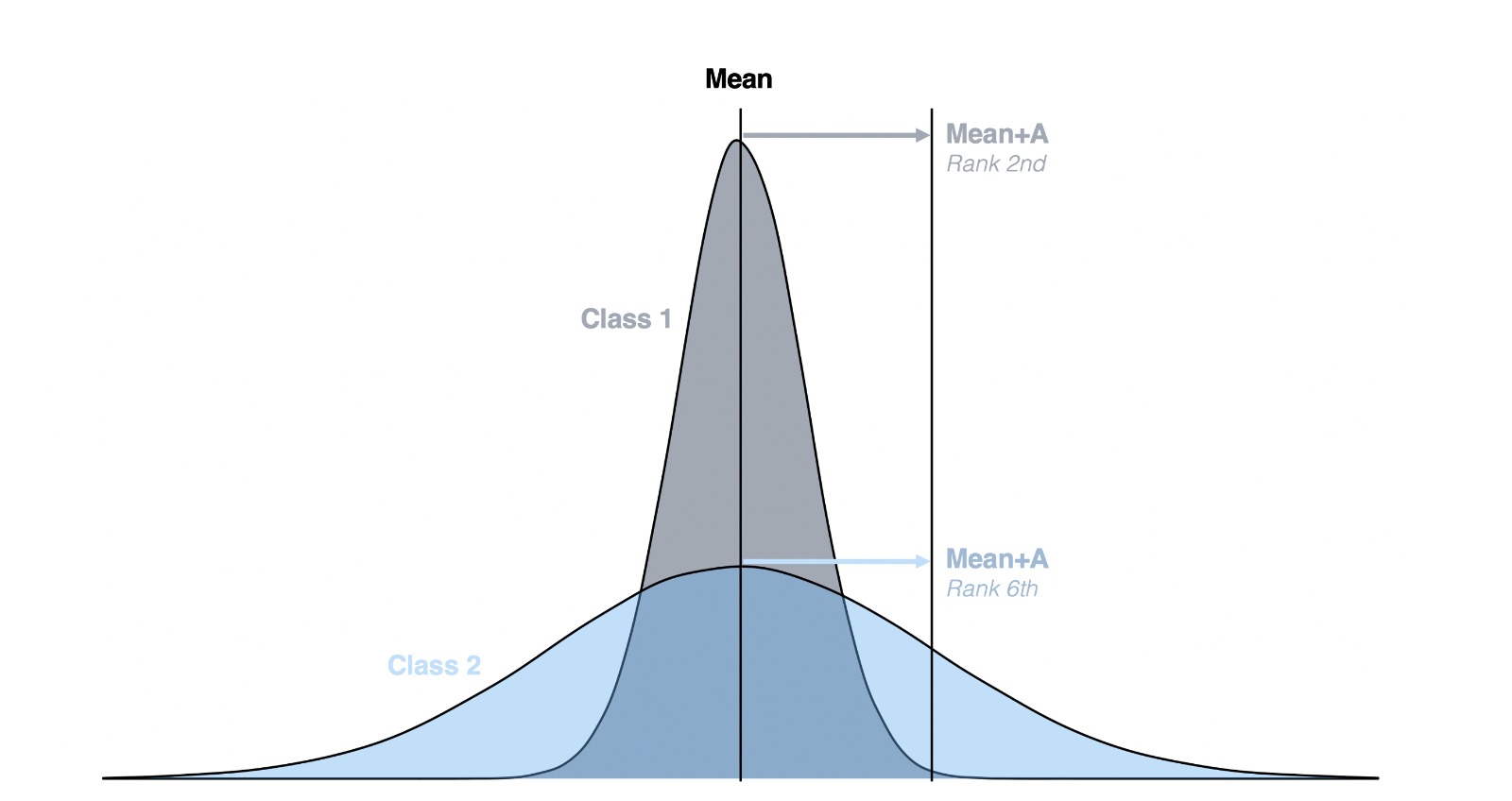}
\end{subfigure}
\begin{subfigure}[b]{.45\textwidth} 
\centering
(b) Below-the-Mean Abilities \\
\includegraphics[width=\textwidth]{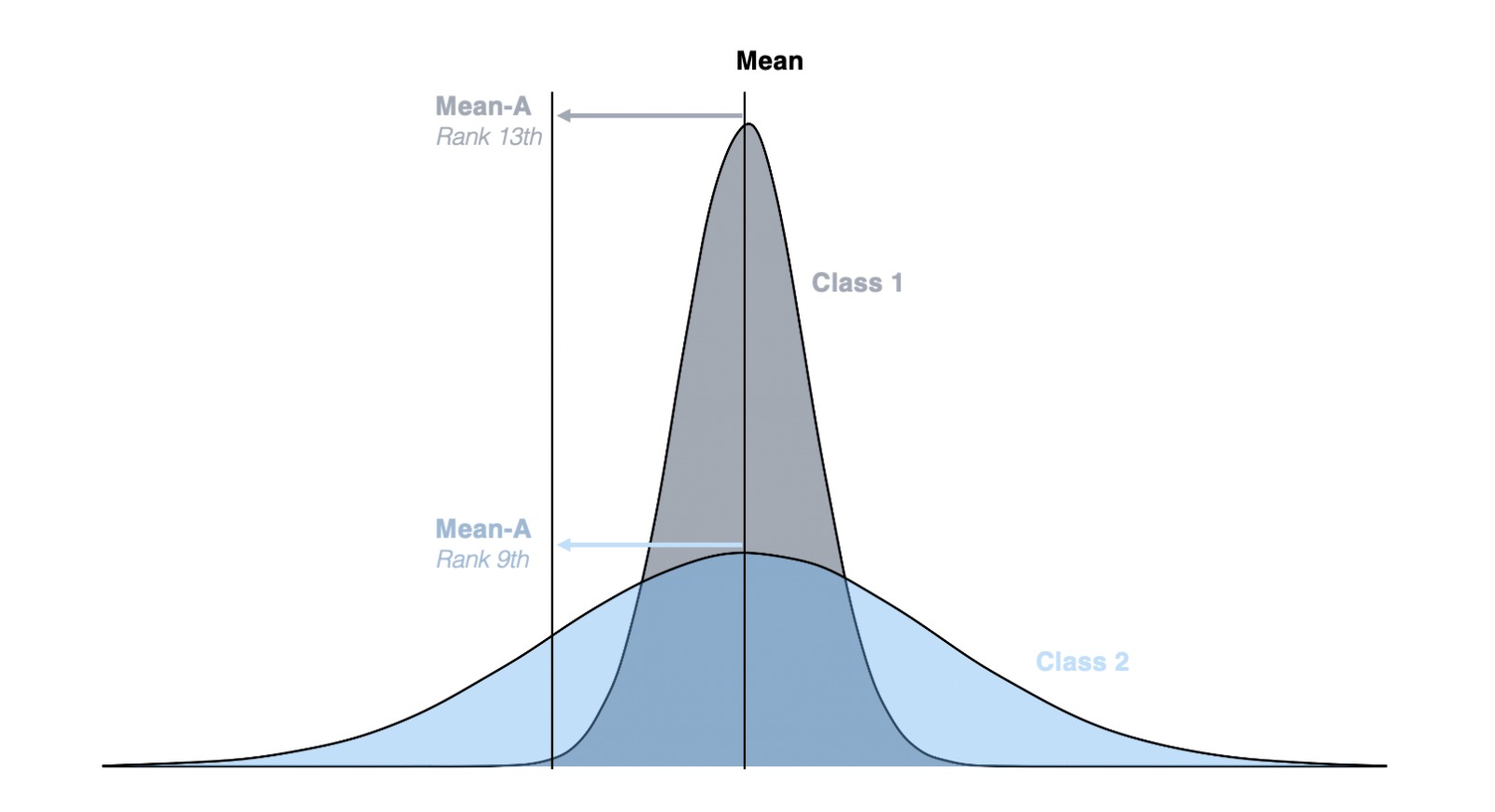}
\end{subfigure}
\caption{\textbf{Higher Order Peer Effects on Rank}}
\label{fig:higher_order_peer_effects}
\vspace{.2cm}
\footnotesize \begin{tabular}{p{14cm}}
\setstretch{1} \textbf{Note}: This Figure illustrates the example developed in Section \ref{sec:empirical_strategy}.
\vspace{.5cm}
\end{tabular}
\end{figure}

Consider, as a toy example, classes 1 and 2. Both have the same mean ability but the variance of 1 is lower than that of 2, as depicted in Figure \ref{fig:higher_order_peer_effects}. Consider students X and Y, respectively in classes 1 and 2, both with ability A points above the mean. Then, the rank of X is higher than that of Y (Figure \ref{fig:higher_order_peer_effects}(a)). On the contrary, if both have ability A points below the mean, X's rank is lower than Y's (Figure \ref{fig:higher_order_peer_effects}(b)). To avoid omitted variable bias, one must account for the effect of the interaction between the variance of the class ability distribution and a student's ability, as it impacts the rank differently depending on ability. However, this does not suffice: this problem arises with any moment of the distribution and can be tackled only by constraining the two distributions to be identical. But holding both individual ability $a_i$ and the class ability distribution $F_c$ fixed, effectively bars identification since this precludes any variation in $R^U_{ic}$ from $R^U_{ic}=g(a_i,F_c)$.

\section{Questionario di Contesto}\label{sec:app_questionnaire}

In this Section, we give the detail of the questions asked to students as part of the Questionnaire. They have to answer on a six-point Likert scale.

\subsection{Subject Interest}
``Let's talk about Italian/Math. Do you agree with the following sentences? \\
1.  Totally disagree; 2. Slightly agree; 3. Moderately agree; 4. Agree; 5. Strongly agree; 6. Totally agree.

Questions:
\begin{itemize}
    \item "In general, I enjoy learning Italian/Math"
\item  "I like to read Italian/Math books"
\item "I am happy to study Italian/Math"
\item "I am interested in learning Italian/Math well"
\item "I like to learn new topics in Italian/Math"	
\item "I can't wait to take Italian/Math lessons"
\end{itemize}

\subsection{Career Prospects}
``Thinking about your future, how true do you think these sentences are? \\
1. Totally false; 2. Slightly true; 3. Moderately true; 4. True; 5. Very True; 6. Totally true.

Questions:
\begin{itemize}
    \item "I will achieve the degree I want".
    \item "I will always have enough money to live".
    \item "In life I will be able to do what I want".
    \item "I'll be able to buy the things I want".
    \item "I will find a good job".
\end{itemize}

\subsection{Self-Confidence}
``Thinking about yourself, how much do you agree with the following sentences?'' \\
1. Totally disagree; 2. Slightly agree; 3. Moderately agree; 4. Agree; 5. Strongly agree; 6. Totally agree. 

Questions:
\begin{itemize}
    \item "I'm able to think fast".
    \item "I think I'm a nice guy".
    \item "In the face of obstacles I work harder".
    \item "I usually have good ideas".
    \item "I learn new things with ease".
    \item "I know how to make others understand my point of view".
\end{itemize}

\subsection{Peer Recognition}
``Thinking about your companions, how much do you agree with the following sentences?'' \\
1. Totally disagree; 2. Slightly agree; 3. Moderately agree; 4. Agree; 5. Strongly agree; 6. Totally agree.

Questions:
\begin{itemize}
    \item "I think my teammates enjoy working with me".
    \item "In class, I feel accepted".
    \item "I can trust my companions".
    \item "I have fun with my companions".
    \item "In school, I have many friends".
\end{itemize}

\subsection{Perception of the School System}

``Think about your experience in school. How true do you think these sentences are? \\ 
1. Totally false; 2. Slightly true; 3. Moderately true; 4. True; 5. Very True; 6. Totally true.

Questions:
\begin{itemize}
    \item "I want to stop going to school as soon as possible".
    \item "Going to school is an effort".
    \item "I feel fine at school".
    \item "I feel like I am wasting time at school".
    \item "At school, I get bored".
    \item "I have no reason to go to school".
    \item "At school I do interesting things".
\end{itemize}

\subsection{Parental Support}

``How much do you agree with the following sentences?'' \\
1. Totally disagree; 2. Slightly agree; 3. Moderately agree; 4. Agree; 5. Strongly agree; 6. Totally agree.

Questions:
\begin{itemize}
    \item "My parents are interested in what I do at school".
    \item  "My parents encourage me to commit to studies".
    \item  "My parents help me when I have difficulties at school".
    \item "My parents encourage me to be self-confident".
\end{itemize}

\end{appendix}

\end{document}